\pdfoutput=1
\documentclass[useAMS,usenatbib]{mn2e}
\usepackage{graphicx}
\usepackage{natbib} 
\usepackage{algorithmicx} 
\usepackage{algpseudocode}
\usepackage{amsmath}

\newcommand{\CH}{{\cal H}} 

\newcommand{\NCanAll}{2566} 

\newcommand{\NCanStrong}{663}

\newcommand{\NCanEROS}{1160}
\newcommand{\NCanMACHO}{2551}
\newcommand{\NMatchesMACHOEROS}{191}
\newcommand{\NMatchesStrongMACHOEROS}{131}
\newcommand{\NMatchesMACHOKPDW}{1148} 
\newcommand{\NMatchesMACHOKPDWHC}{491} 
\newcommand{\NunExistEROSinMACHODW}{332} 

\title[Quasar detection method in EROS-2 and MACHO LMC]{An improved quasar detection method in EROS-2 and MACHO LMC datasets}

 \author[K. Pichara, P. Protopapas, D.-W. Kim, J.-B. Marquette and P. Tisserand]{K. Pichara,$^{1,3}$
   P. Protopapas,$^{2,3}$ D.-W.Kim,$^{2,3}$ J.-B. Marquette,$^{4}$ and P. Tisserand$^{5}$\\
$^{1}$Computer Science Department, Pontificia Universidad Cat\'olica de Chile, Santiago, Chile.\\
$^{2}$Harvard-Smithsonian Center for Astrophysics, Cambridge, MA, USA\\
$^{3}$Institute for Applied Computational Science, Harvard University, Cambridge, MA, USA\\
$^{4}$UPMC-CNRS, UMR7095, Institut d'Astrophysique de Paris, F-75014, Paris, France\\
$^{5}$Research School of Astronomy and Astrophysics, Australian National University, Canberra, Australia}

\begin{document}

\date{}

\pagerange{\pageref{firstpage}--\pageref{lastpage}} \pubyear{2012}

\maketitle

\label{firstpage}

\begin{abstract}

We present a new classification method for quasar identification in the EROS-2 and MACHO datasets based on a boosted version of Random Forest classifier. 
We use a set of variability features including parameters of a continuous auto regressive model. We prove that continuous auto regressive parameters are very important discriminators in the classification process.
We create two training sets (one for EROS-2 and one for MACHO datasets) using known quasars found in the LMC. Our model's accuracy in both EROS-2 and MACHO training sets is  about 90\% precision  and  86\% recall, improving the state of the art models accuracy in quasar detection.
We apply  the model on the complete, including 28 million objects,  EROS-2 and MACHO LMC datasets, finding \NCanEROS{} and \NCanMACHO{}  candidates respectively. 
To further  validate our  list of  candidates, we crossmatched our list with a previous \NCanStrong{} known strong candidates
, getting 
74\% of matches for MACHO and 40\% in EROS.
The main difference on matching level is because  EROS-2 is a slightly shallower survey  which translates to significantly  lower signal-to-noise ratio lightcurves.

\end{abstract}

\begin{keywords}
Magellanic Clouds -- methods: data analysis -- quasars: general
\end{keywords}

\section{Introduction}

Given the immense amount of data being produced by current deep-sky surveys
such as Pan-STARRS \citep{Kaiser2002SPIE},  and future surveys such as LSST \citep{Matter:2007} and SkyMapper \citep{Keller:2007}, 
astronomy is facing new challenges on how to analyze {\em{big data}}
and thus on how to search or predict events/patterns of interest.

The size of the data has already exceeded the capability
 of manual examination  or the capability of standard  data analysis tools.
LSST will produce 15 terabytes of data per night,
which is even beyond the capacity of typical data storage today.

Thus in order to analyze such a huge amounts of data
and detect interesting events or patterns with minimum false positives,
innovative and novel data analysis methods are crucial
for the success of such surveys.

In our previous works \citep{Kim:2011,Kim:2012} we developed classification 
models for the selection of quasars from large photometric
databases using variability characteristics as the main discriminators. 
In particular we used a supervised classification model trained using 
a set of variability features calculated from MACHO lightcurves \citep{Alcock2000ApJ}.
We applied the trained model to the entire MACHO database
consisting of $\sim$40 million lightcurves and selected few thousands of quasar candidates. 
In this paper,
we present an improved classification model used to detect quasars on MACHO \citep{Alcock2000ApJ} and EROS-2 dataset \citep{Tisserand:2007}. The new model
which works over an extended set of variability features, 
substantially decreases false positive rate and increases efficiency.

The actual model improvement is a result of an improvement in the machine learning classification model and the lightcurve features we use. Machine learning classification methods have been very popular for many decades. These methods are data analysis models that learn to predict a categorical variable from a set of other variables (of any type).
Most known classification models are: decision trees \citep{Quinlan:1993}, naive Bayes \citep{Duda:Hart:1973}, Neural Networks \citep{Rumelhart:1986}, Support Vector Machines \citep{Cortes:1995} and Random Forest \citep{Breiman:2001}. There are some meta-models to improve classification results like Boosting methods \citet{Freund:1997} and Mixtures of Experts \citep{Jordan:1994}, among others. In general more recent classifiers are a result of research focused on building models able to search for patterns within high dimensional datasets, where the combinatorial number of possible projections of data is large.

Many machine learning classifiers have been applied to the analysis of astronomical data in particular to classify transients and variable stars from time series data \citep{Bloom2:2011,  Richards:2011, Bloom:2011, Debosscher:2007, Wachman:2009, Wang:2010, Kim:2011, Kim2011ApJ}.  \citep{Wang:2010}  proposed an algorithm to fit phase-shifted periodic time series using a mixture of Gaussian processes. \citep{Debosscher:2007}  used many machine learning classifiers to learn a model that classifies variable stars in a sample from Hipparcos and OGLE databases.  \citep{Richards:2011} used Random Forest classifier to classify between pulsational variables and eclipsing systems used in Milky Way tomography. In \citet{Bloom2:2011} they used machine learning algorithms to classify transients and variable stars from the Palomar Transient Factory (PTF) survey \citep{Raw:2009}. In  \citet{Wachman:2009} they used cross correlation as a phase invariant feature to be used as a similarity indicator in a kernel function.

In this work we used a Random Forest classifier \citep{Breiman:2001} boosted with the AdaBoost algorithm \citep{Freund:1997}. The Random Forest classifier comes from the well-known decision tree model \citep{Quinlan:1993} and Baggging techniques \citep{Breiman:1996}, where the model randomly explores several subsets of features while analyzing samples of training data. This model performs very well in many machine learning domains \citep{Breiman:2001}. AdaBoost algorithm \citep{Freund:1997} is a boosting technique which fits a sequence of classification models (in this case a sequence of many Random Forests) to different subsets of data objects (in our case lightcurves),
 generating a mixture of classifiers each one specialized in smaller areas of the feature space. We call these classifiers as ``weak classifiers" or ``simpler classifiers". This is a nice property for quasar classification, given that there are only a few known training quasars compared with the amount of non-quasars lightcurves. Having some weak classifiers that take care of some areas with no training quasars helps to filter out many non-quasars, while other specialized classifiers perform well near the quasar areas in the feature space.

Besides improving the classification model, we added new features as descriptors of lightcurves. These features correspond to the parameters of the continuous auto regressive (CAR(1)) model \citep{Belcher:1994}  fitted to the lightcurves. Previous work shows that describing quasars using CAR(1) fitting parameters gives suitable results to differentiate them from other classes of lightcurves \citep{Kelly:2009}. In this work they did not use machine learning classifiers to automatically detect quasars, they use CAR(1) model to fit 100 quasars lightcurves in order to find correlations between CAR(1) parameters and luminosity characteristics. 

In our work we show that by adding CAR(1) features to our previous set of features (used in \citet{Kim:2011}), we can learn more accurate models for quasar detection. Given that our model is built  to find quasars over dozens of millions of stars, we need to be very efficient in the estimation of the optimal parameters in order to make the process feasible within a considerable amount of time. Unfortunately,  methods such as Metropolis Hastings or Gibbs Sampling are not suitable for our purposes, given the computational cost they involve. 

To gain efficiency, we reduce the problem by approximating one of the parameters (the mean value of the light curve) and optimizing the remaining parameters (the amplitude and time scale of the variability) using a multidimensional unconstrained nonlinear minimization \citep{Nelder:1965}. Once we get the optimal parameters we use them as features of the object corresponding to the lightcurve. Besides the CAR(1) features we also used time series features as in our previous work \citep{Kim2011ApJ}, in section \ref{sec:Features} we give details about all the features we extracted.

  To check the fitting accuracy of our model we first calculate the training accuracy of our classifier using 10-fold cross validation over a training set, which consists of about six thousand known light curves corresponding to different kinds of variable stars, non-variable stars and confirmed quasars; one set corresponding to the MACHO database and another to the EROS-2 database. In the MACHO case we substantially improve our training accuracy compared with our previous work \citep{Kim2011ApJ}, increasing 14.3\% in precision and 3.6\% in recall for the MACHO database. In  EROS-2 training database, we get about the same training efficiency as in the MACHO case but we could not  compare to our previous work because this is the first time we attempt to classify in EROS-2 database. As an extra test for our candidates, we crossmatch them with the previous set of strong candidates found in \citet{Kim2011ApJ}, details are presented in section \ref{sec:Results}.\\

Using parallel computing we decrease the processing time 
to allow us to select quasar candidates from the entire database within three days.
Note that the data analysis schema used in this work
can be applied to any of the ongoing and future synoptic sky surveys such
as Pan-STARRS, LSST, and SkyMapper, among others. \footnote{Our main computer resource is \href{http://hptc.fas.harvard.edu/}{The Odyssey cluster} supported by the FAS Research 
Computing Group at \href{http://harvard.edu/}{Harvard}. }

If confirmed the selected quasars from the MACHO database will
provide critical information for galaxy evolution, black hole growth, large scale structure, etc. 
(\citealt{Heckman2004ApJ, Bower2006MNRAS, Trichas2009MNRAS, Trichas2010MNRAS}).
Moreover the resulting quasar lightcurves will be a valuable dataset for
quasar time variability studies, (e.g. time scale, blackhole mass, type i and ii variability) since MACHO and EROS lightcurves are well-sampled over 7.4 years \citep{Alcock2000ApJ}. 

    The paper is organized as follows, in section \ref{sec:EROS2} we present details about EROS-2 database, in section \ref{sec:Methodology} we describe in details the classification model we use, including the Random Forest Model and AdaBoost, in section \ref{sec:Features} we describe the features we use to describe the lightcurves, in section \ref{sec:Results} we describe the experimental results for the MACHO and EROS-2 dataset.
  


\section{EROS-2 Dataset}
\label{sec:EROS2}
The EROS-2 collaboration made use of the MARLY telescope, a one meter diameter Ritchey-Chr\'etien ($f$/5.14) instrument dedicated to the
survey. It was operated between July 1996 and March 2003 at La Silla Observatory (ESO, Chile). It was equipped with two wide angle CCD
cameras which are located behind a dichroic beam-splitter. Each camera is a mosaic of 8 CCDs, 2 along right ascension and 4 along
declination. Each CCD has $ 2048 \times 2048 $ pixels of $15 \times 15~\mu$m$^{2}$ individual size, corresponding to a $ 0.6 \times
0.6~$arcsec$^{2}$ pixel surface on the sky. The size of the field of view is $ 0.7^{\circ} $ along right ascension and $ 1.4^{\circ} $
along declination. The dichroic beam-splitter allowed simultaneous imaging in two broad non-standard passbands, $ B_{E} $ in the range
4200-7200  (the so-called \lq \lq blue" channel), and $ R_{E} $ in the range 6200-9200  (the so-called \lq \lq red" channel). The blue filter is
intermediate between the standard $ V $ and $ R $ standard passbands, while the red filter is analogous to $ I_{c} $. The normalized
transmission curve of these filters, compared to standard ones, is given by \citet{2004HamadachePhD}\footnote{Available at URL:
\texttt{http://tel.archives-ouvertes.fr}} on Fig. 3.3. \citet{Tisserand:2007} give in Eq. (4) the equations to transform EROS-2
magnitudes into $ V $ and $ I_{c} $ ones within an accuracy of 0.1 magnitude. 



The light curves of individual stars were constructed from fixed positions on templates using PEIDA, a software specifically developed
for the photometry of EROS 2 images \citep{1996VA.....40..519A}. The nomenclature of objects is
defined as in \citet{2002A&A...389..149D}.

\section{Methodology}
\label{sec:Methodology}

 To train a model that learns to detect quasars, we propose to use a combination of classifiers. Combination of multiple classifiers was first proposed by  \citet{Suen:1992}. In that work, they  proved that combining multiple classifiers overcome many of the individual classifiers limitations. In many pattern recognition problems, such as character recognition, handwritten text recognition and face recognition \citep{Zhao:2003, Plamondon:2000}, combination of multiple classifiers obtain much better classification performance. One effective way to combine classifiers is the AdaBoost algorithm, proposed in  \citet{Freund:1997}. 
 
 The AdaBoost algorithm consists of a set of base classifiers that are trained sequentially, such that each classifier is trained on the instances where the previous classifier obtained a bad performance (learn what your partners could not learn). In \citet{Freund:1997}, they show that if the training set used for each classifier depends on the goodness of fit of the previous classifier, then the performance of the whole system improves. To make that the base classifiers focus on different subsets of the training set, we assign weights to training data instances. The lower the weight for an instance, the less the classifier focuses on it (see section \ref{sec:adaboost} for further details).  
  
   One of the advantages of boosting methods is that after the model fitting phase is completed, each of the base classifiers become an expert in some subset of data objects. This is one of the main reasons that motivate us to use a previous boosting step. Given that we have a very small amount of known quasars in our training set compared with the amount of non quasars, training a set of base classifiers that just learn how to filter out some of the non quasars would be very helpful for the next base classifier used in the sequential process. We now present a detailed description of the boosting method we use in this work, the AdaBoost algorithm \citep{Freund:1997}.
  
  \subsection{AdaBoost Algorithm} \label{sec:adaboost}

AdaBoost, short for adaptive boosting, is a machine learning algorithm proposed by  Freund and Schapire \citep{Freund:1997}. It is a meta-algorithm because it combines many learning  algorithms to perform classification. AdaBoost is adaptive in the sense that subsequent classifiers built are tweaked in favor of those instances misclassified by previous classifiers. Although AdaBoost is sensitive to noisy data and outliers, it is less susceptible to overfitting \citep{Dietterich:1995} than most learning algorithms.

In the context of lightcurve-classification, suppose we have a training (labeled) set of $n$ lightcurves and $q$ features describing each lightcurve. Each lightcurve in the training set has a known given label (e.g. quasar or non-quasar). Let $\left[\mathbf{x}_{1},\ldots,\mathbf{x}_{n} \right]$ 
be a set of $n$ descriptors where each $\mathbf{x}_{i} \quad i \in [1 \ldots n]$ is a vector associated to the lightcurve $i$ where its descriptor (features) values are $\{x_{i1},\ldots,x_{iq}\}$ where $q$ is the number of features.  Let $\{y_{1},\ldots,y_{n}\}$ be the labels such that  $y_{i} = 1$ if the lightcurve $i$ is a quasar and $y_{i} = -1$ otherwise.

Let $H$ be the set of $m$ classifiers $\{h_1,\ldots,h_m\}$, where $h_i: X \rightarrow Y$  and $D^{(t)}$ be the distribution of weights on classifiers at iteration $t$. Define $m$ to be the number of classifiers and a constant $T$  to be the number of times to iterate in the AdaBoost algorithm. 

\vspace{1cm}

\noindent\textbf{Initialization:}
\begin{algorithmic}
\State $X = \left[ x_1, x_2, \ldots, x_n\right]$
\State $Y = \left[ y_1, y_2, \ldots, y_n \right]$
\State $D^{(1)} = \left[ d^{(1)}_1, d^{(1)}_2, \ldots, d^{(1)}_n\right] := \left[ \frac{1}{n}, \frac{1}{n}, \ldots, \frac{1}{n} \right]$
\State $T \leq n$
\end{algorithmic}

\vspace{.15in}

\noindent\textbf{Algorithm:}
\begin{algorithmic}
\For{$t = 1$ to T} 
\For{$j = 1$ to m} 
\State $\epsilon_j := \sum_{i = 1}^n d^{(t)}_i \, (1-\delta_{y_i,h_j(x_i)}) $   

\EndFor
\State $\epsilon_t := \text{min} \,\epsilon_j$
\If{$\epsilon_t \geq 0.5$} \State \textbf{break} \EndIf
\State $h_t := \underset{h_j \in H} {\text{argmin}} \, \{ \epsilon_j \}$ 
\State $\alpha_t := \frac{1}{2}\text{ln}((1 - \epsilon_t)/\epsilon_t)$
\For{$i= 1$ to n} 
\State $d^{(t+1)}_i := d^{(t)}_i \, \text{exp}(-\alpha_t  \, y_i \,  h_t(x_i))/Z_t$
\EndFor
\EndFor
\State $\CH(X) := \left[ \CH(x_1), \CH(x_2), \ldots, \CH(x_n) \right]$, such that \[\CH(x_i) = \text{sign}\, \bigg(\sum_{t=1}^T \alpha_t \, h_t(x_i)\bigg)\].
\end{algorithmic}

\noindent\textbf{Notes:} 
\vspace{-0.2cm}\begin{itemize}
\item $\delta_{i,j}$ is the  Kronecker delta.
\item $Z_t$ is a normalization factor  \[ Z_t =\sum_{i=1}^n d^{(t)}_i \exp(-\alpha_t \, y_i \, h_t(x_i)\]\\
The equation to update the classifier weight distribution is constructed so that $-\alpha \, y_i \, h_t(x_i) < 1$ when $y_i = h_t(x_i)$ and $-\alpha\, y_i \, h_t(x_i) > 1$ when $y_i \neq h_t(x_i)$. Thus, after selecting an optimal classifier $h_t$,  for the distribution $D_t$, the objects $x_i$ that classifier $h_t$ classified correctly are given less weight and those that it identified incorrectly are given more weight. Hence, when the algorithm proceeds to test the classifiers on $D^{(t+1)}$, it is more likely to select a classifier that better classifies the objects that $h_t$ missed. Adaboost minimizes the training error (exponentially fast) if each weak classiÞer performs better than random guessing ($\epsilon_t < 0.5$).

\end{itemize}

  The base classifier we used in this work is the Random Forest classifier \citep{Breiman:2001}, a very strong classifier that has shown very good results in many different domains. The following section shows details about the Random Forest classifier.
 
\subsection{Random Forest Classifier}

Random Forests (RF) is a popular and very efficient algorithm   based on decision tree models \citep{Quinlan:1993} and Bagging for classification problems \citep{Breiman:1996, Breiman:2001}.
 It belongs to the family of ensemble methods,
appearing in machine learning literature at the end of nineties \citep{Dietterich:2000} and has been used recently in the astronomical journals \citep{Carliles:2010,Richards:2011}.
The process of training or building a Random Forest given training data
is as follows:\\

\begin{itemize}
  \item Let $P$ be the number of trees in the Forest and $F$ the number of features on each tree, both values are model parameters.
  \item Build $P$ sets of $n$ samples taken with replacement from the training set; this is called bagging. Note that each of the $P$ bags has the same number of elements from the training set but less different examples, given that the samples are taken with replacement.
   \item For each of the $P$ sets, train a decision tree using a random sample of $F$ features from the set of $q$ possible features.         
\end{itemize}

  The Random Forest classifier creates many linear separators inside many feature-subsets until it gets suitable separations between objects from different classes. Linear separations come from each decision tree, each of the feature-subsets come from the random feature selection process on each tree. The bagging procedure is very useful to estimate the error of the classifier during the training process. This error can be estimated using out-of-the-bag procedure, which means, \lq \lq evaluate the performance of each tree using the objects not selected in the bag which belong to the tree" (see \citet{Breiman:2001} for further details). 

  After training the Random Forest, to classify a new unknown lightcurve descriptor, 
  one uses each of the decision trees already trained with the Random Forest to classify the new unknown instance
and the final decision is the most voted class among the set of $P$ decision trees (see \citet{Breiman:2001} for more details).  
 In \citet{Breiman:2001} they show that as the number of trees tend to infinity the classification error of the RF becomes bounded and the classifier does not overfit the data.

\section{Feature Extraction} 
\label{sec:Features}
 We extracted 14 features per each band for each lightcurve. Those features correspond to 11 time series features used in our previous work \citep{Kim2011ApJ} and 3 features corresponding to the CAR(1) process. 
 
 \subsection{Time Series features}
 Here we very briefly summarize  the 11 time series features used in our previous work \citep{Kim2011ApJ}. 
 
 \begin{itemize}  
  \item  $N_{above},N_{below}$: Is the number of points above/below the upper/lower bound line calculated as points that are $\pm4 \sigma$ over the average of the autocorrelation functions.
  

%
%
   
   \item   Stetson $K_{AC}$: Is the variability index derived based on the autocorrelation function of each lightcurve \citep{Stetson1996PASP}.  
   


   
   \item  $R_{cs}$: Is the range of the cumulative sums (starting from 1 to the number of observations) of each lightcurve \citep{Ellaway1978}. 
   
%

   \item  $\sigma / \bar{m}$:  The ratio of the standard deviation, $\sigma$, to the mean magnitude, $\bar{m}$.

   \item  Period and Period S/N: Using  Lomb-Scargle algorithm \citep{Lomb1976ApSS, Scargle1982ApJ} we 
   used the period with the highest value in the periodiogram along with the signal to noise of the best period.
   

   
   \item  Stetson $L$:  Is a variability index \citep{Stetson1996PASP} that describes the synchronous variability of different bands. 
   
%
%
%
%
%
   \item  $\eta$:  Is the ratio of the mean of the square of successive differences to the variance of data points. 
   
%
%
     
   \item  $B-R$:  Average color for each lightcurve
   
%
%
%
   \item  $Con$:  Is the number of three consecutive data points that are brighter or fainter than 2$\sigma$ and normalized the number by $N-2$. 
                                     
\end{itemize}

 \subsection{Continuous Auto Regressive Process Features}
 
 We use continuous time auto regressive model (CAR(1)) to model irregular sampled time series in MACHO and EROS-2 
 lightcurves. CAR(1) process has three parameters, it provides a natural and consistent way of estimating a characteristic time scale and variance of lightcurves. CAR(1) process is described be the following stochastic differential equation \citep{Brockwell:2002}
   
\begin{eqnarray}
  dX(t)  &=  & -\frac{1}{\tau}X(t)dt +  \sigma_C \sqrt{dt} \, \epsilon(t) + b \,dt,  \\ 
    & & \rm{for} \quad  \tau,\sigma_C,t \geq 0 \nonumber
\end{eqnarray}

\noindent where the mean value of  the lightcurve $X(t)$ is $b \, \tau$ and the variance is $\frac{\tau \, \sigma_{C}^2}{2}$. 
$\tau$ is the relaxation time of the process $X(t)$, it can be interpreted as describing the variability amplitude of the time series.
$\sigma_C$ can be interpreted 
as describing the variability of the time series on time scales shorter than $\tau$. $\epsilon(t)$ is a white noise process with zero mean and variance equal to one. 
 The likelihood function of a CAR(1) model for a lightcurve with observations  $\mathbf{x} = \{x_{1}, \ldots , x_{n}\}$ observed at times $\{ t_{1}, \ldots , t_{n}\}$ with measurement error variances
$\{ \delta^2_{1}, \ldots , \delta^2_{n} \}$ is: \\

\begin{eqnarray}
  p(\mathbf{x} | b , \sigma_C,\tau) &= &  \prod_{i=1}^{n} \frac{1}{ [2 \pi (\Omega_i + \delta^2_i)]^{1/2}}  \exp{ \left\{  -\frac{1}{2} \frac{(\hat{x}_i - x^{*}_{i} )^2}{\Omega_i + \delta_i^2}  \right\} }\\ 
                                             x^{*}_{i}     &=&  x_{i} - b \, \tau\\
                                             \hat{x}_0 &=& 0\\
                                             \Omega_0 &=& \frac{\tau \sigma_{C}^2}{2}\\
                                             \hat{x}_i &=& a_i   \hat{x}_{i-1} + \frac{a_i \Omega_{i-1}}{\Omega_{i-1} + \delta^2_{i-1}} (x^{*}_{i-1} + \hat{x}_{i-1})\\
                                             \Omega_i &=& \Omega_0 (1-a_i^2)  \nonumber \\ 
                                              & & \quad \quad\quad +  a_i^2 \Omega_{i-1} \left(1 - \frac{\Omega_{i-1}}{\Omega_{i-1} + \delta^2_{i-1}} \right)\\
                                             a_i &=& e^{-(t_i - t _{i-1})/\tau}
\end{eqnarray}  
 
  To find the optimal parameters we maximize the likelihood with respect to $\sigma_C$, $b$ and $\tau$. Given that the likelihood does not have an analytical solution, we can solve it with a statistical  sampling method such Metropolis Hastings \citep{Metropolis:1953}. Given that we extract features for all the lightcurves in EROS-2 and MACHO datasets (about 28 and 40 millions of stars respectively), performing a statistical sampling process to determine the optimal parameters would be feasible only in cases where stable solutions are found in a reasonable amount of time. We consider that less than 3 seconds is reasonable given our hardware resources. Unfortunately we could not get stable solutions considering that restriction.  
  To overcome this situation we simplify the optimization problem by reducing the number of parameters to be estimated. Instead of estimating 
  $\sigma_C$, $b$ and $\tau$, we just estimate $\sigma_C$ and $\tau$  and then we calculate $b$ as the mean magnitude of the lightcurve divided by $\tau$. To check that this estimation works well, we use a sample of 250 lightcurves  and compare the reduced Chi-square error using two and three parameters optimization, getting differences smaller than 2.5\% in average.  
    
\begin{figure}
\includegraphics[width=10cm,height=4.9cm] {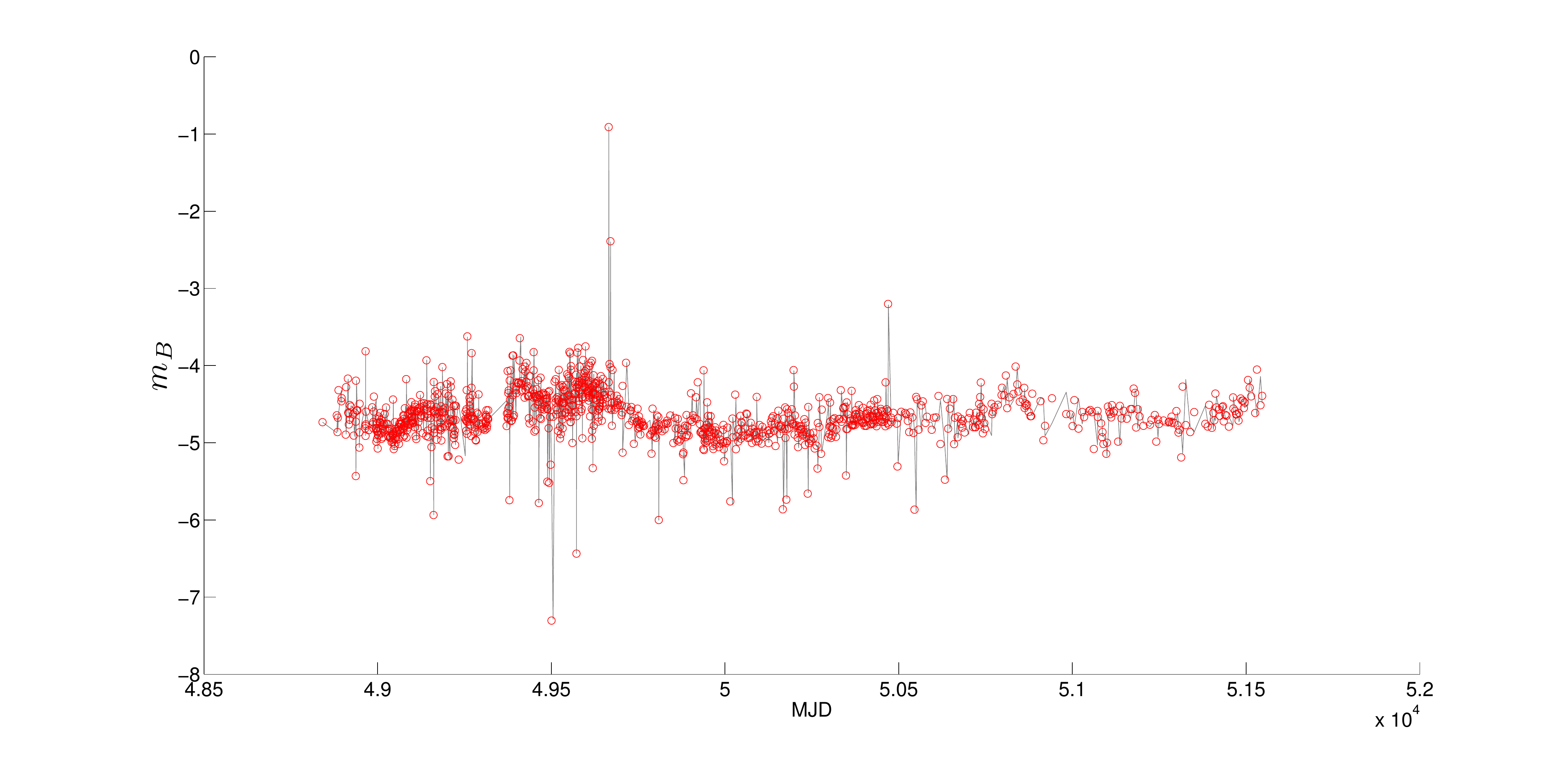}
\includegraphics[width=10cm,height=4.9cm] {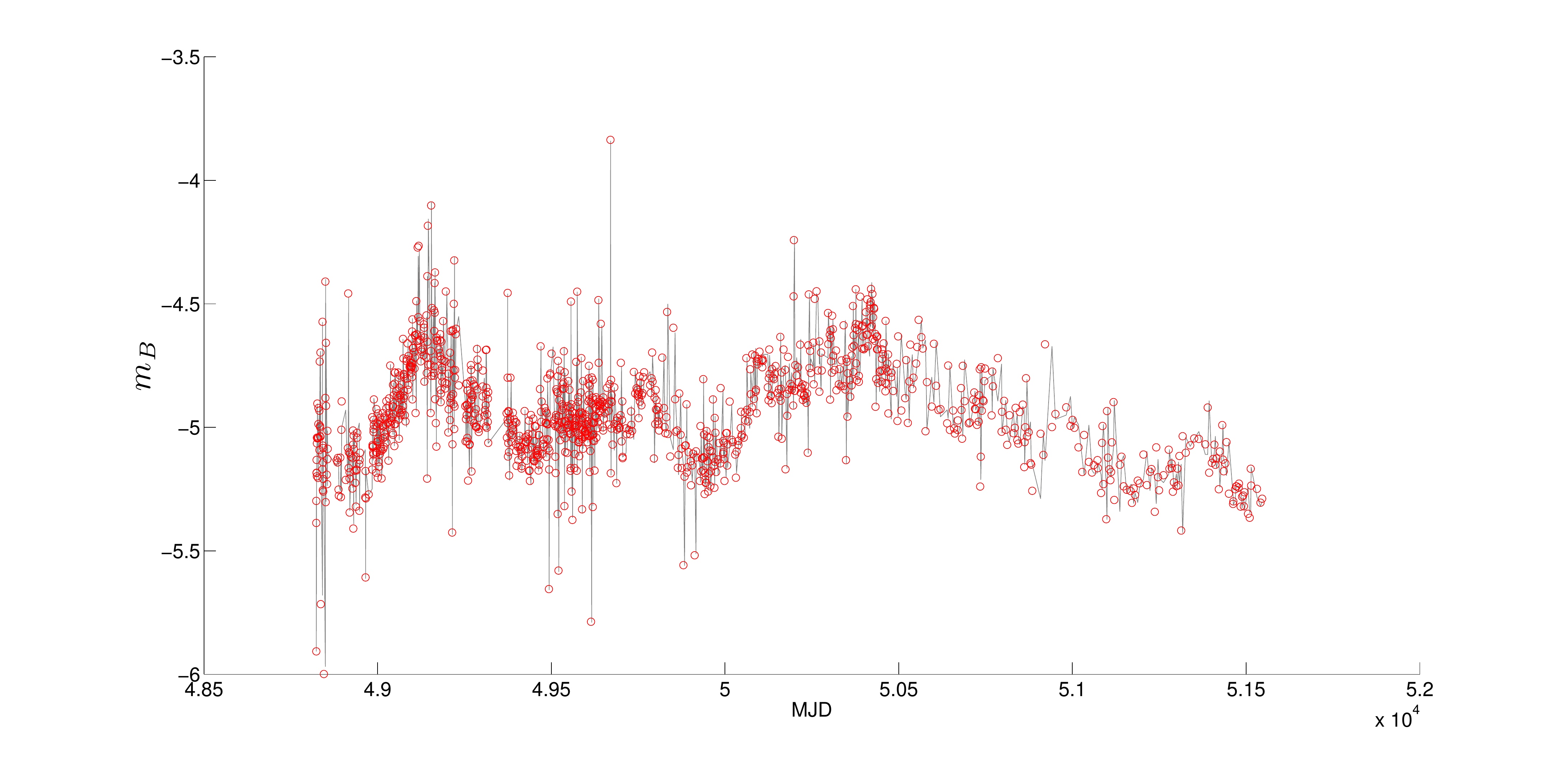}
\includegraphics[width=10cm,height=4.9cm] {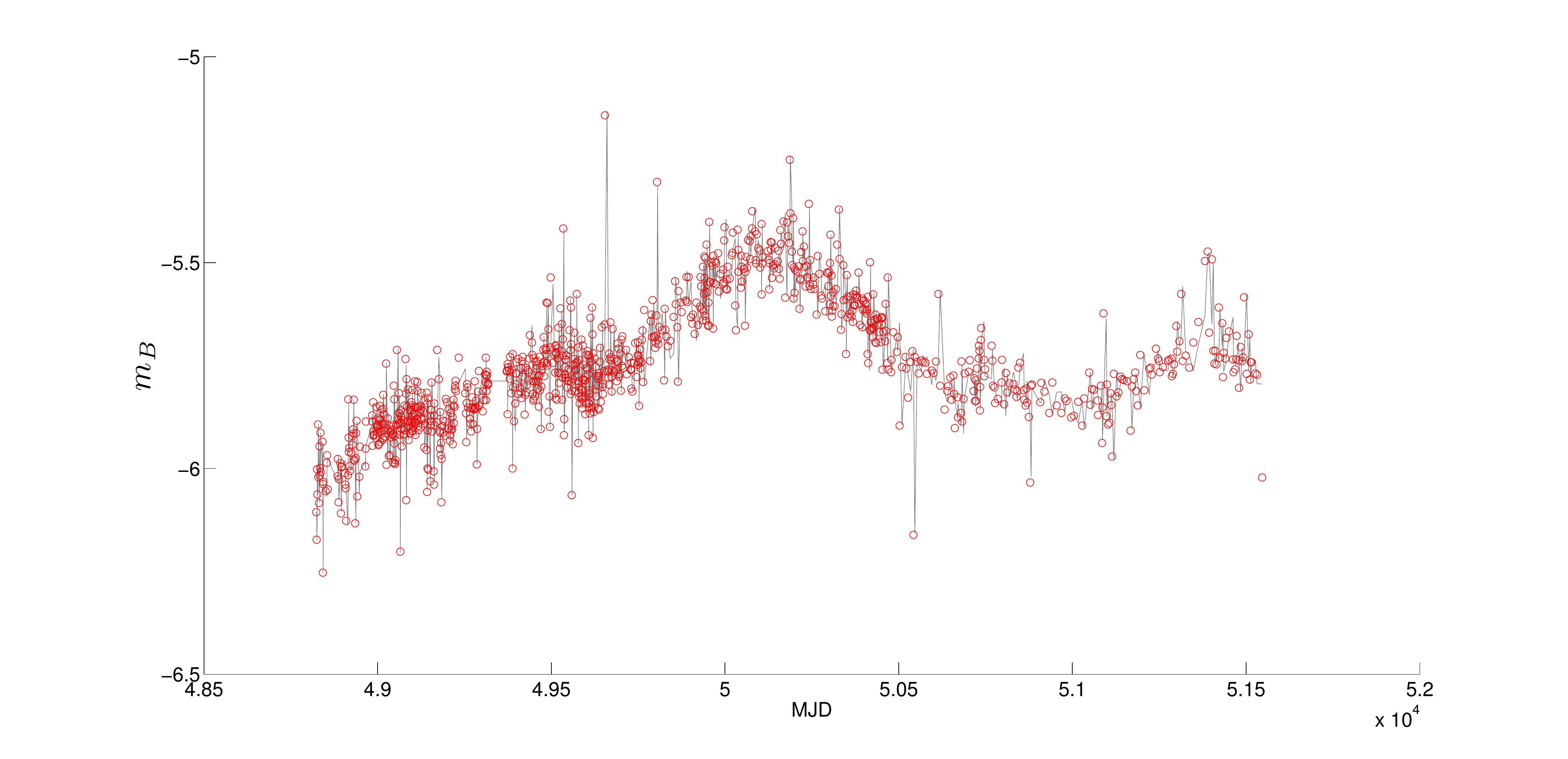}
\includegraphics[width=10cm,height=4.9cm] {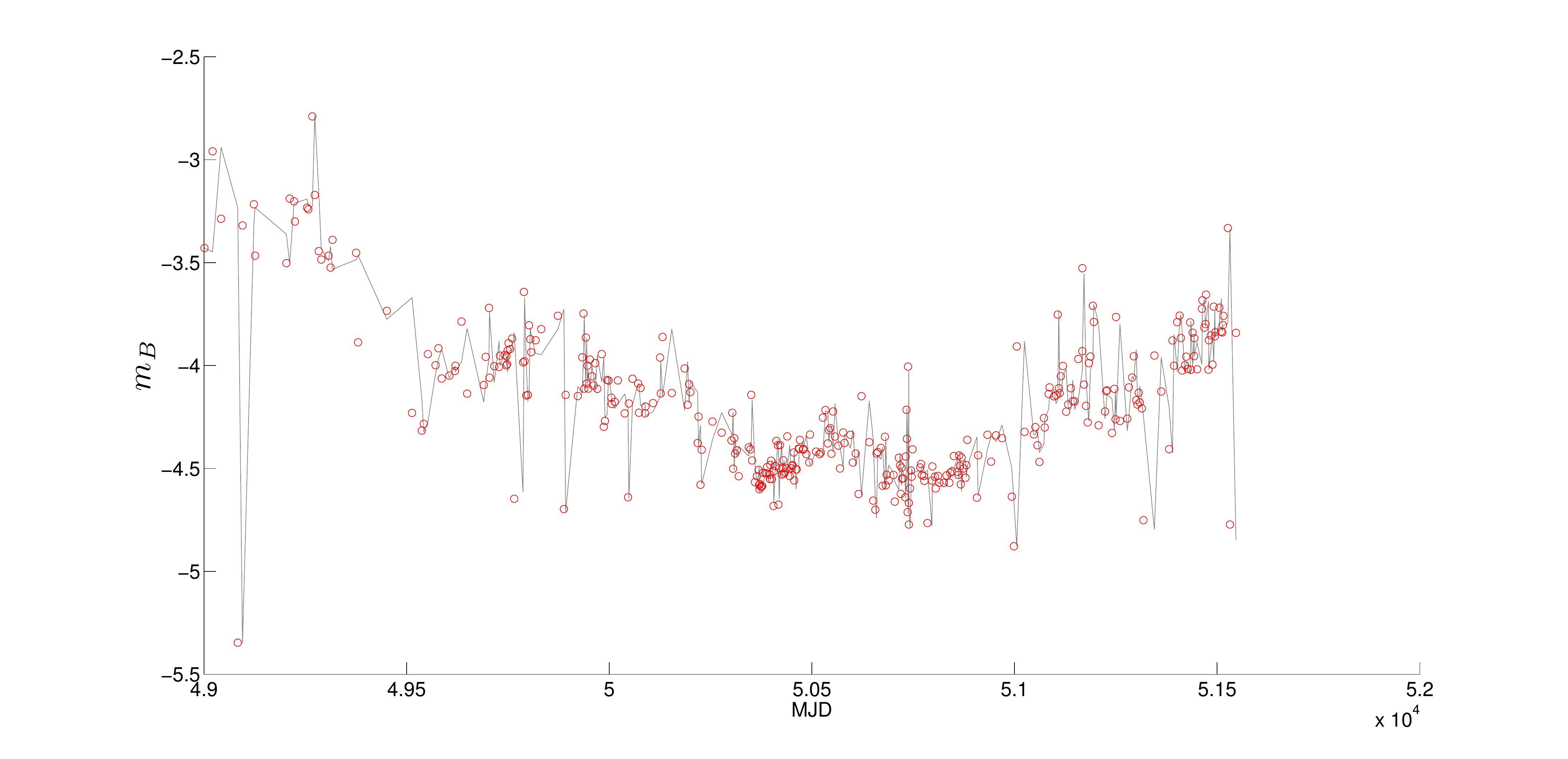}
\caption{Quasar lightcurves (red circles) fitted with optimal CAR(1) model (gray lines) using Nelder-Mead.}
\label{Fig:QSO_fit}
\end{figure}

  This approximation allows us to perform a two dimensional optimization which can be solved with a regular numerical method in less than one second per lightcurve. We used the Nelder-Mead multidimensional unconstrained nonlinear optimization \citep{Nelder:1965} to find the optimal parameters. Figure \ref{Fig:QSO_fit} shows the fitting of three quasar lightcurves with the resulting CAR(1) coefficients  using the Nelder-Mead algorithm.   
 Note that instead of using $b$ directly as a feature, we use the mean magnitude of the lightcurve ($\overline{m}$), in order to have a cleaner feature ($b$ is calculated from $\tau$, which is already used as a feature).

\section{QSO candidates on EROS-2 and MACHO datasets} \label{sec:Results} 
 
 \subsection{EROS-2 dataset}
  To train a model able to find quasars in EROS-2 we create a training set composed of 65 known quasars, 67 Be stars, 330 Long Periodic stars, 5829 non-variable stars, 1727 RR Lyrae, 406 Cepheids, and 488 EB stars. We get these stars cross-matching the EROS-2 dataset with MACHO known stars using positional matching with 3 arcsec of accuracy. We extracted features in bands R and B. Figures \ref{Fig:EROS_Train_Features_1} and \ref{Fig:EROS_Train_Features_2} show projections of the training set on different sets of features containing CAR(1) features. In many cases is easy to get a natural separation between quasars and the variable stars, but usually quasars overlap many of the non variable stars (ex. $\sigma_C$ with B $-$ R, $\sigma_C$ with $\tau$, $\overline{m}$ with $\tau$). Fortunately, there are many projections where quasars and non-variable stars are mostly separated, (ex. $\sigma_C$ with $Con$, $\sigma_C$ with $\overline{m}$,  $\sigma_C$ with Stetson $K_{AC}$ , $\tau$ with $R_{cs}$, $\tau$ with Stetson $K_{AC}$.)

\newcommand{\anchoFeats}{0.48}

\begin{figure*}
 \fbox{\includegraphics[width=0.8\textwidth]{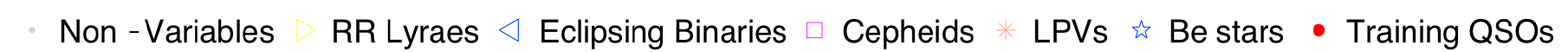}} 
  \begin{minipage}[b]{\anchoFeats \textwidth}
    \centering
    \includegraphics[width=7cm]{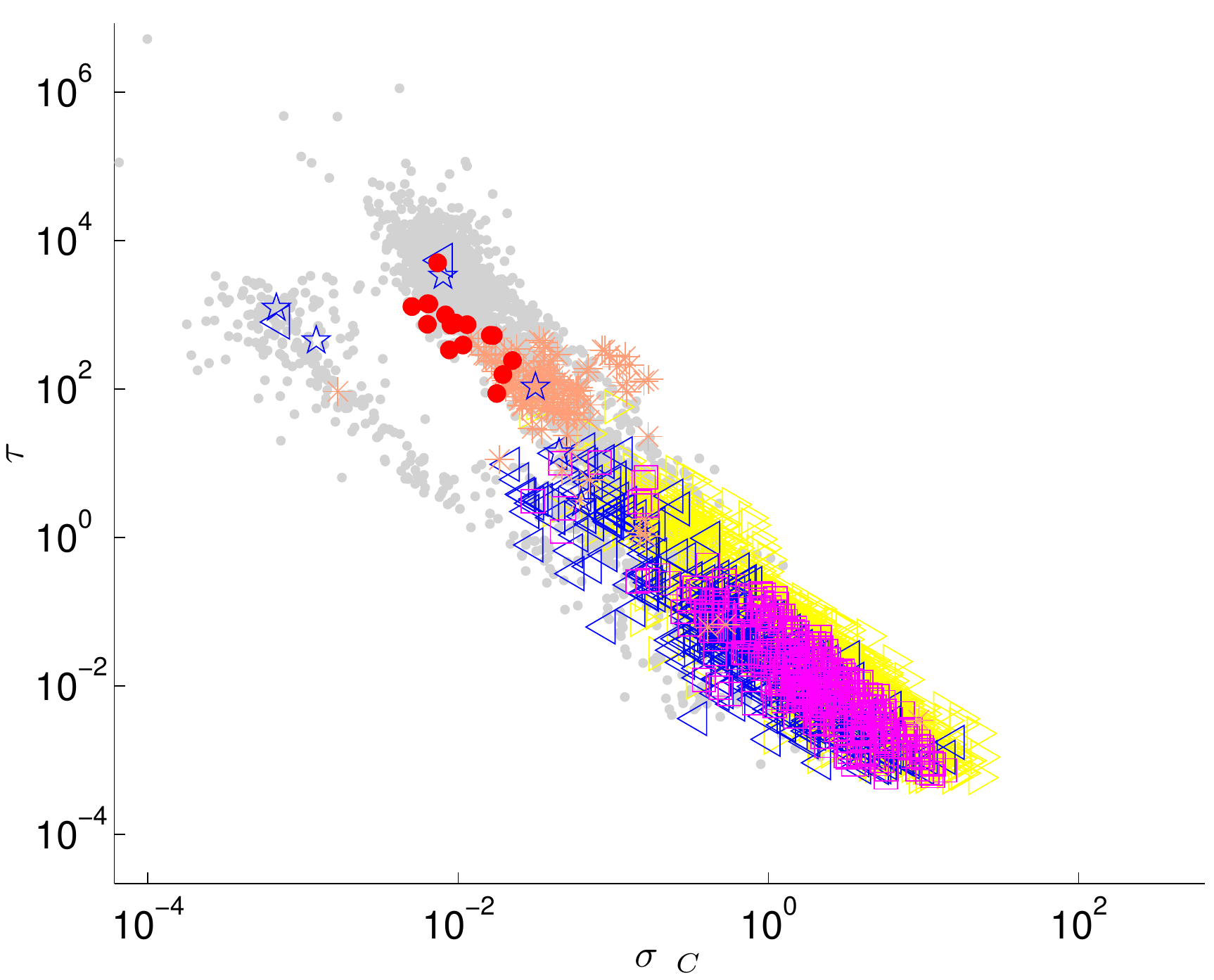} 
  \end{minipage}
  \hspace{0.5cm}
  \begin{minipage}[b]{\anchoFeats \textwidth}
    \centering
    \includegraphics[width=7cm]{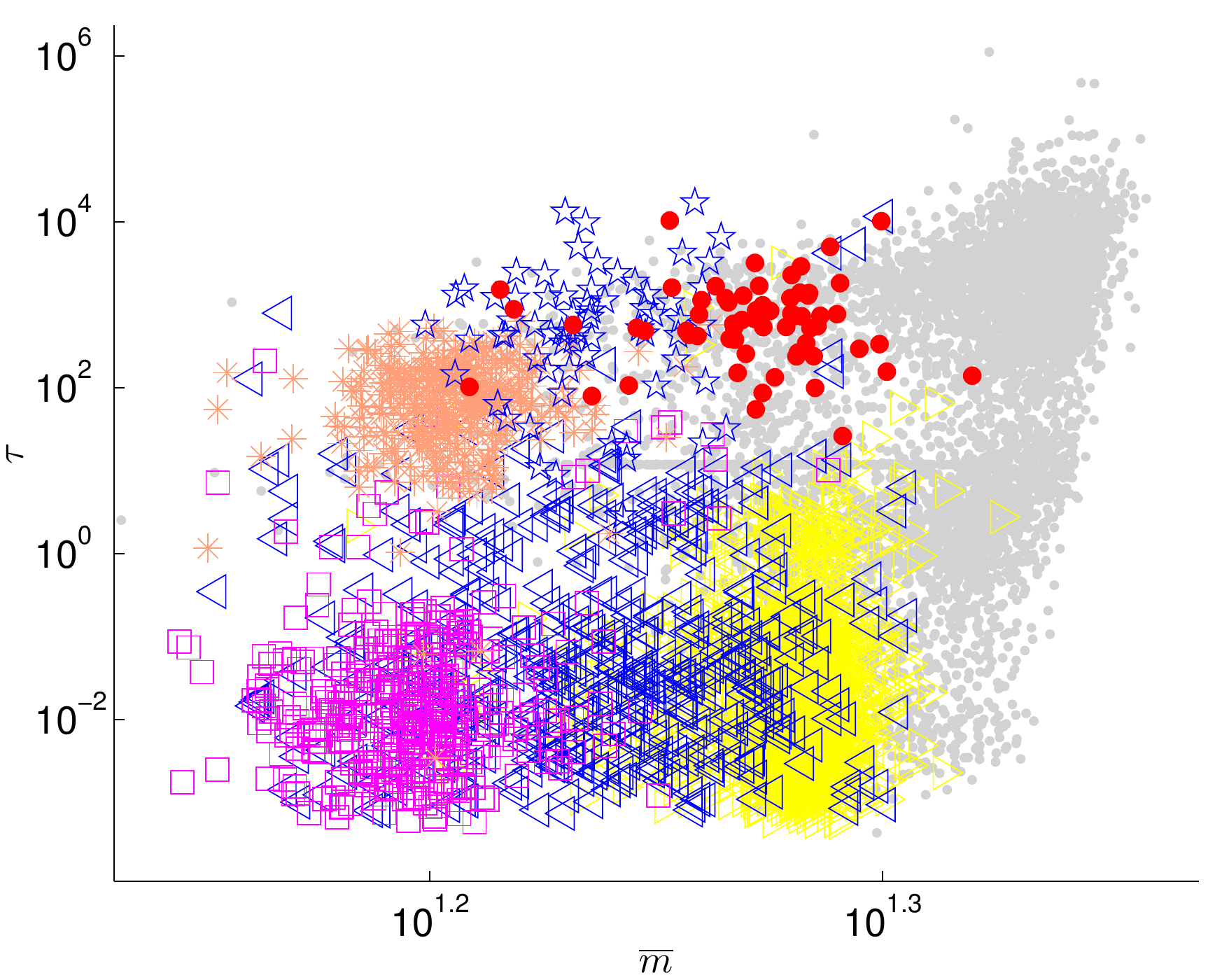}
   \end{minipage}
  \begin{minipage}[b]{\anchoFeats \textwidth}
    \centering
    \includegraphics[width=7cm]{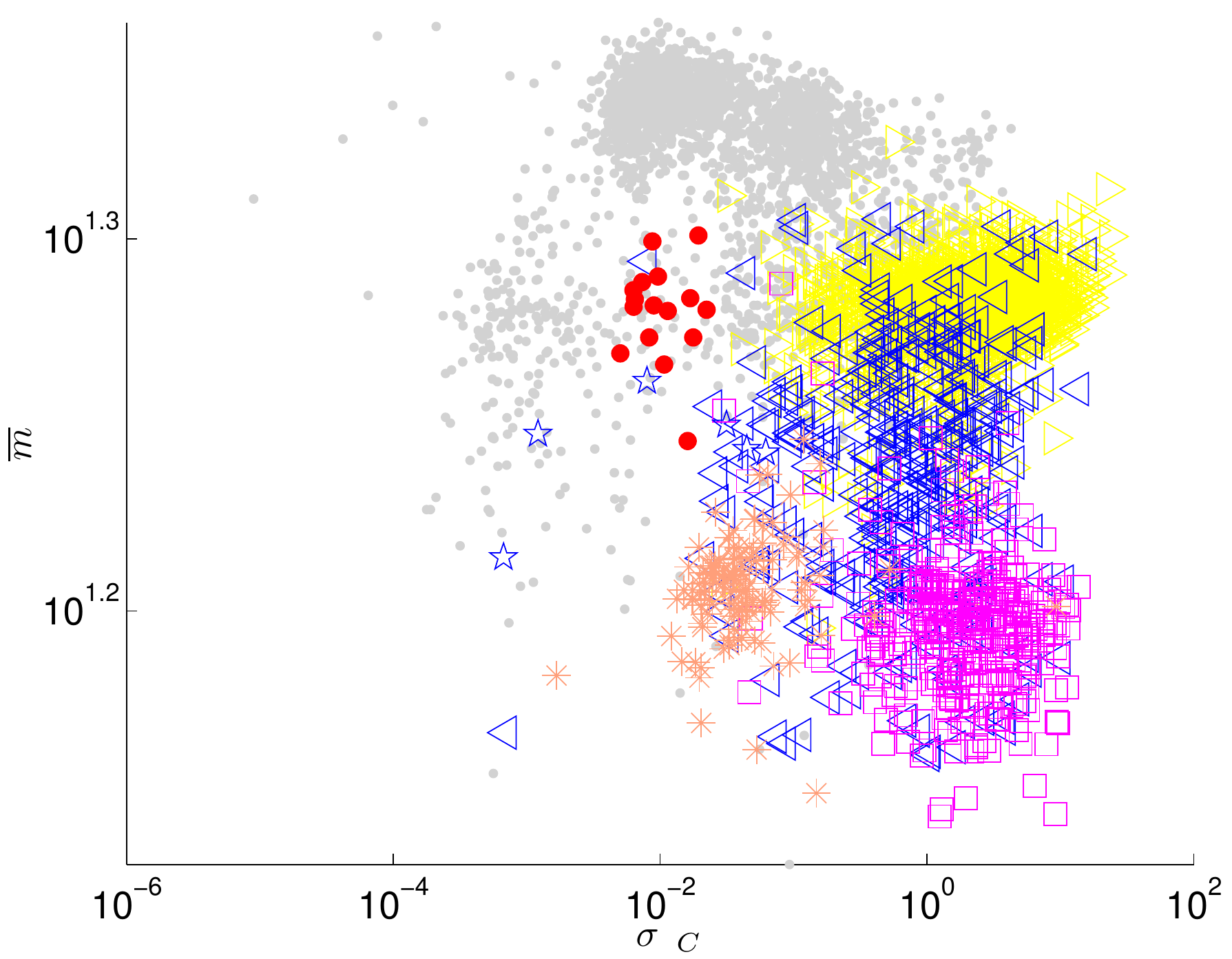}
  \end{minipage}
     \caption{Projections on different pairs of CAR(1) features for EROS-2 training data}
      \label{Fig:EROS_Train_Features_1}  
\end{figure*}

\begin{figure*}
 \fbox{\includegraphics[width=0.8\textwidth]{./Plots/EROS/Symbols_EROS_Train-eps-converted-to.pdf}} 
  \begin{minipage}[b]{\anchoFeats \textwidth}
    \centering
    \includegraphics[width=7cm]{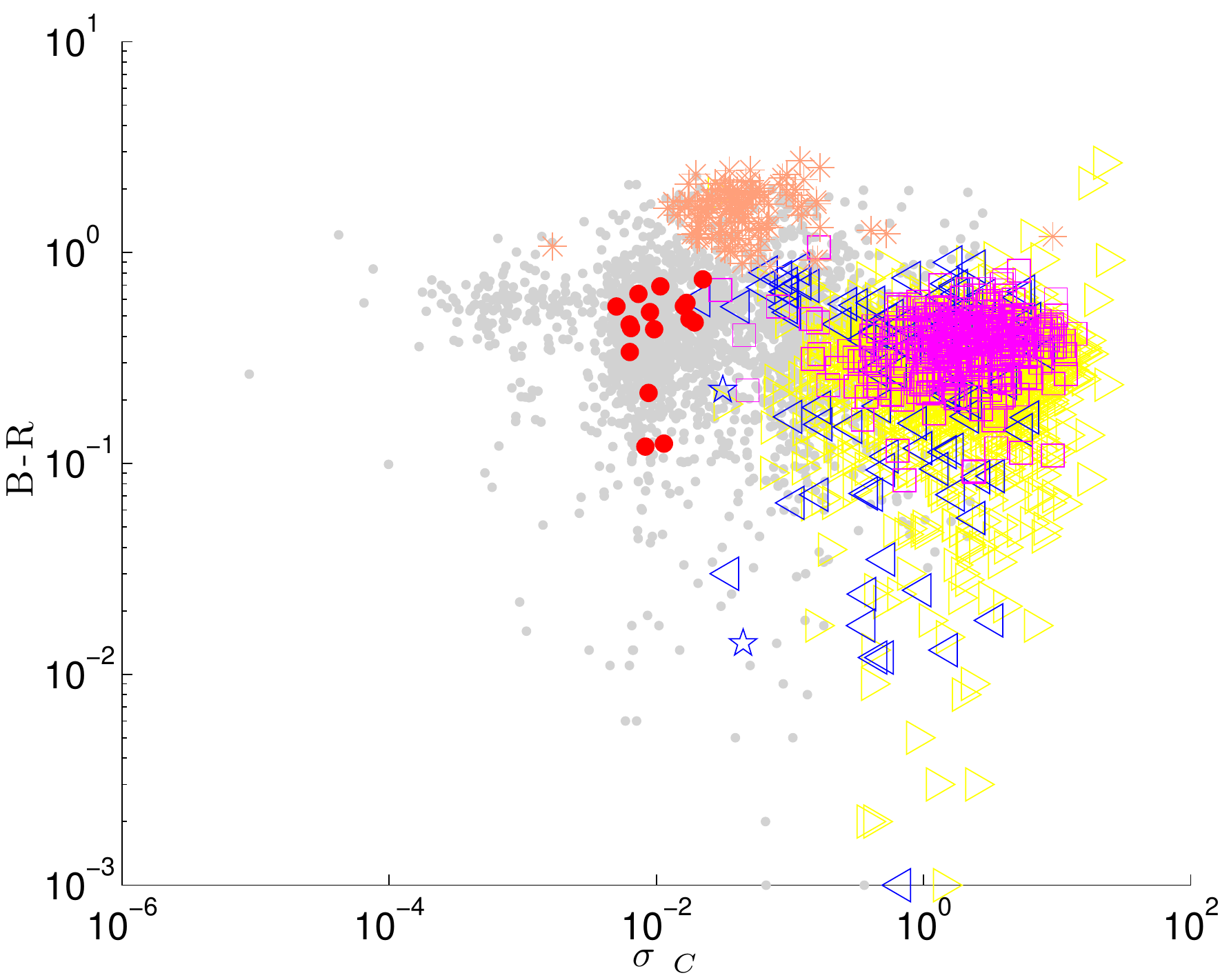}
  \end{minipage}
  \begin{minipage}[b]{\anchoFeats \textwidth}
    \centering
    \includegraphics[width=7cm]{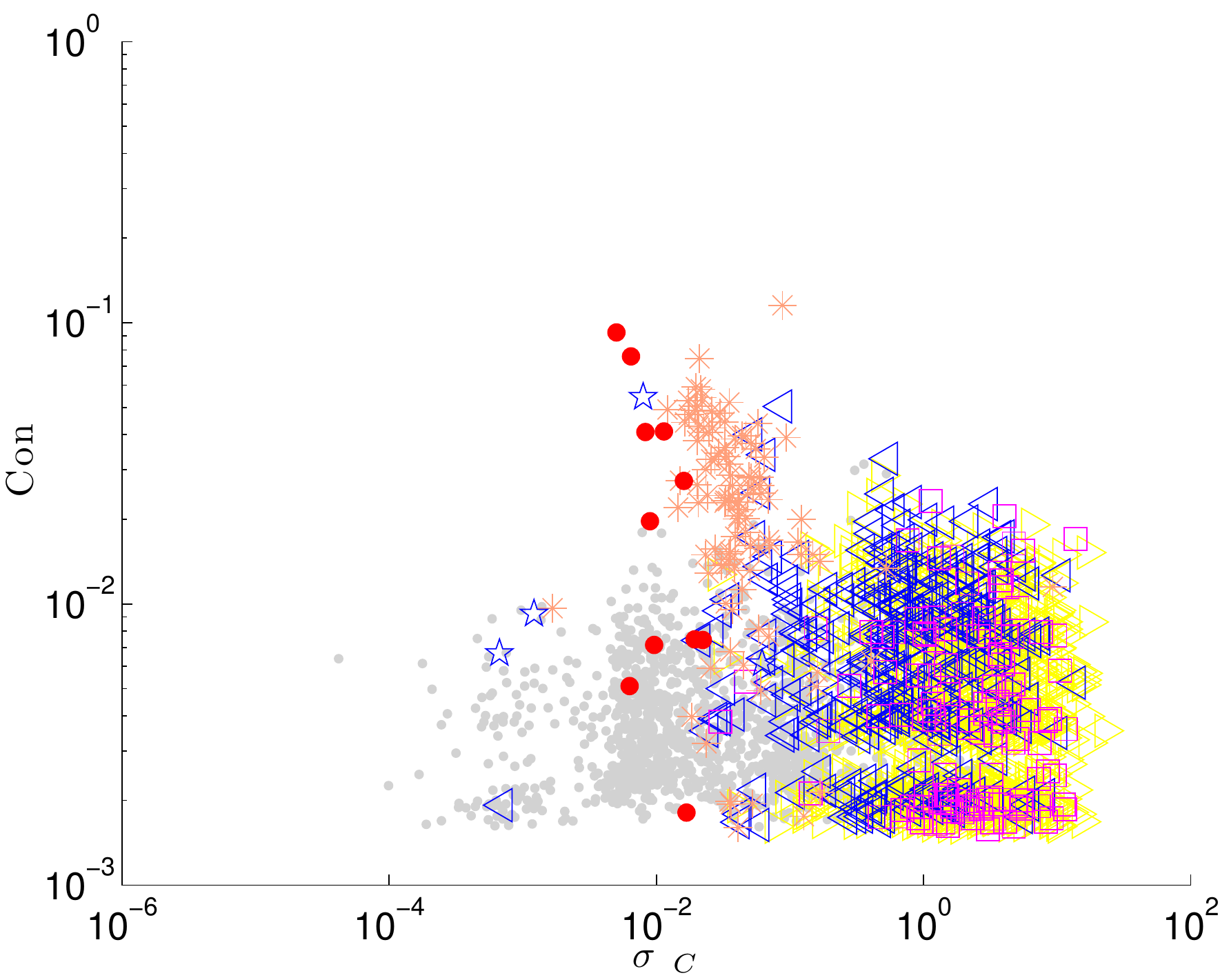}
  \end{minipage}
  \begin{minipage}[b]{\anchoFeats \textwidth}
    \centering
    \includegraphics[width=7cm]{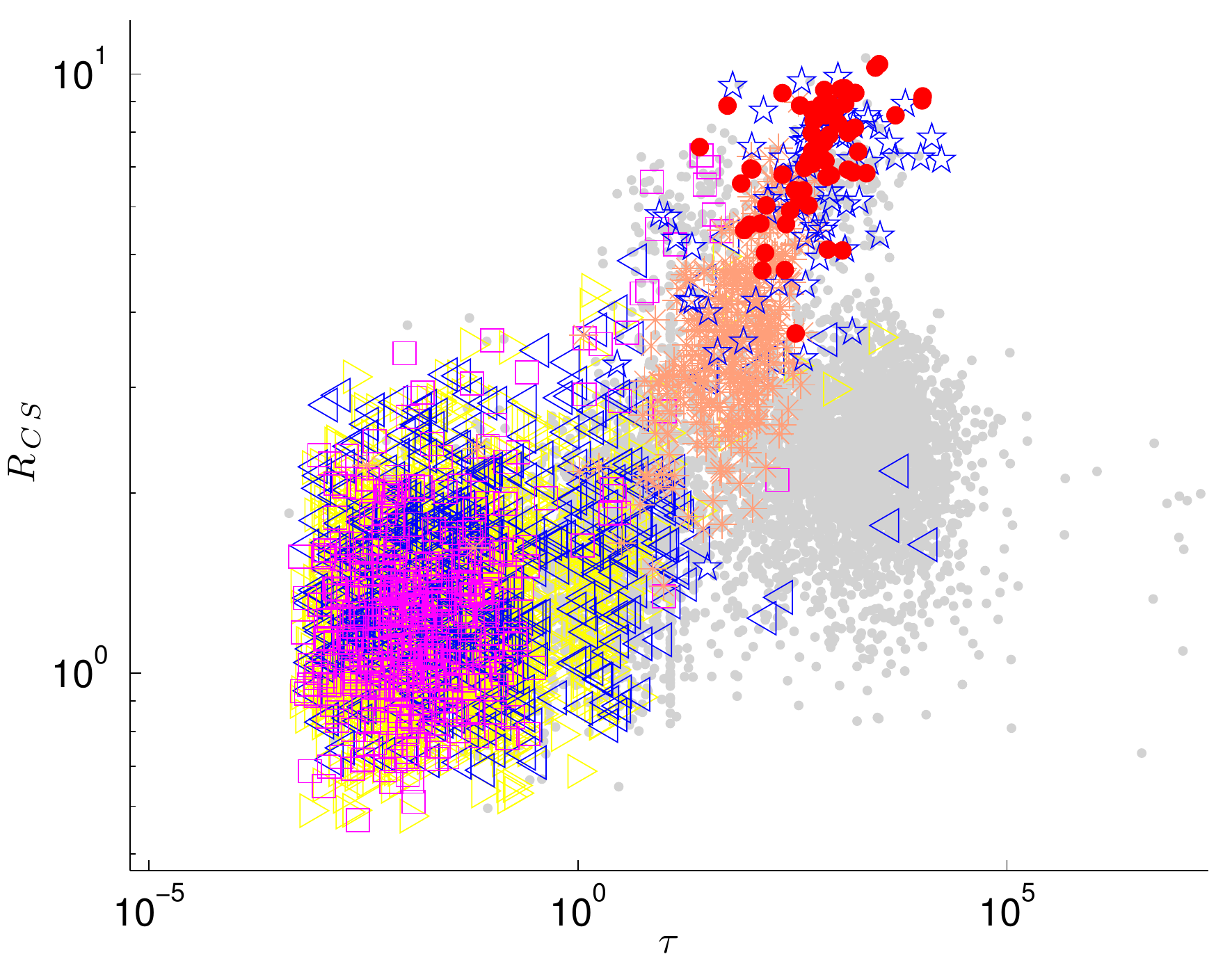}
  \end{minipage}
  \begin{minipage}[b]{\anchoFeats \textwidth}
    \centering
    \includegraphics[width=7cm]{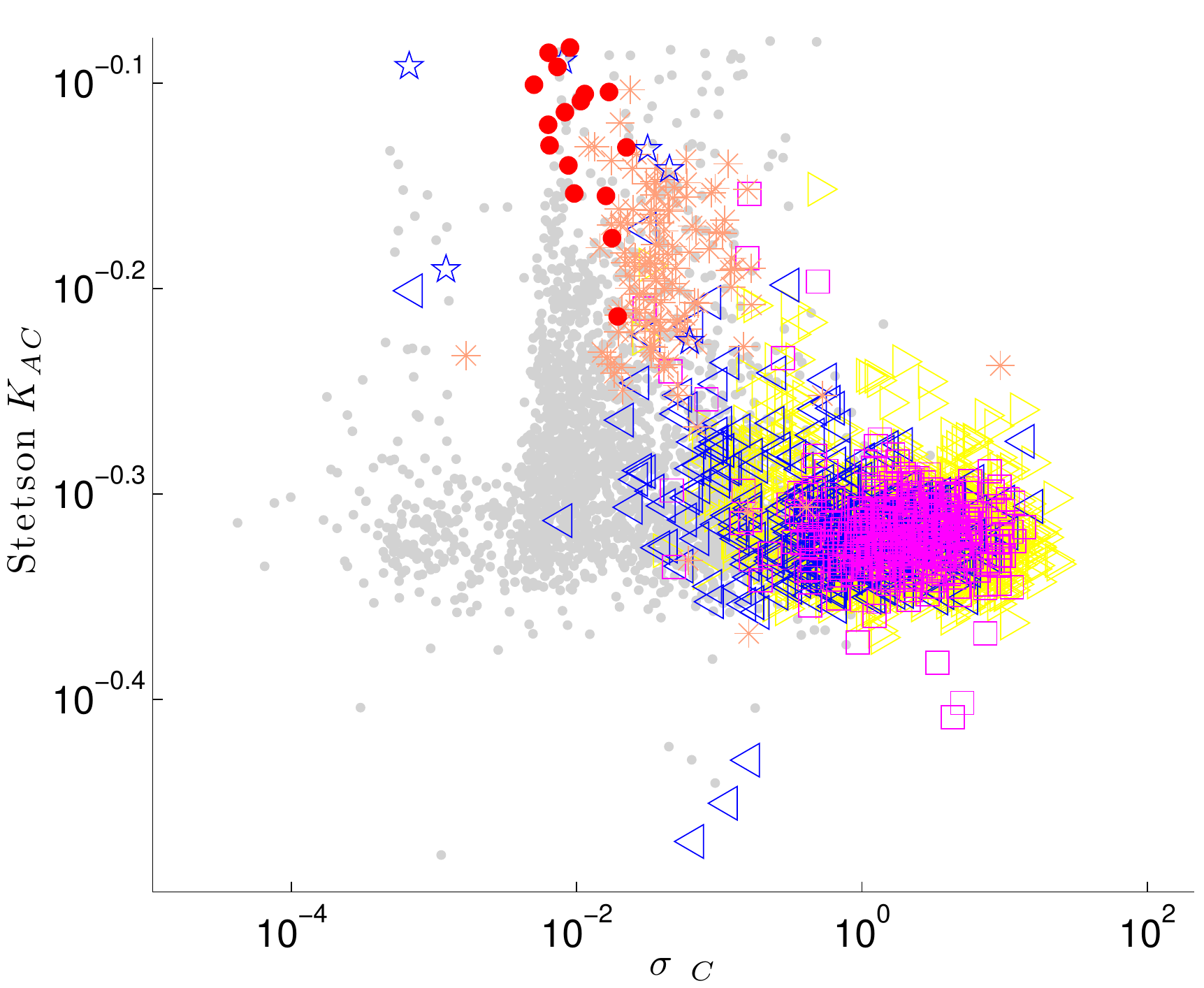}
  \end{minipage}  
  \begin{minipage}[b]{\anchoFeats \textwidth}
    \centering
    \includegraphics[width=7cm, height=5.5cm]{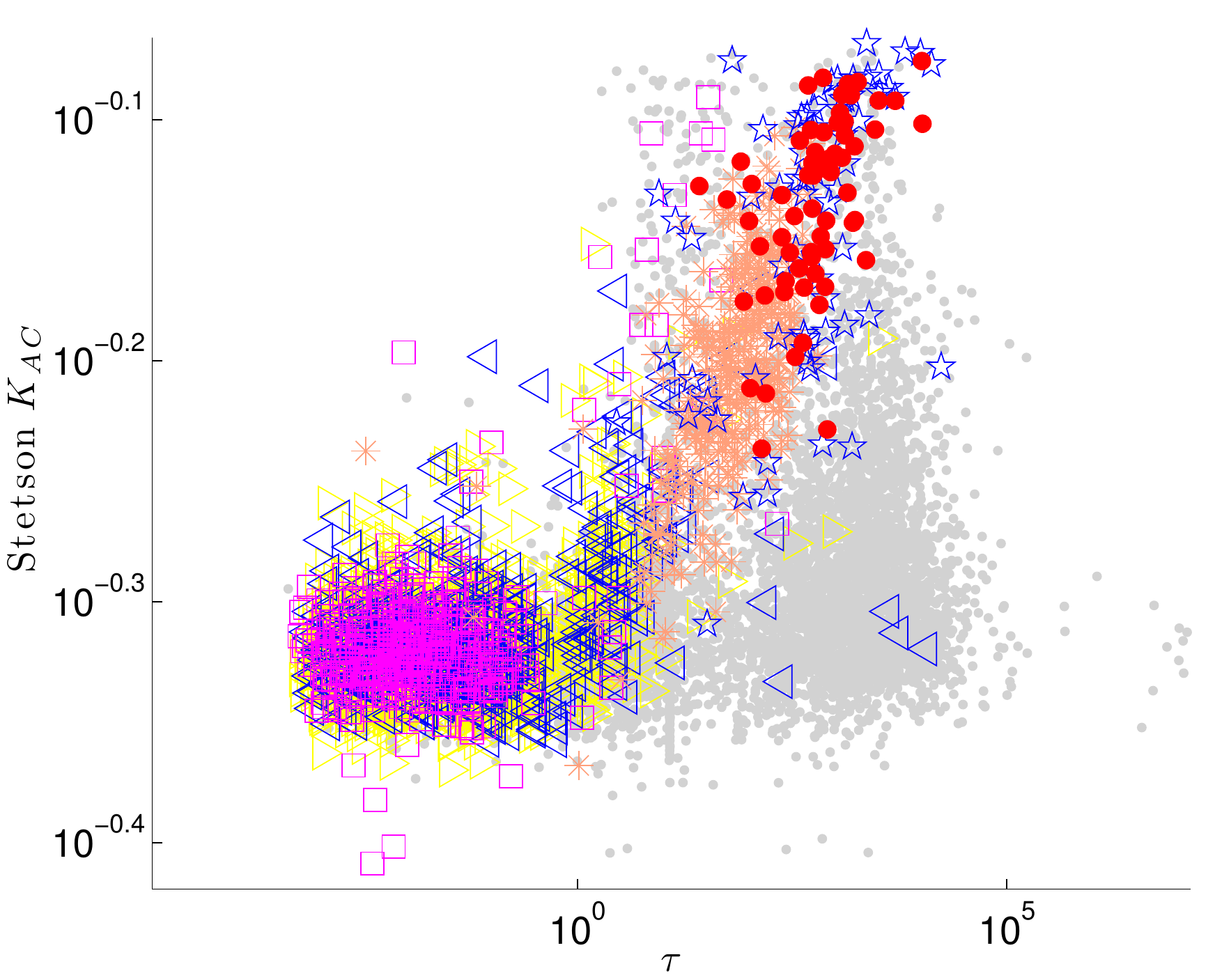}
 \end{minipage}
     \caption{Projections on different pairs of features, combining CAR(1) features with time series features for EROS-2 training data}
     \label{Fig:EROS_Train_Features_2} 
\end{figure*}

\begin{figure*}
  \fbox{\includegraphics[width=0.8\textwidth]{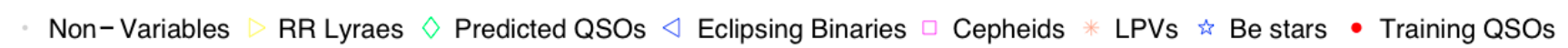}} 
  \begin{minipage}[b]{\anchoFeats \textwidth}
    \centering
    \includegraphics[width=7cm]{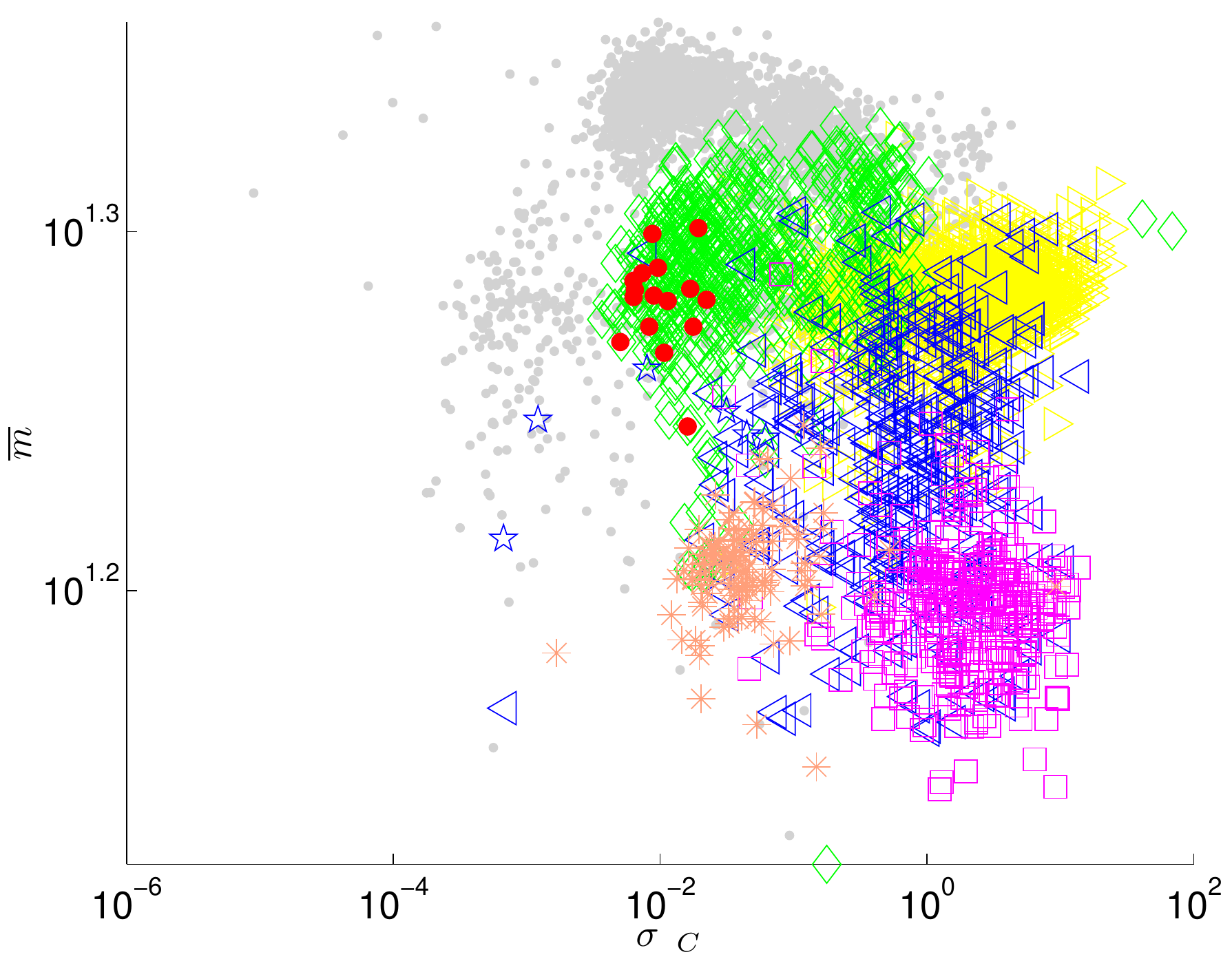}
  \end{minipage}
  \hspace{0.5cm}
  \begin{minipage}[b]{\anchoFeats \textwidth}
    \centering
    \includegraphics[width=7cm]{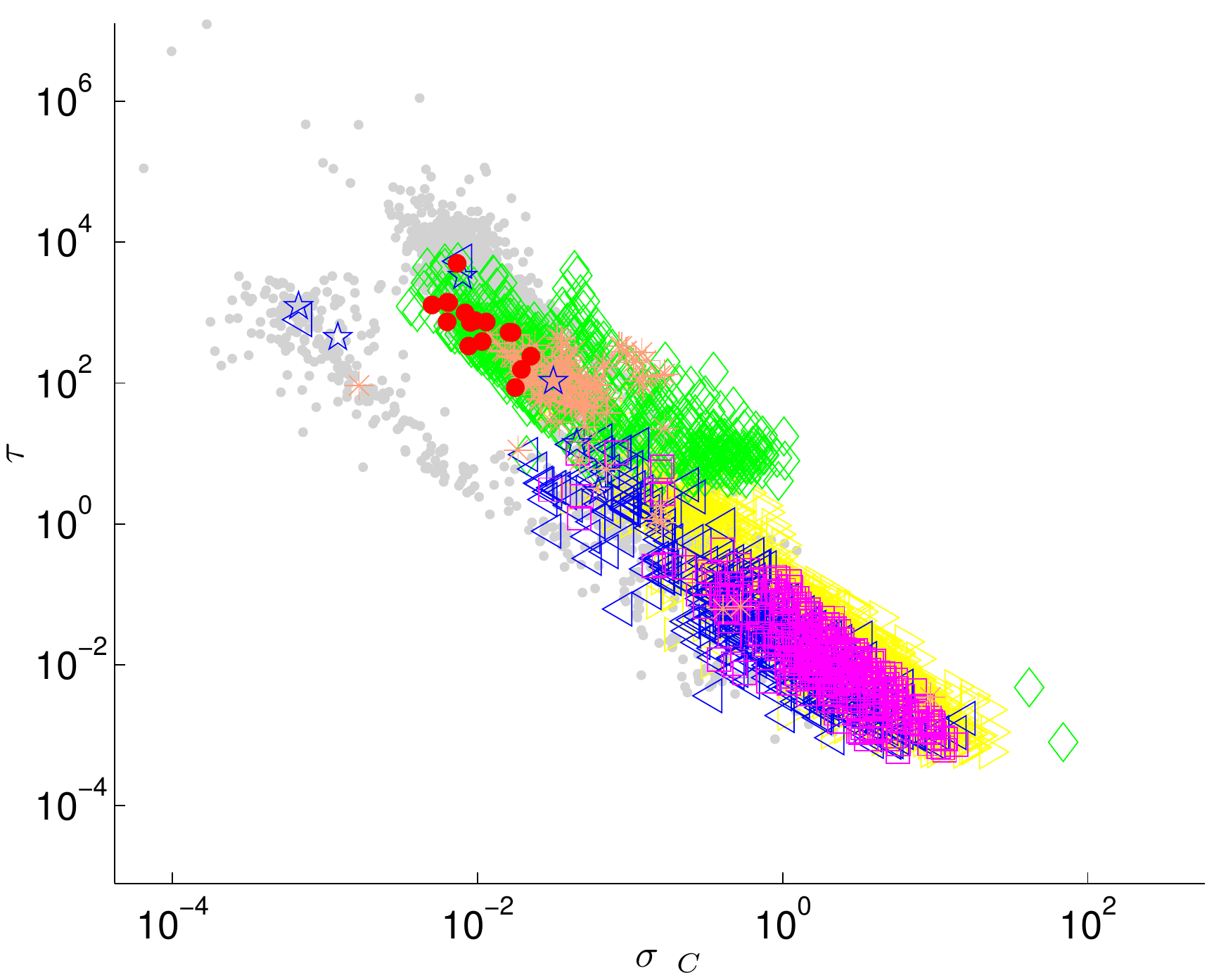}
  \end{minipage}
  \begin{minipage}[b]{\anchoFeats \textwidth}
    \centering
    \includegraphics[width=7cm]{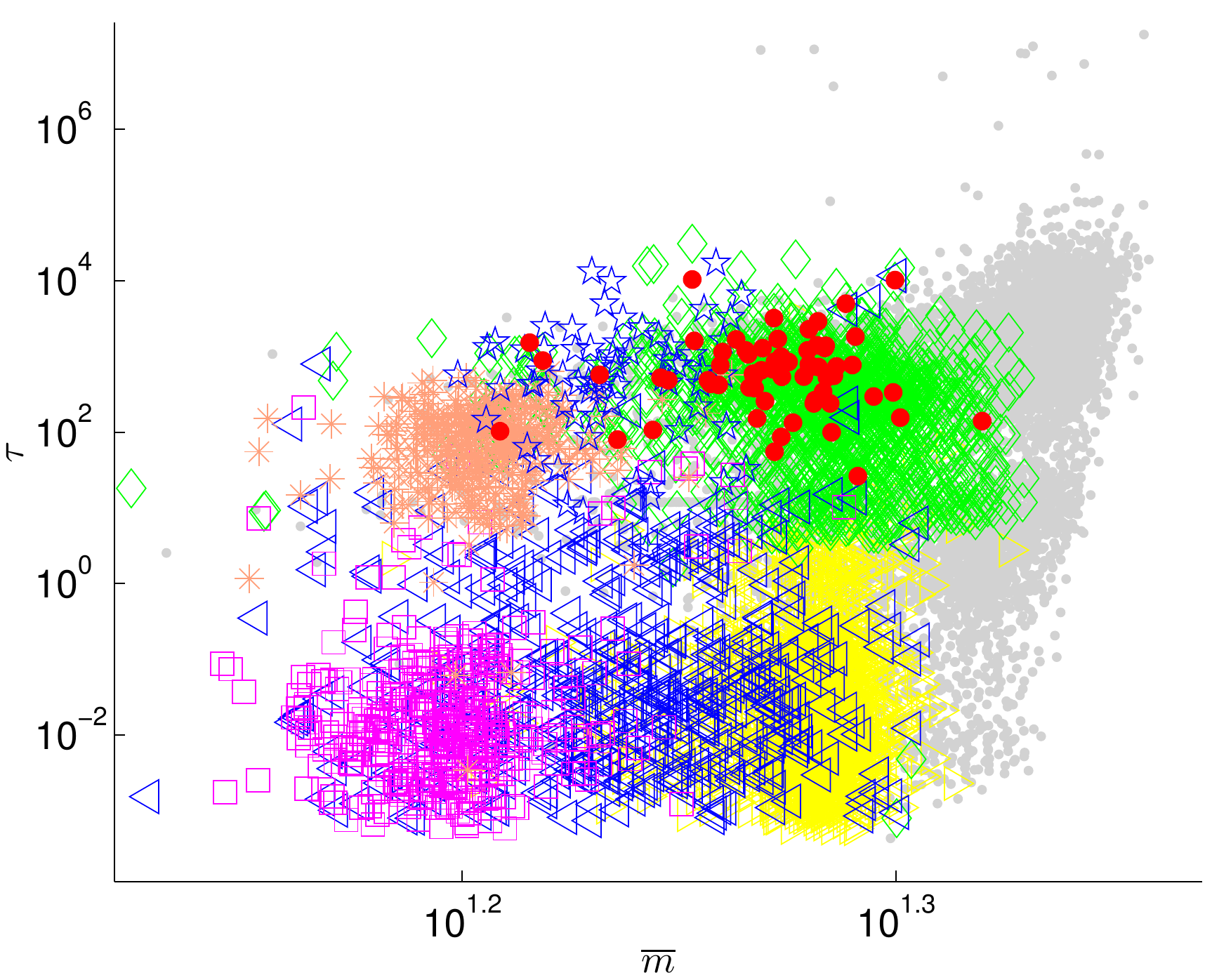}
  \end{minipage}
     \caption{Predicted quasars and Training stars distributions projected on different pairs of CAR(1) features for EROS-2 data.}
     \label{Fig:EROS_Pred_Features_1}  
 \end{figure*}

\begin{figure*}
  \fbox{\includegraphics[width=0.8\textwidth]{./Plots/EROS/Symbols_EROS_Pred-eps-converted-to.pdf}} 
  \begin{minipage}[b]{\anchoFeats \textwidth}
    \centering
    \includegraphics[width=7cm]{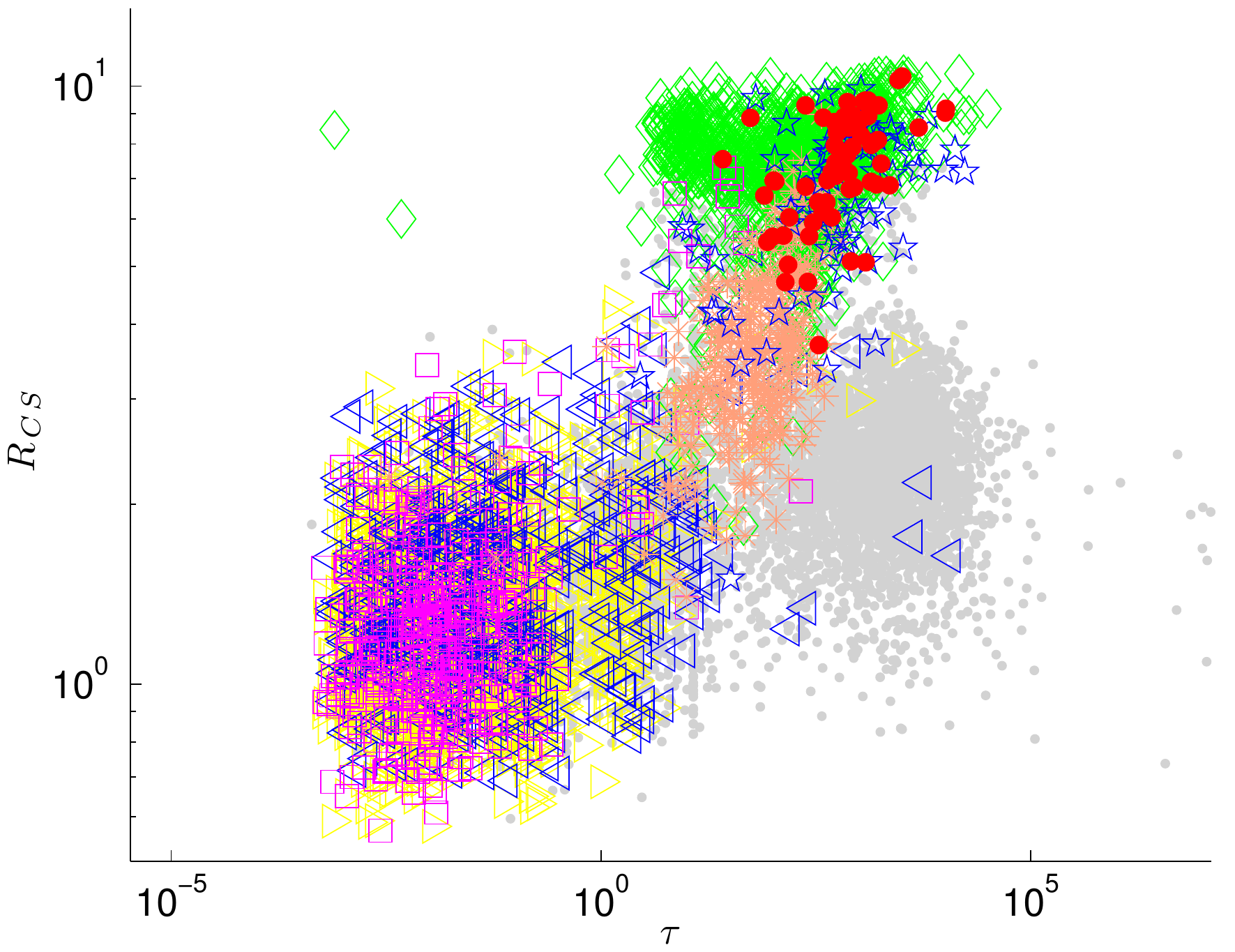}
  \end{minipage}
  \begin{minipage}[b]{\anchoFeats \textwidth}
    \centering
    \includegraphics[width=7cm]{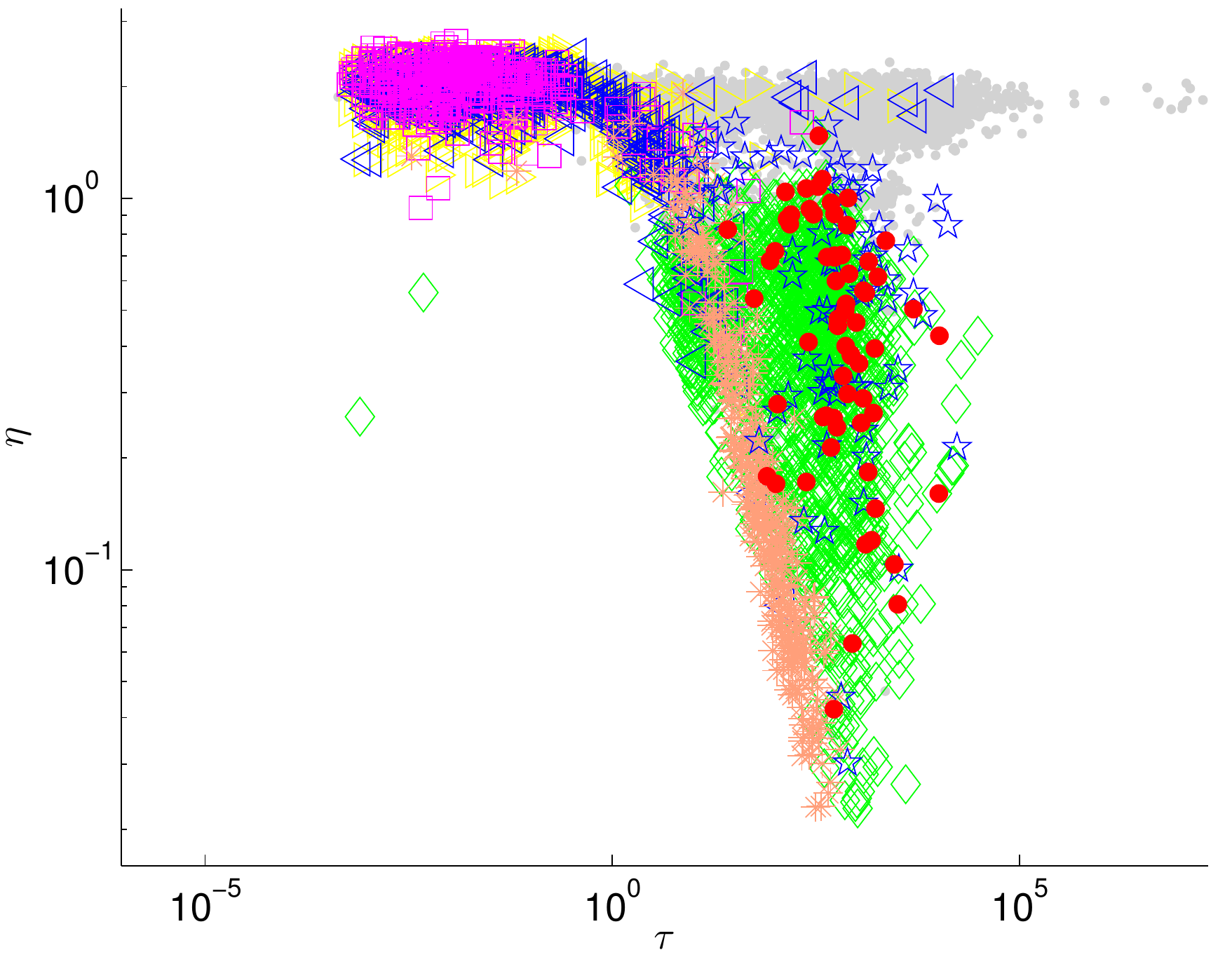}
  \end{minipage}
  \begin{minipage}[b]{\anchoFeats \textwidth}
    \centering
    \includegraphics[width=7cm]{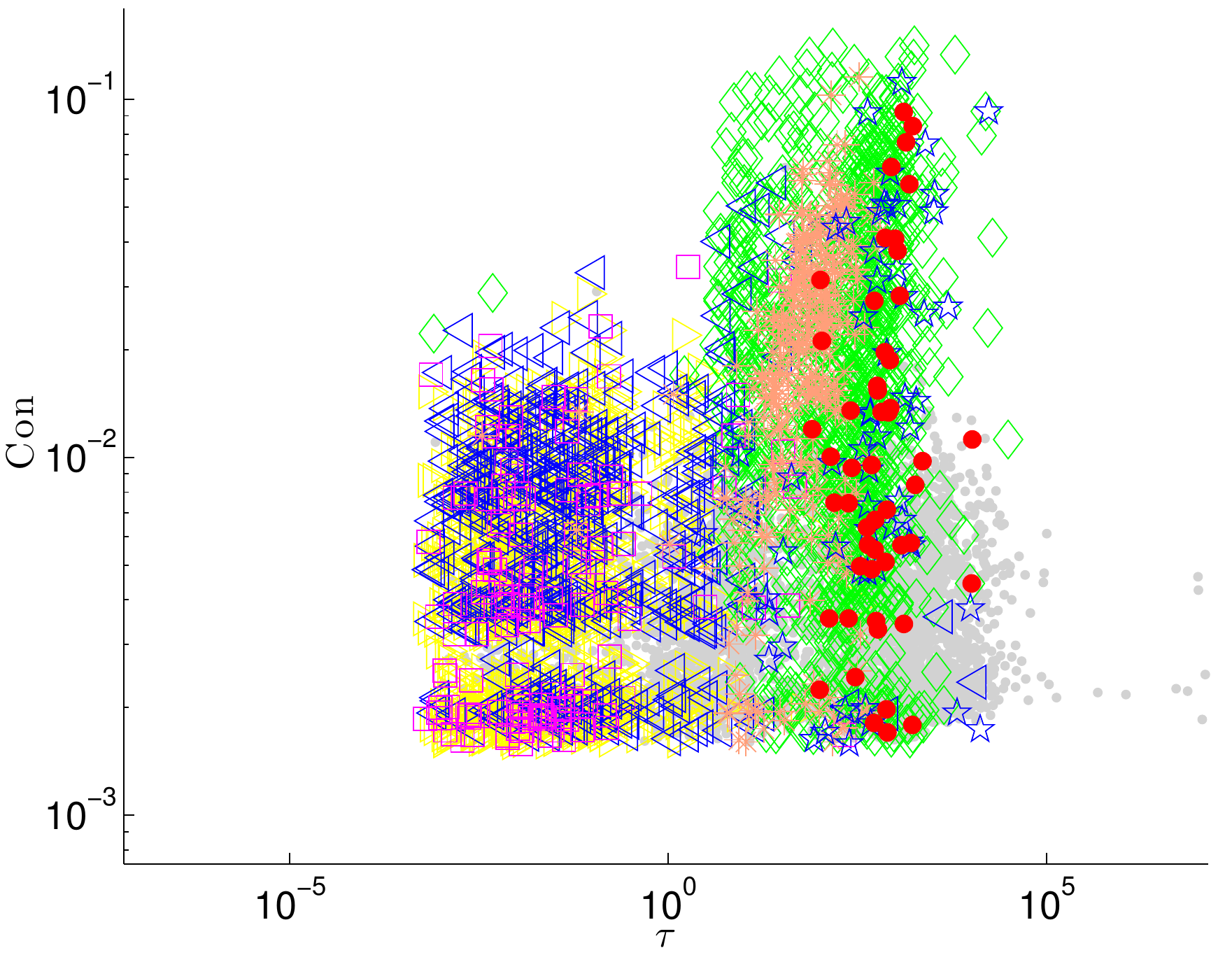}
  \end{minipage}
  \begin{minipage}[b]{\anchoFeats \textwidth}
    \centering
    \includegraphics[width=7cm]{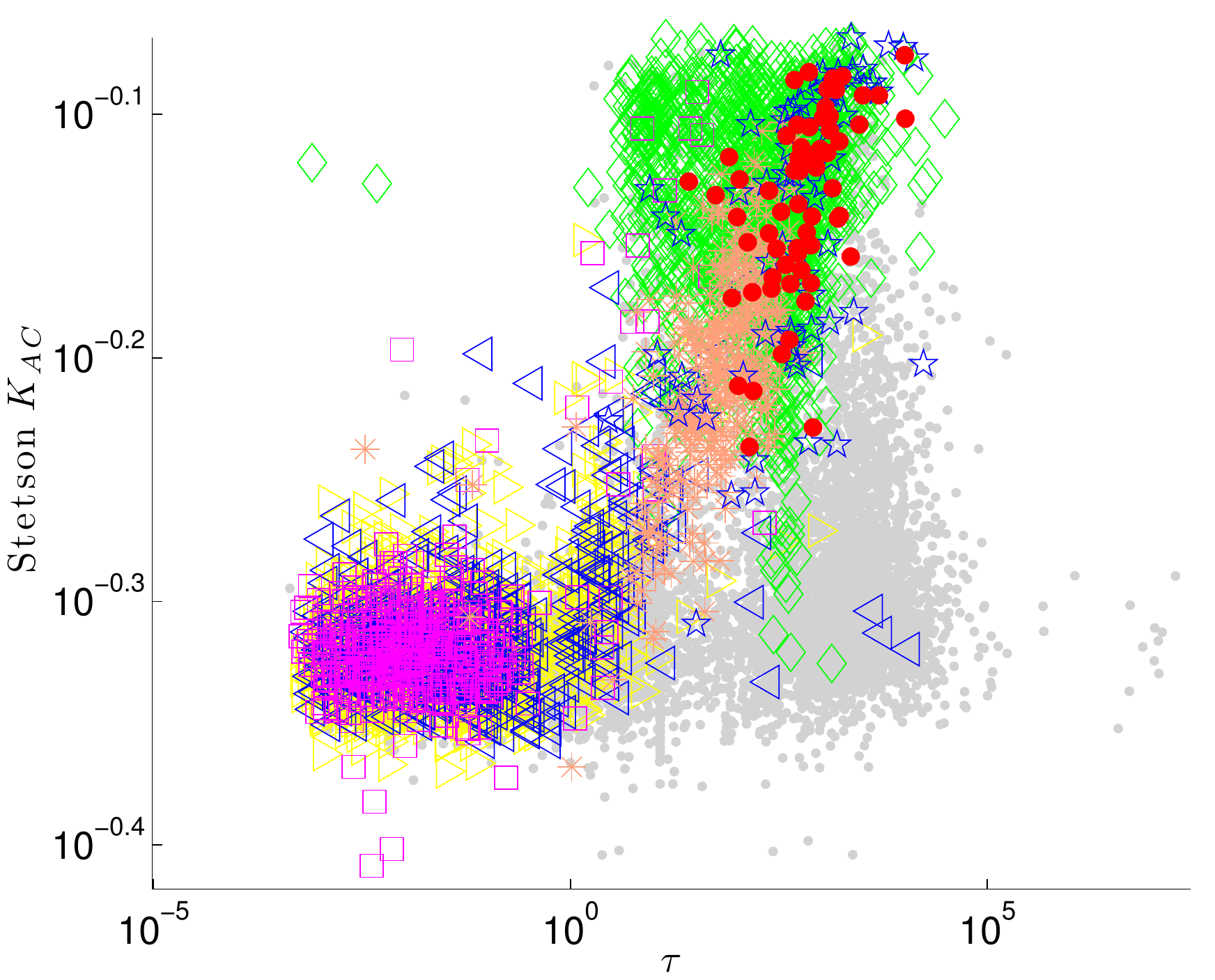}
  \end{minipage}
  \caption{Predicted quasars and training stars distributions for $\tau$ combined with three time series features for EROS-2 data.}
  \label{Fig:EROS_Pred_Features_2}  
\end{figure*}

  To compare the distribution of the objects predicted as quasars with the training quasars and other variable stars, we plot our EROS-2 training data plus the predicted quasars projected on many different pairs of features (Figs. \ref{Fig:EROS_Pred_Features_1} , \ref{Fig:EROS_Pred_Features_2}). We can see that in most of the cases Predicted quasars and Training quasars have very similar distributions regardless of the small amount of training quasars we use. Main differences between both distributions are in general because of the big difference in size comparing training and testing data, resulting in a set of predicted quasars $\sim$ 20 times bigger than the training quasars set.

 To get an indicator of the accuracy in the training set on EROS-2 dataset, we run a 10 fold cross validation. This validation method consists of partitioning the dataset in 10 folds (subsets) of the same size, we iterate 10 times, on iteration $k$ we train the classifier with all the folds but the fold $k$, then we test the performance on the fold $k$ (the one which the model did not see during the training). The process returns the model prediction for the entire dataset (the union of the 10 testing folds is equal to the data).    
  We measure the accuracy using the {\em F-score} indicator. This indicator is calculated as the harmonic mean of precision and recall:\\
  
   F-Score $ =  2 \times \frac{precision \times recall}{precision + recall}$\\   
   
   Where precision and recall are defined as:\\
   
    $precision = \frac{tp}{tp + fp}    \qquad   recall = \frac{tp}{tp + fn}$\\  
  
   Where $tp,fp$ and $fn$ are the number of true positives, false positives and false negatives respectively.\\  

   Table \ref{table:ClassifiersTrainingEROS-2} show the results for the boosted version of Random Forest, regular Random Forest and SVM (classifier used in our previous work \citep{Kim2011ApJ} ) with and without CAR features.

\begin{table}
  \caption{{\em F-score} for the EROS-2 training set using 10-fold cross validation for different classification models. Each classifiers is tuned with the optimal set of parameters. We can see that the boosted version of Random Forest with CAR features outperforms other classification models. In all cases using CAR features improves the result of the corresponding classifier.}    
  \begin{center} 
  \begin{tabular}{c|c|c|c|c|c|}
     \hline 
      SVM     & SVM   &   RF        &    RF  & AB+RF  & AB+RF\\
      No CAR & CAR   & No CAR &  CAR &  No CAR & CAR\\
     \hline 
     0.74 & 0.855 & 0.787 & 0.813 & 0.81 & 0.868\\ 
      \hline 
     \end{tabular}
     \end{center}
          \label{table:ClassifiersTrainingEROS-2}
\end{table}
 
 We find \NCanEROS{} candidates in the EROS-2 dataset. To validate our candidates we crossmatch them with the list of \NCanStrong{} MACHO strong candidates in \citet{Kim2011ApJ}. From that list, only \NunExistEROSinMACHODW{} objects exists in EROS-2 dataset, we find \NMatchesMACHOEROS{} matches between our EROS-2 candidates and those \NunExistEROSinMACHODW{} objects. Figure \ref{Fig:QSO_Cand_EROS_pics} shows some of the lightcurves of the quasar candidates for the EROS-2 dataset. 

  Regarding the efficiency in the extraction of the CAR(1) features and the time series features, we implemented parallel processing in order to perform the features extraction and classification in a reasonable amount of time. EROS-2 and MACHO databases are stored as a set of thousands of folders where each folder contains thousands of lightcurves of a given field. The feature extraction process runs as a set of parallel threads that run over different compressed files at the same time, extracting them and processing the lightcurves to get the features. Once the features are calculated they are written into a common file related to a particular folder, so each compressed file has a corresponding data file that stores the feature values of all the lightcurves within the folder. After the feature extraction process we run a classification process that runs in parallel over the thousands of data feature files calculated in the previous step. 

\newcommand{\anchoEROS}{5}

\begin{figure*}
\includegraphics[width= \anchoEROS cm]{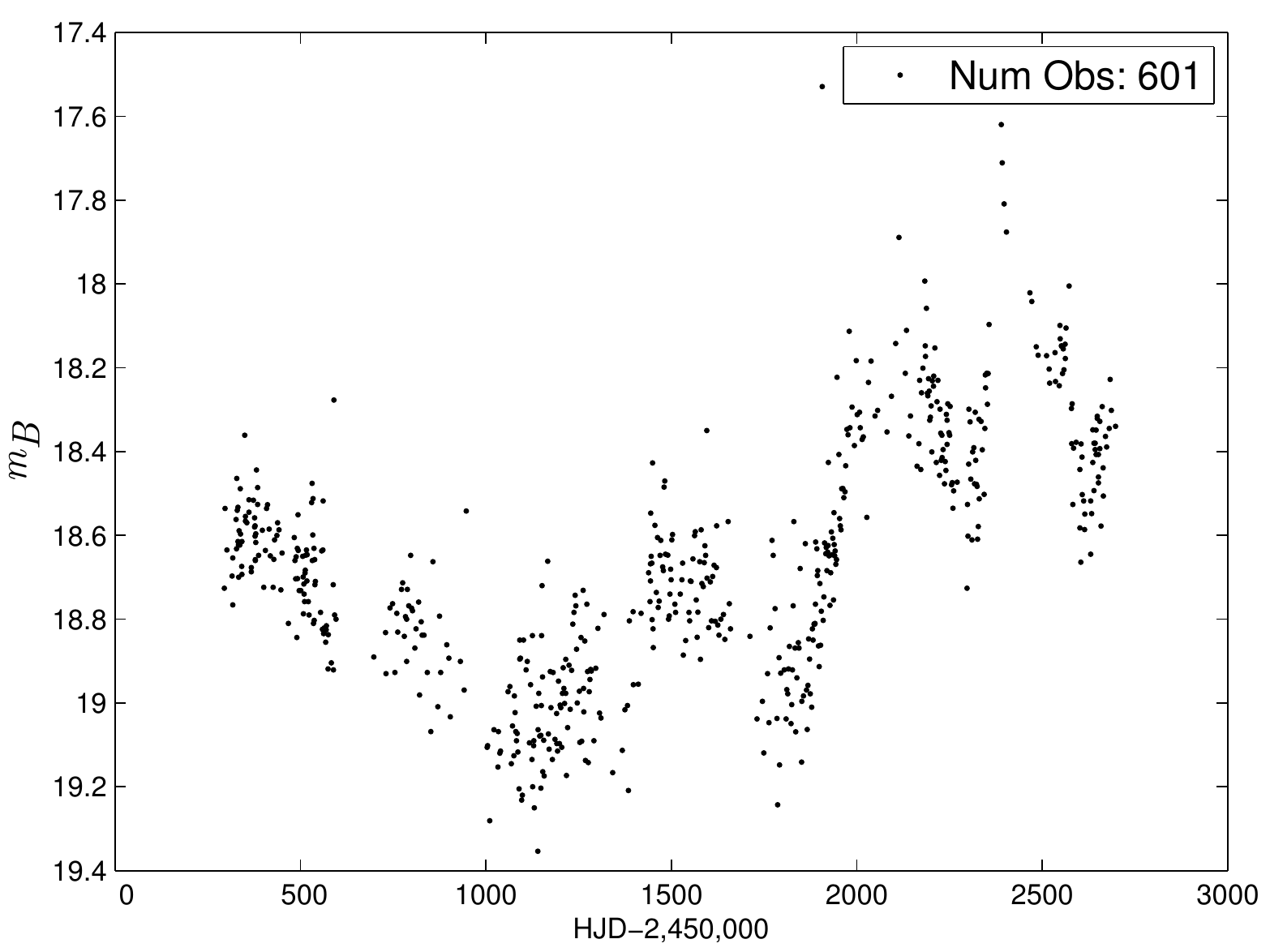}
\includegraphics[width= \anchoEROS cm]{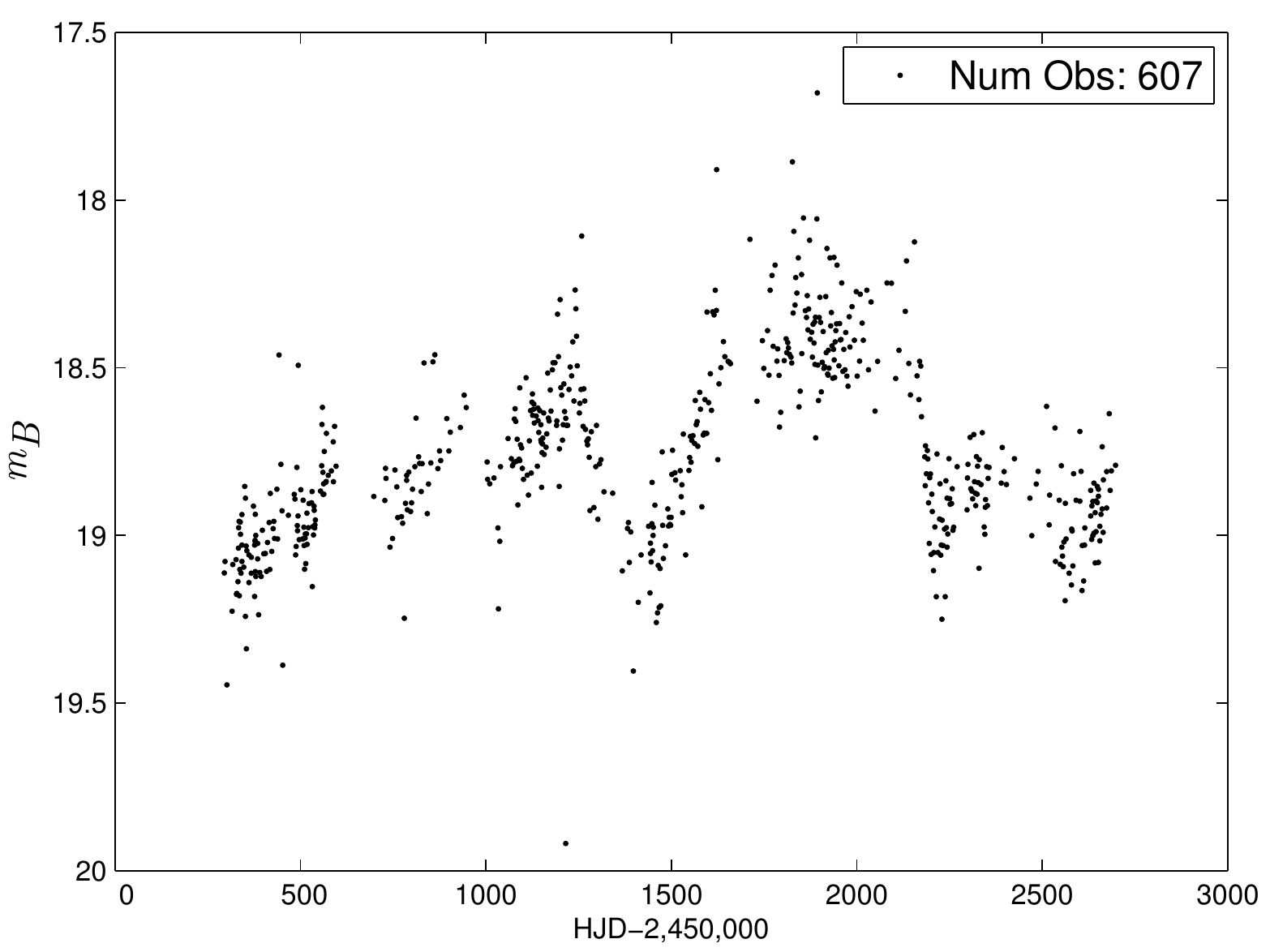}
\includegraphics[width= \anchoEROS cm]{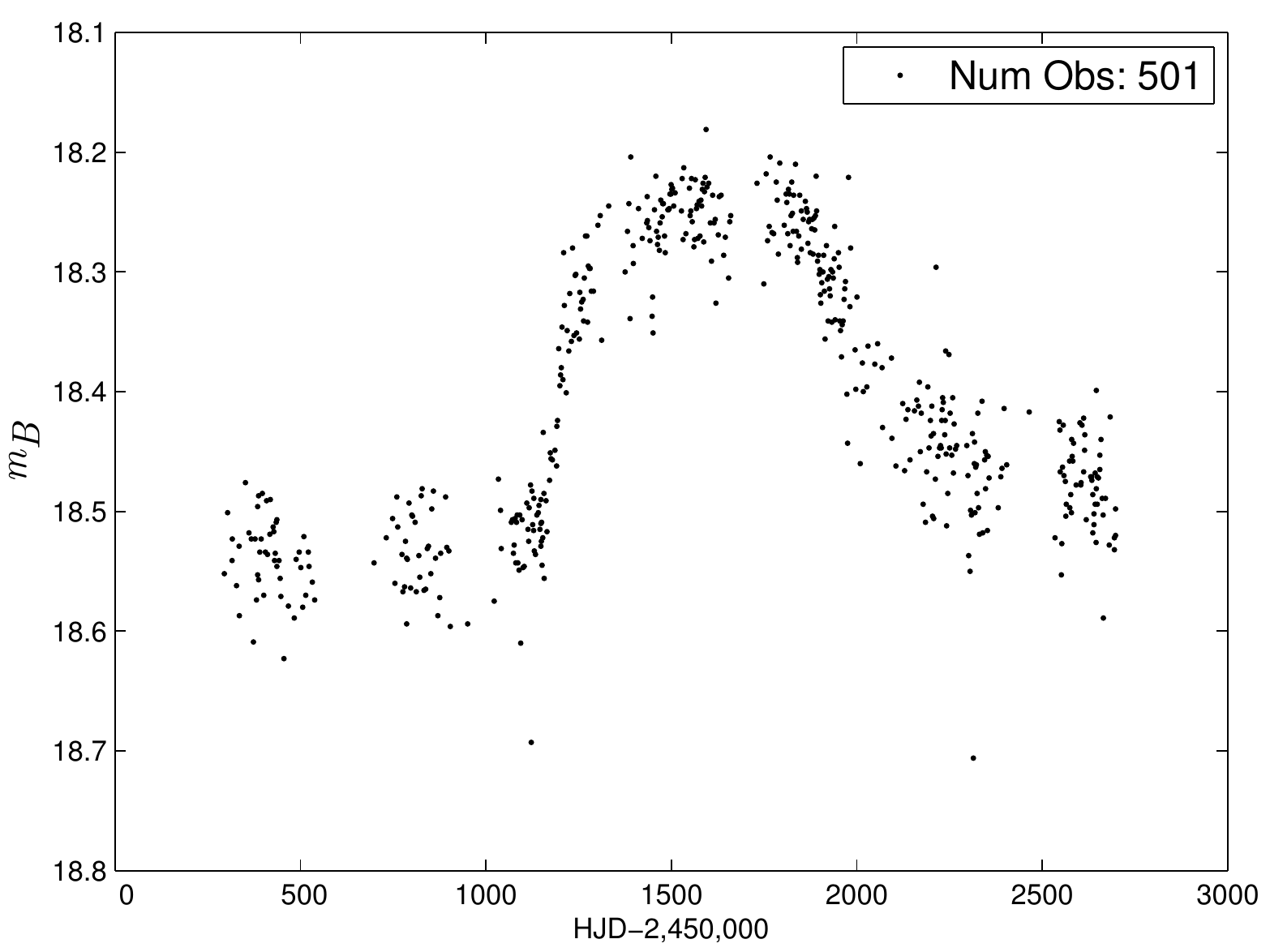}
\includegraphics[width= \anchoEROS cm]{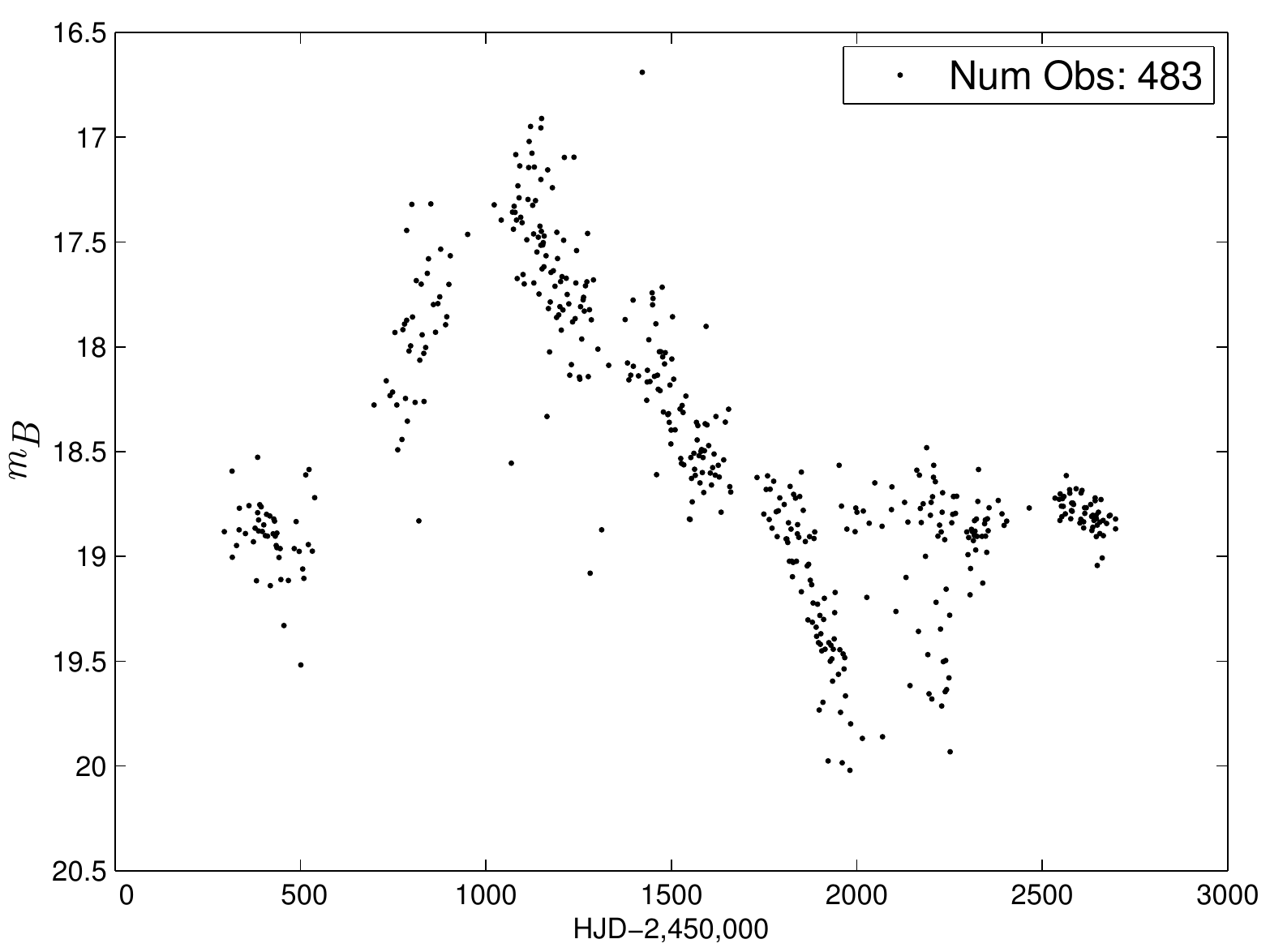}
\includegraphics[width= \anchoEROS cm]{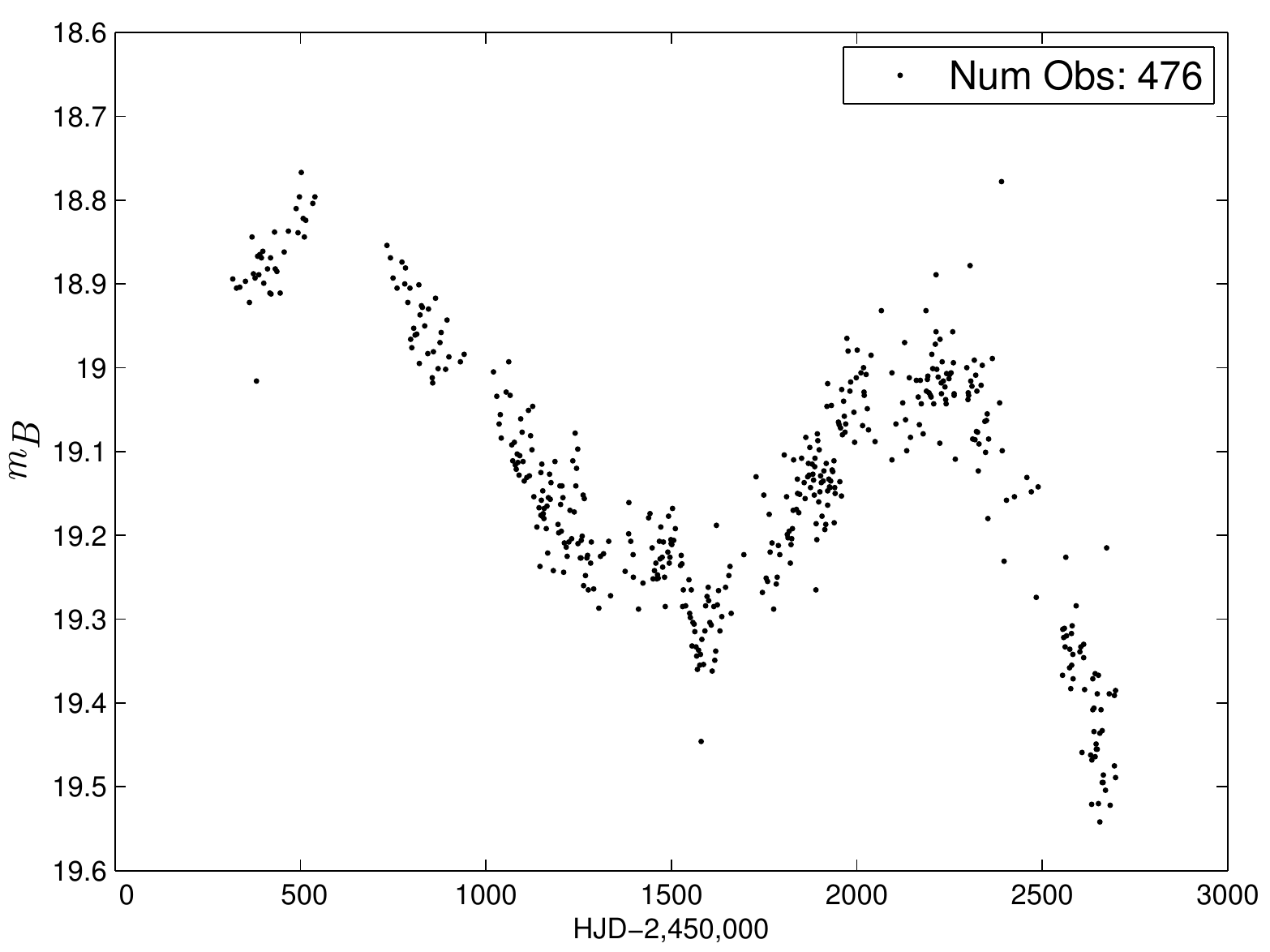}
\includegraphics[width= \anchoEROS cm]{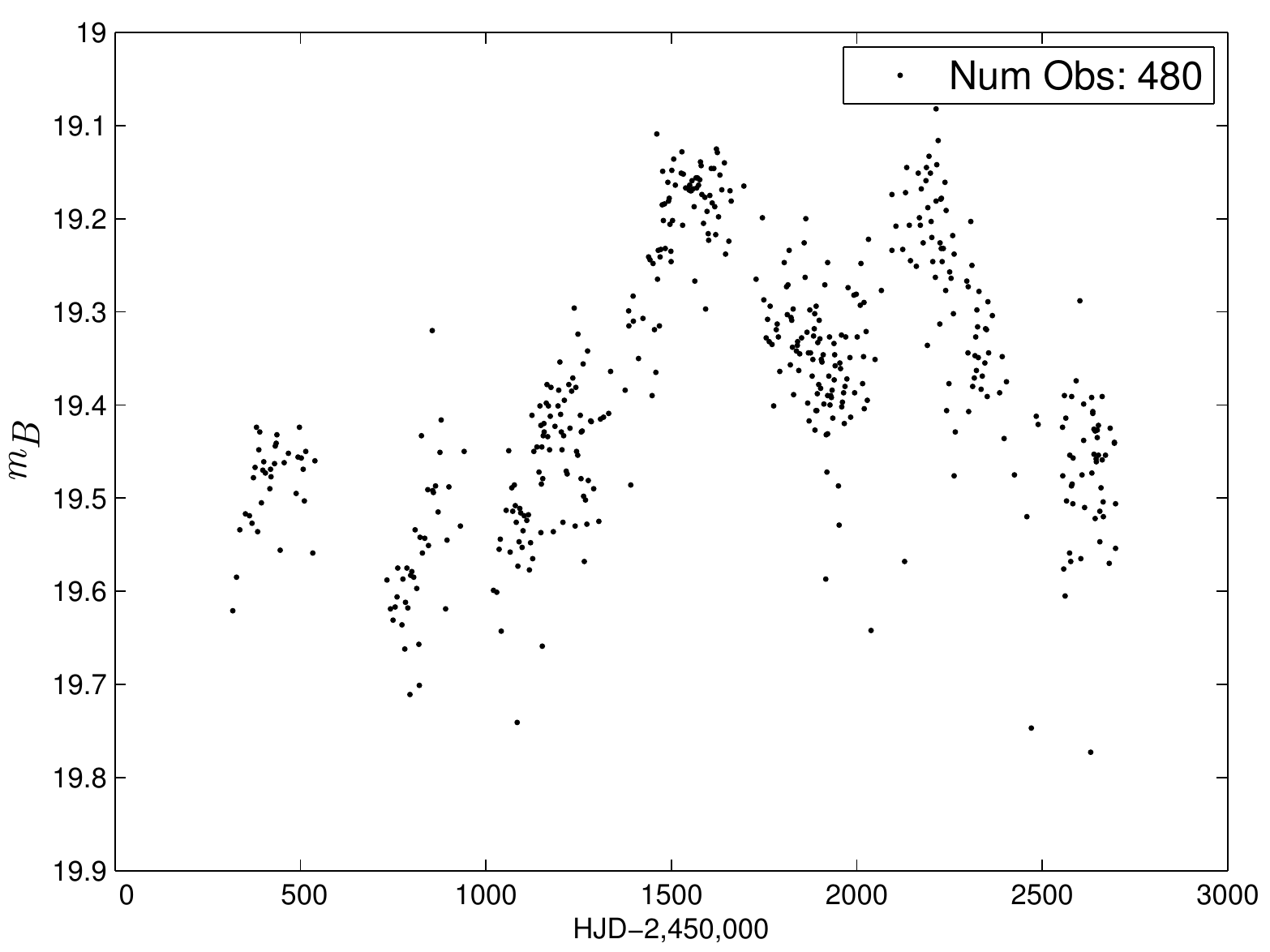}
\includegraphics[width=  \anchoEROS cm]{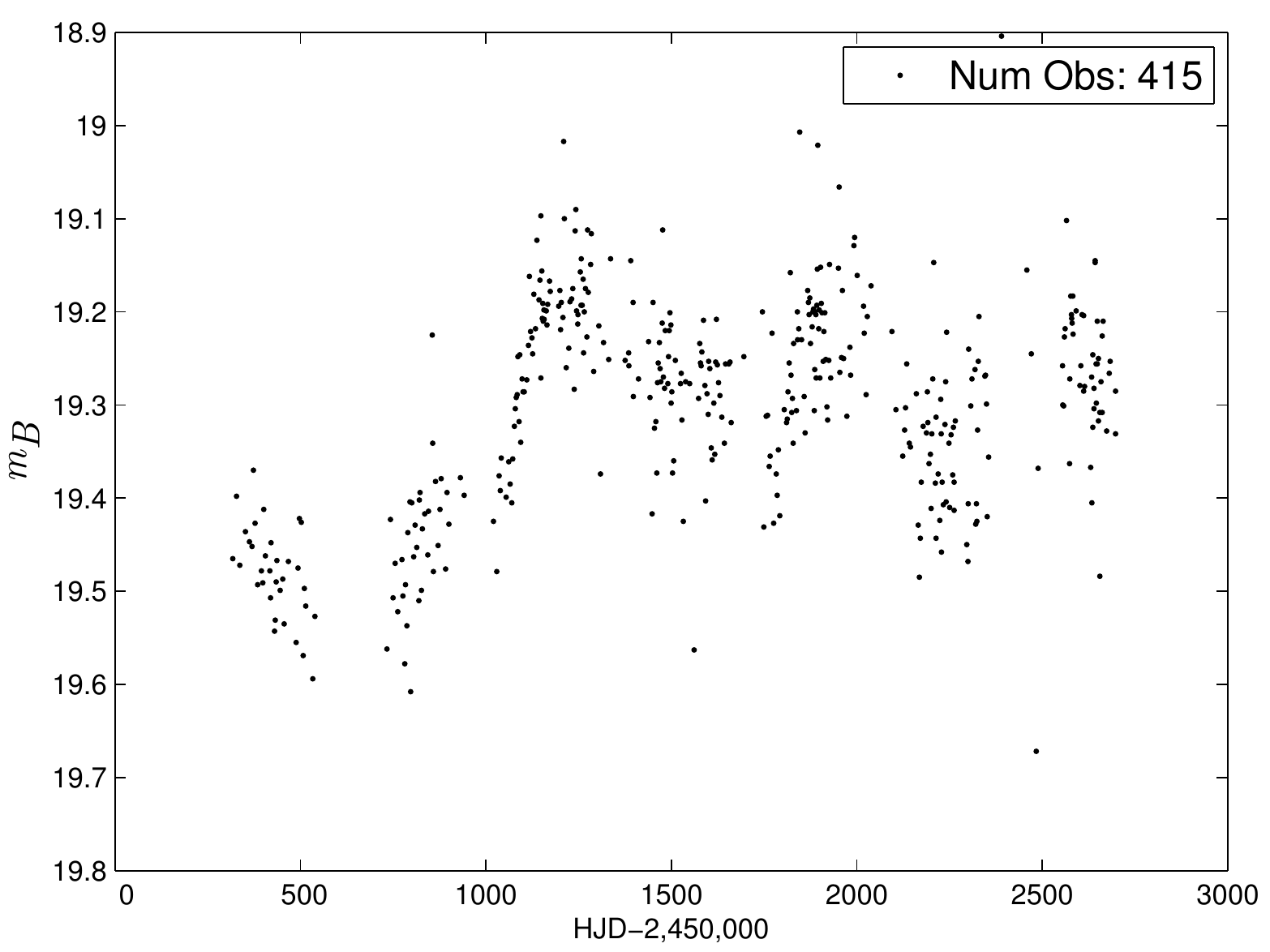}
\includegraphics[width=  \anchoEROS cm]{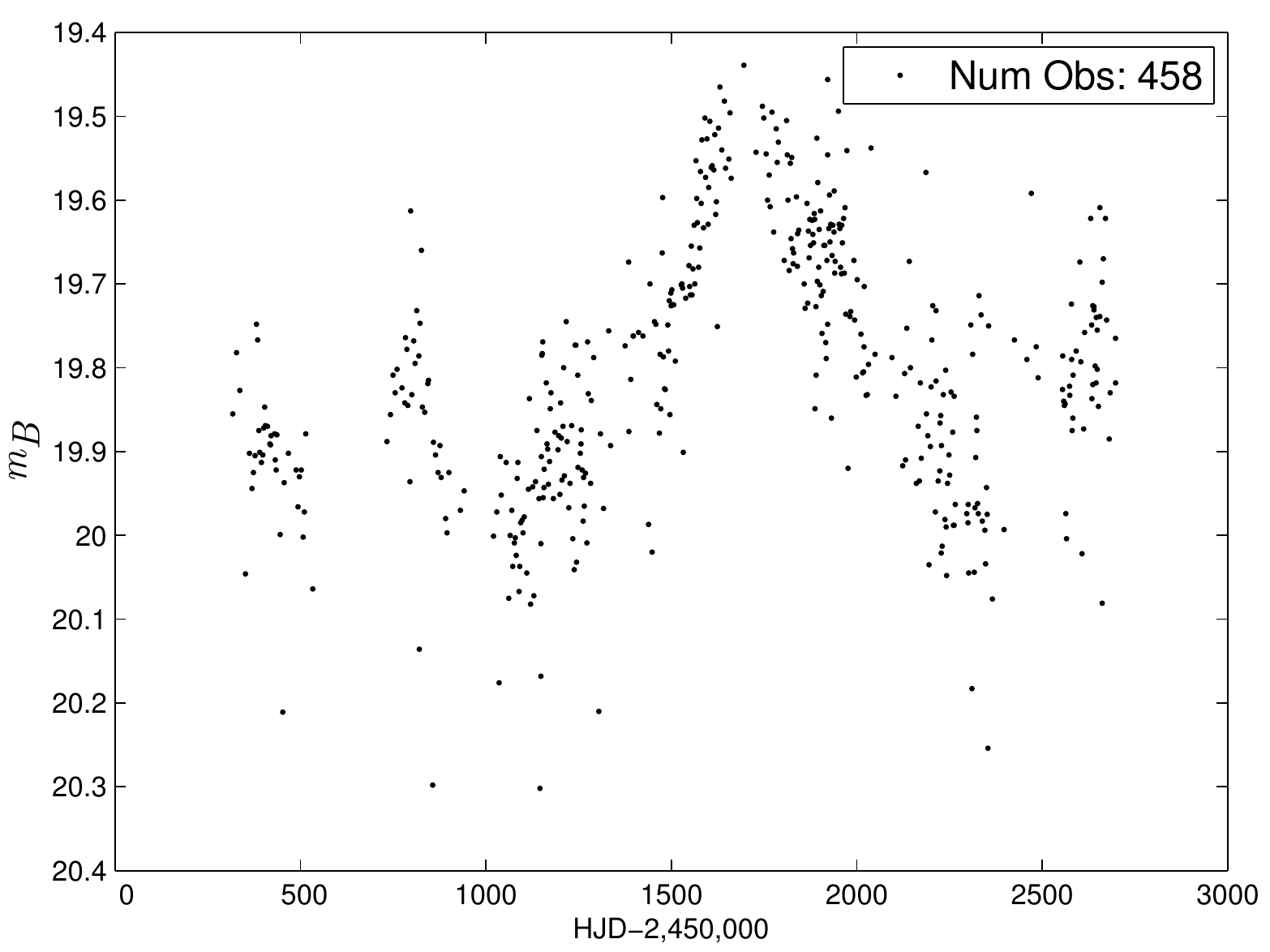}
\includegraphics[width=  \anchoEROS cm]{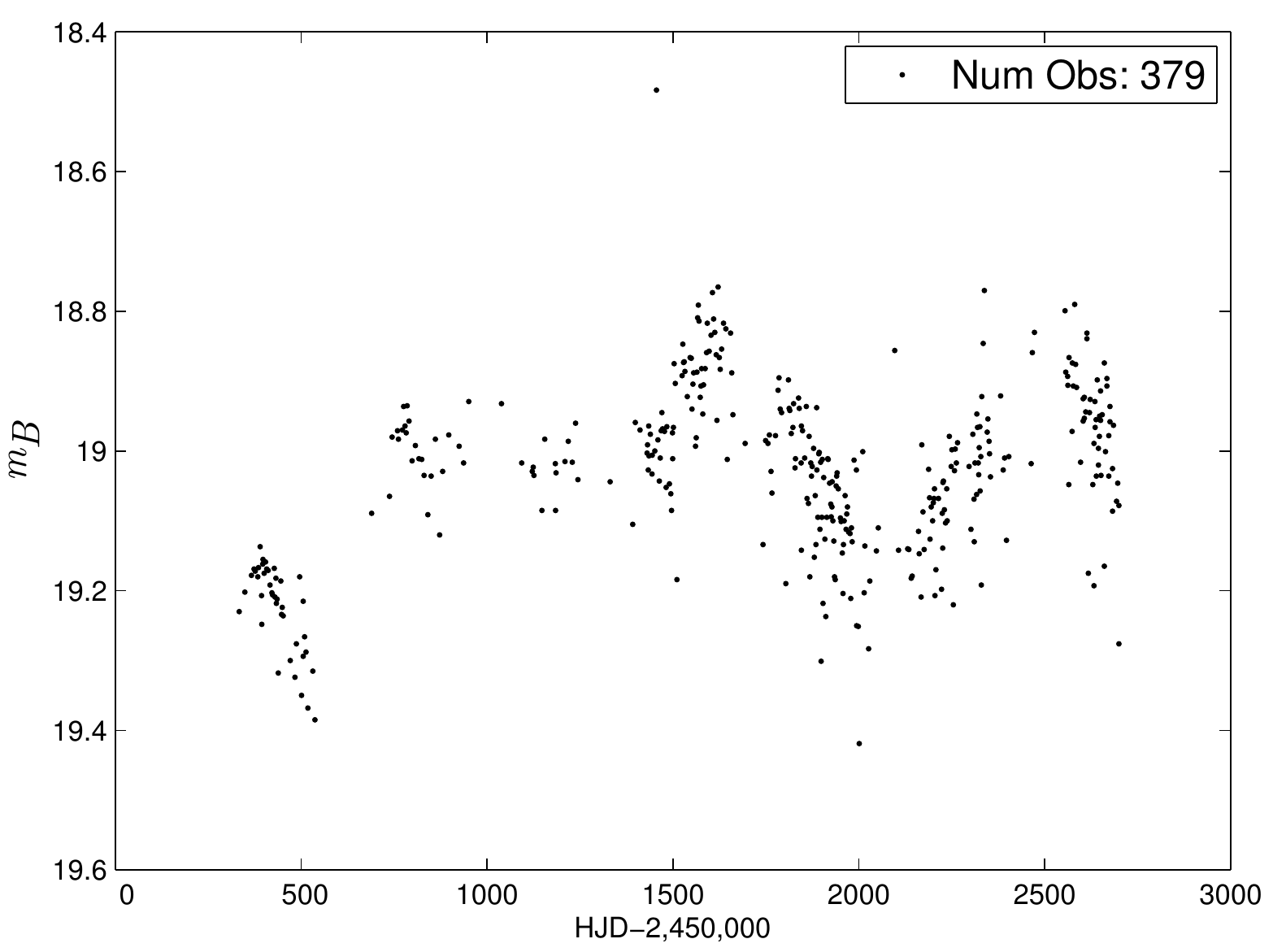}
\includegraphics[width=  \anchoEROS cm]{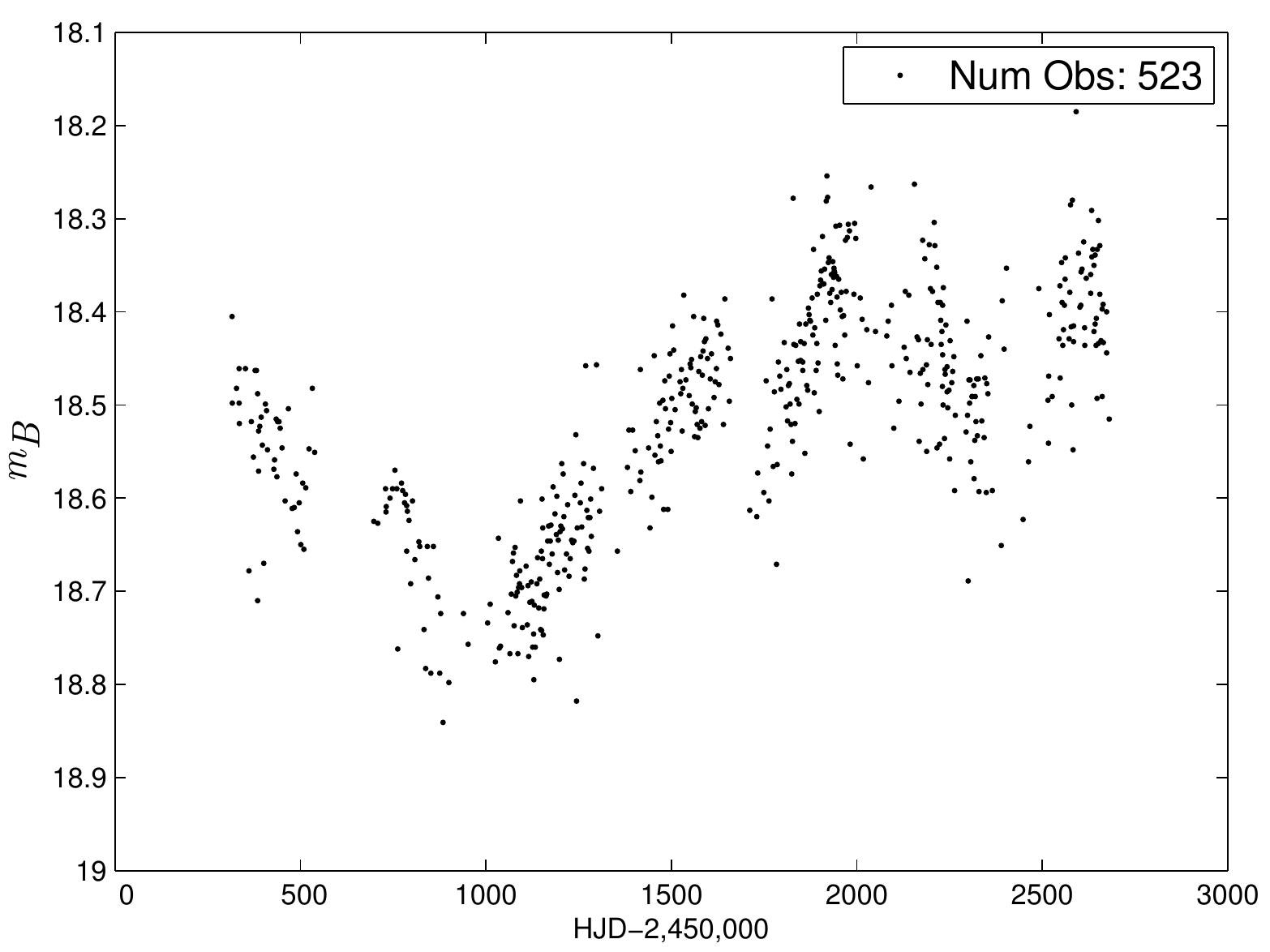}
\includegraphics[width=  \anchoEROS cm]{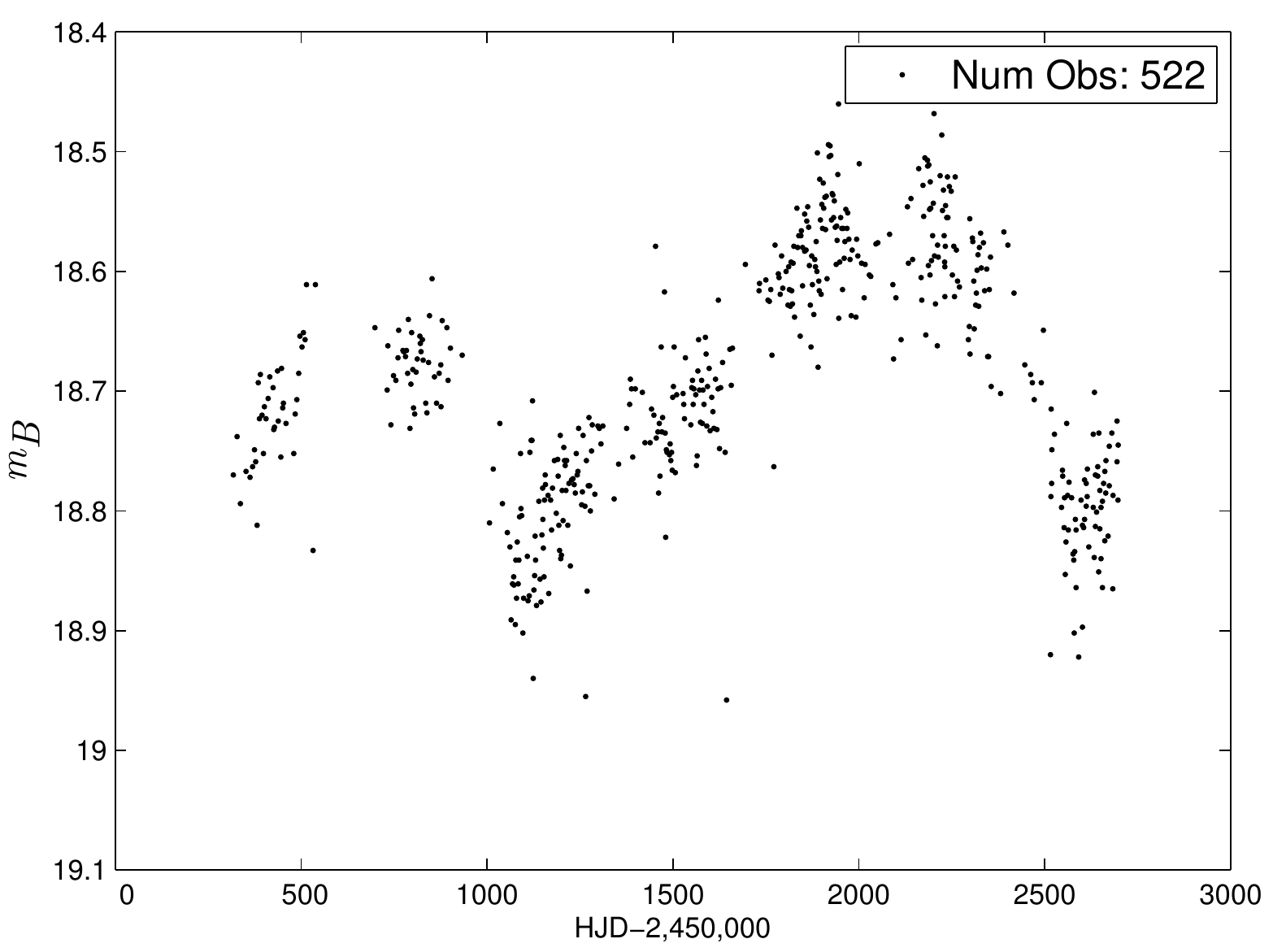}
\includegraphics[width=  \anchoEROS cm]{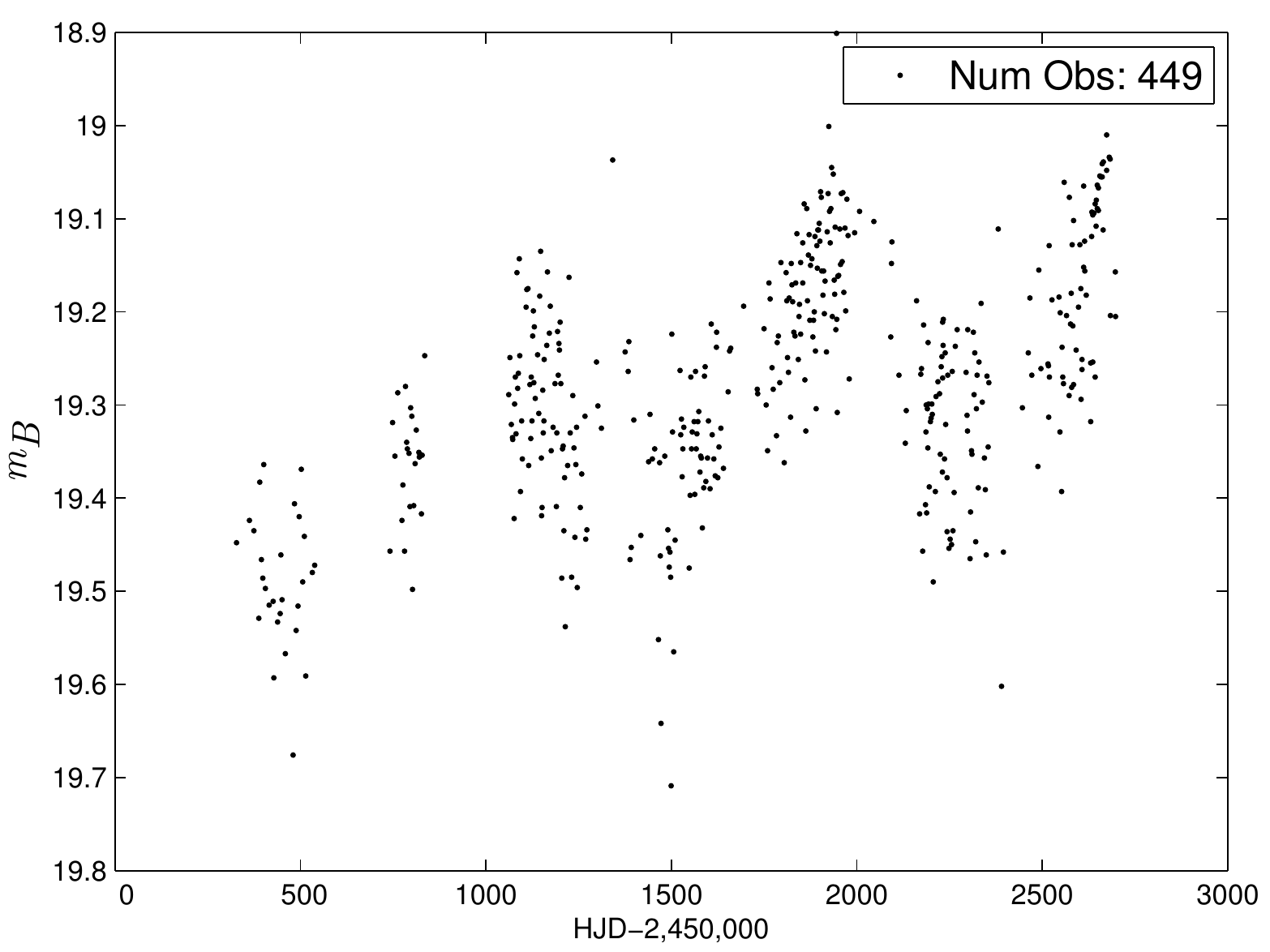}
\caption{Lightcurves of quasar candidates predicted on EROS-2 dataset}
\label{Fig:QSO_Cand_EROS_pics}
\end{figure*}

\subsection{MACHO dataset}
 MACHO was a survey 
 which observed the sky starting in July 1992 and ending in 1999
 to detect microlensing events produced by Milky Way halo objects.
 Several tens of millions of stars where observed in the Large Magellanic Cloud (LMC), Small Magellanic Cloud (SMC) and Galactic bulge \citep{Alcock2000ApJ}.
  
  For the MACHO dataset we built a training set composed of 3969 non-variable stars, 127 Be stars, 78 Cepheids, 193 eclipsing binaries, 288 RR Lyrae, 574 microlensing, 359 long-period variables, and 58 quasars. We get the variable stars from the list of known MACHO variable sources extracted from \href{http://simbad.u-strasbg.fr/simbad/}{SIMBAD}'s 
MACHO variable catalog\footnote{\href{http://vizier.u-strasbg.fr/viz-bin/VizieR?-source=II/247}{http://vizier.u-strasbg.fr/viz-bin/VizieR?-source=II/247}}
\citep{Alcock2001} and also from several other literature sources \citep{Alcock1997ApJL, Alcock1997ApJa, Wood2000PASA, Keller2002AJ, Thomas2005ApJ}.
 To get the non variable stars, we randomly chose a subset of MACHO lightcurves from a few MACHO LMC fields and removed all the known MACHO variables from the subset.

  Each lightcurve is described as a feature vector which contains 28 features, 14 features for band B and 14 features for band R as described in section \ref{sec:Features}.

\begin{figure*}
  \begin{center}
  \fbox{\includegraphics[width=0.8\textwidth]{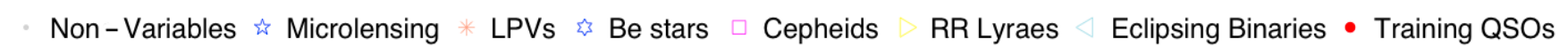}}
  \begin{minipage}[b]{0.48\textwidth}
    \centering
    \includegraphics[width=7cm]{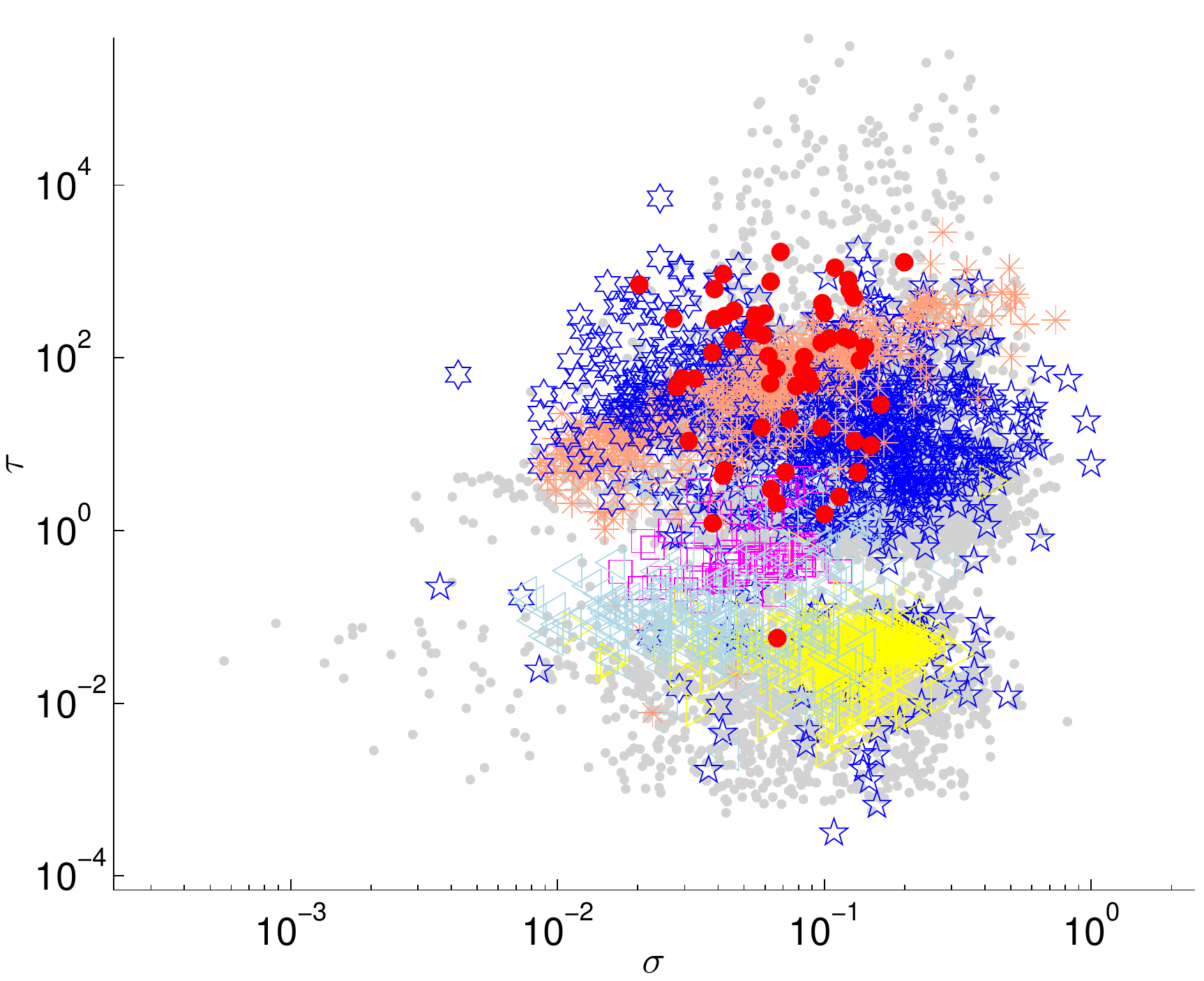}
  \end{minipage}
  \hspace{0.5cm}
  \begin{minipage}[b]{0.48\textwidth}
    \centering
    \includegraphics[width=7cm]{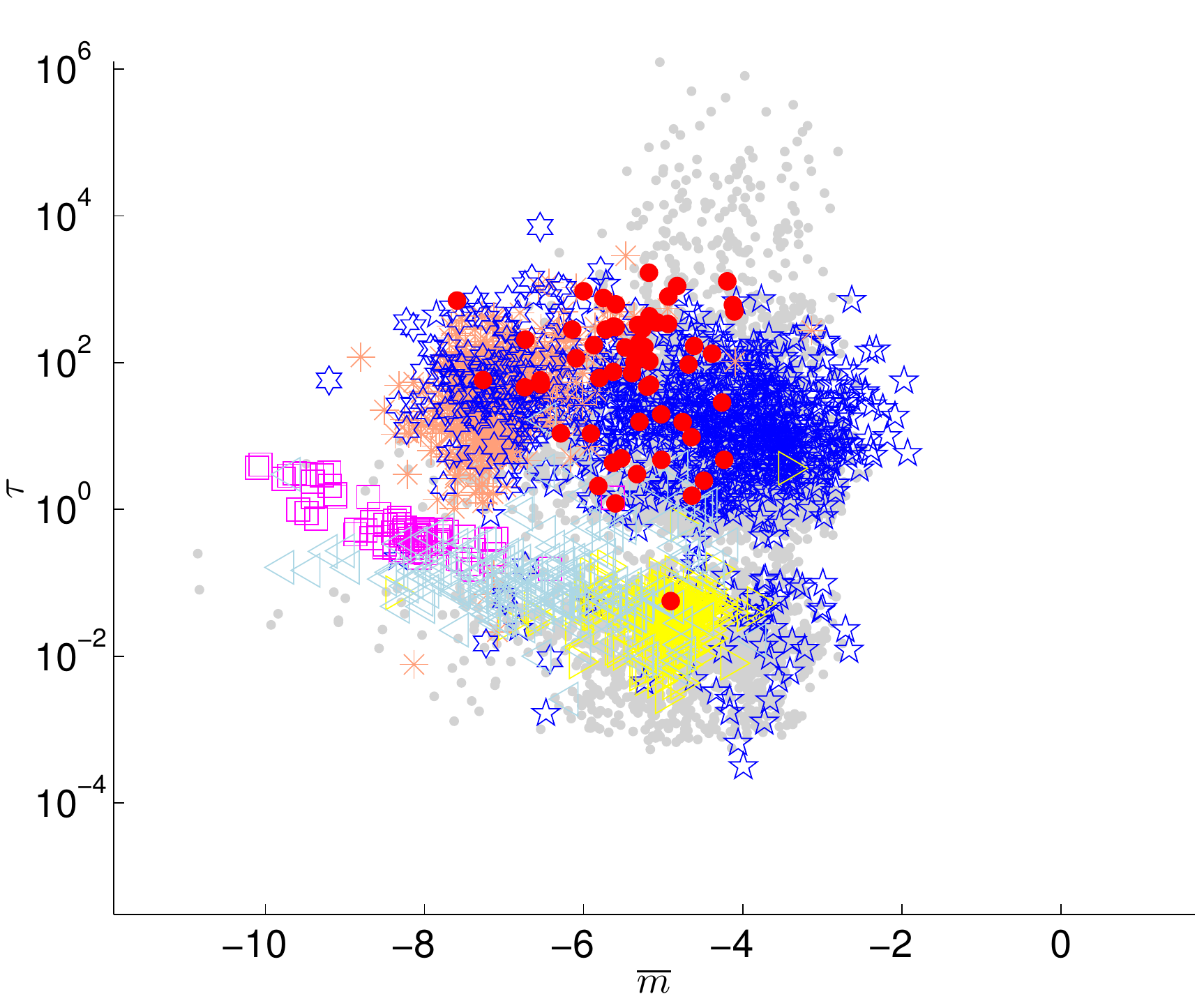}
  \end{minipage}
  \begin{minipage}[b]{0.48\textwidth}
    \centering
    \includegraphics[width=7cm]{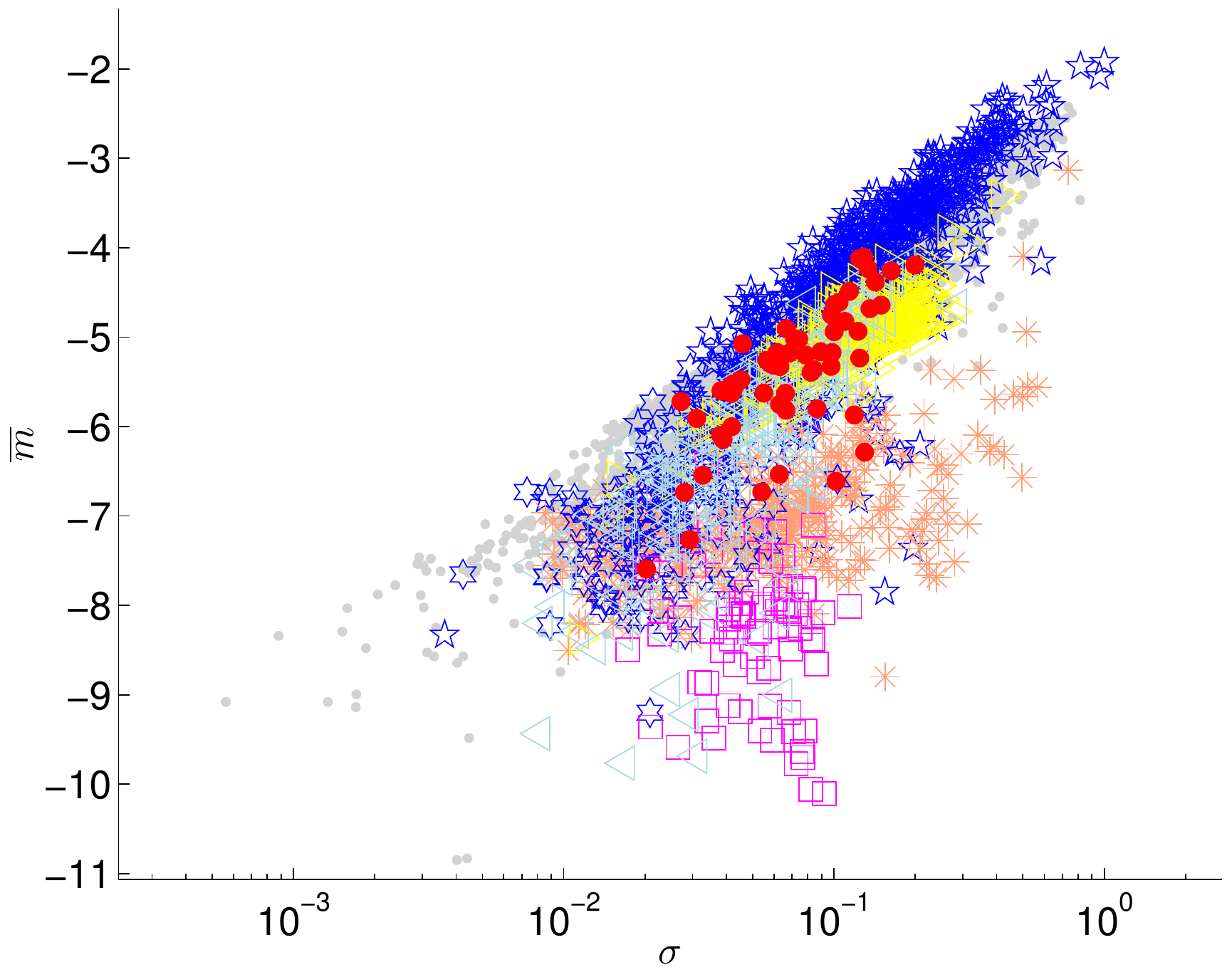}
  \end{minipage}
  \end{center}
  \caption{Training set projected on different pairs of CAR(1) features  for MACHO data.}
  \label{Fig:MACHO_Training_Features_1}  
 \end{figure*}

\begin{figure*}
  \begin{center}
 \fbox{\includegraphics[width=0.8\textwidth]{./Plots/MACHO/Symbols_MACHO_Train-eps-converted-to.pdf}} 
  \begin{minipage}[b]{0.48\textwidth}
    \centering
    \includegraphics[width=7cm]{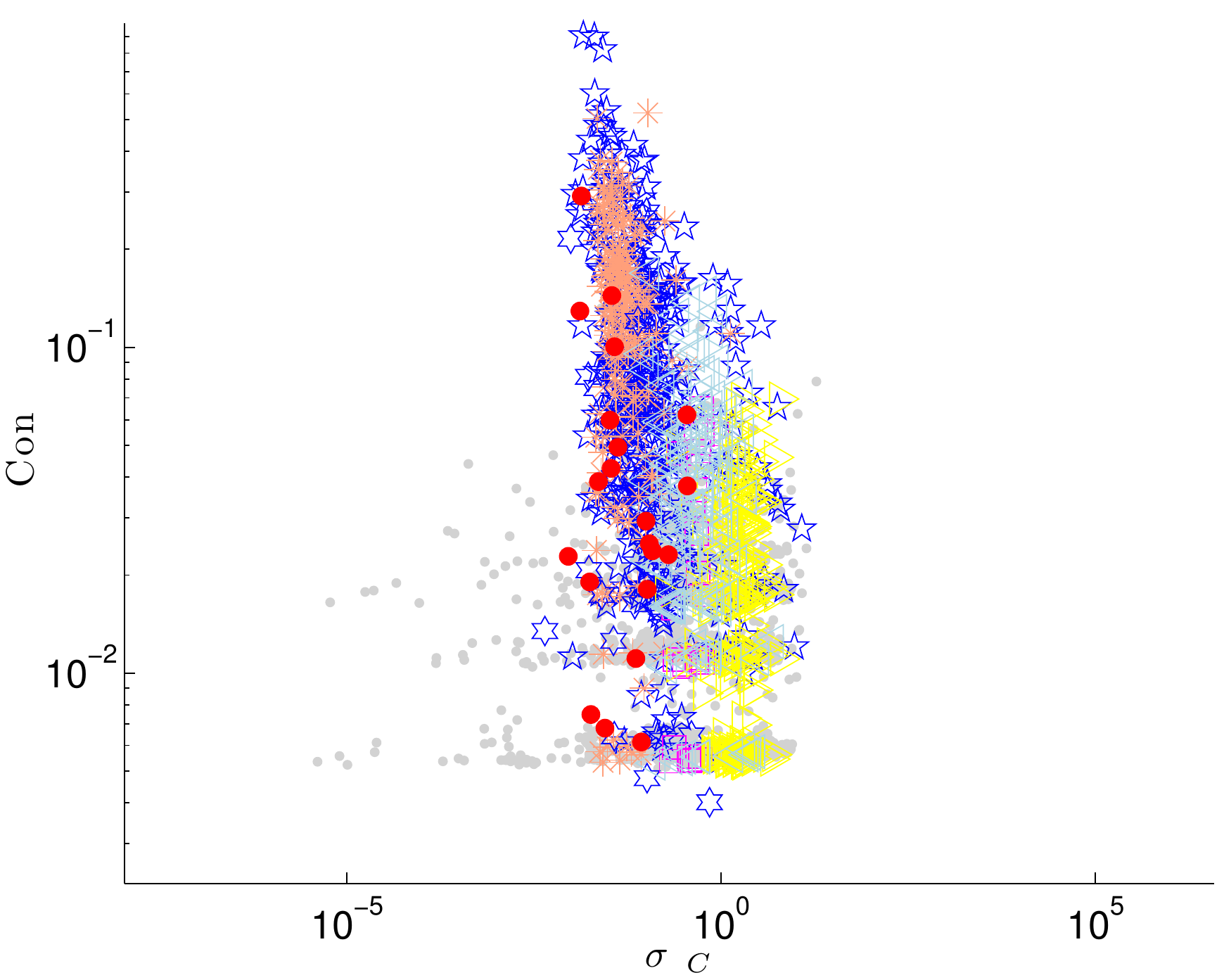}
  \end{minipage}
  \begin{minipage}[b]{0.48\textwidth}
    \centering
    \includegraphics[width=7cm]{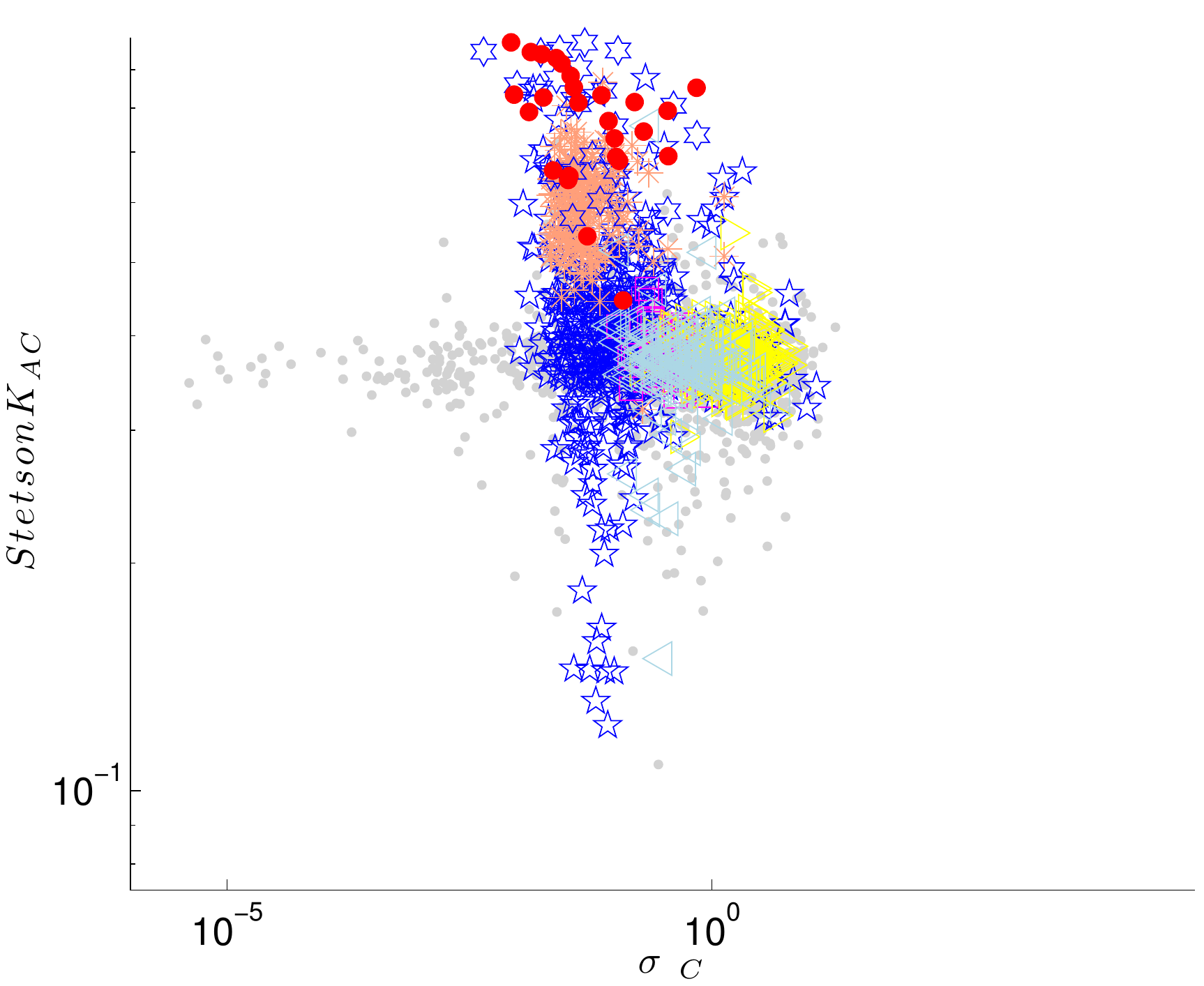}
  \end{minipage}
  \begin{minipage}[b]{0.48\textwidth}
    \centering
    \includegraphics[width=7cm]{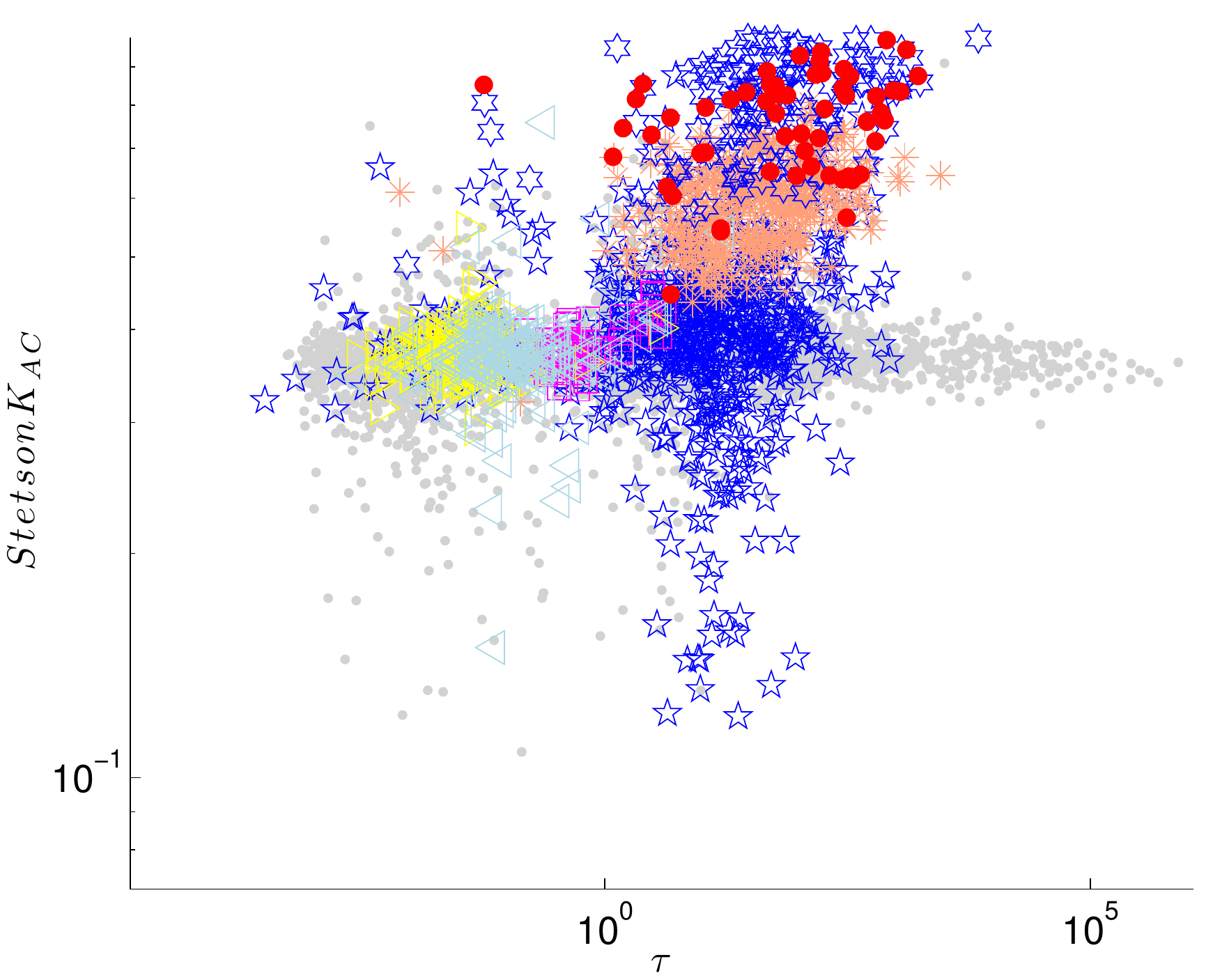}
  \end{minipage}
  \begin{minipage}[b]{0.48\textwidth}
    \centering
    \includegraphics[width=7cm]{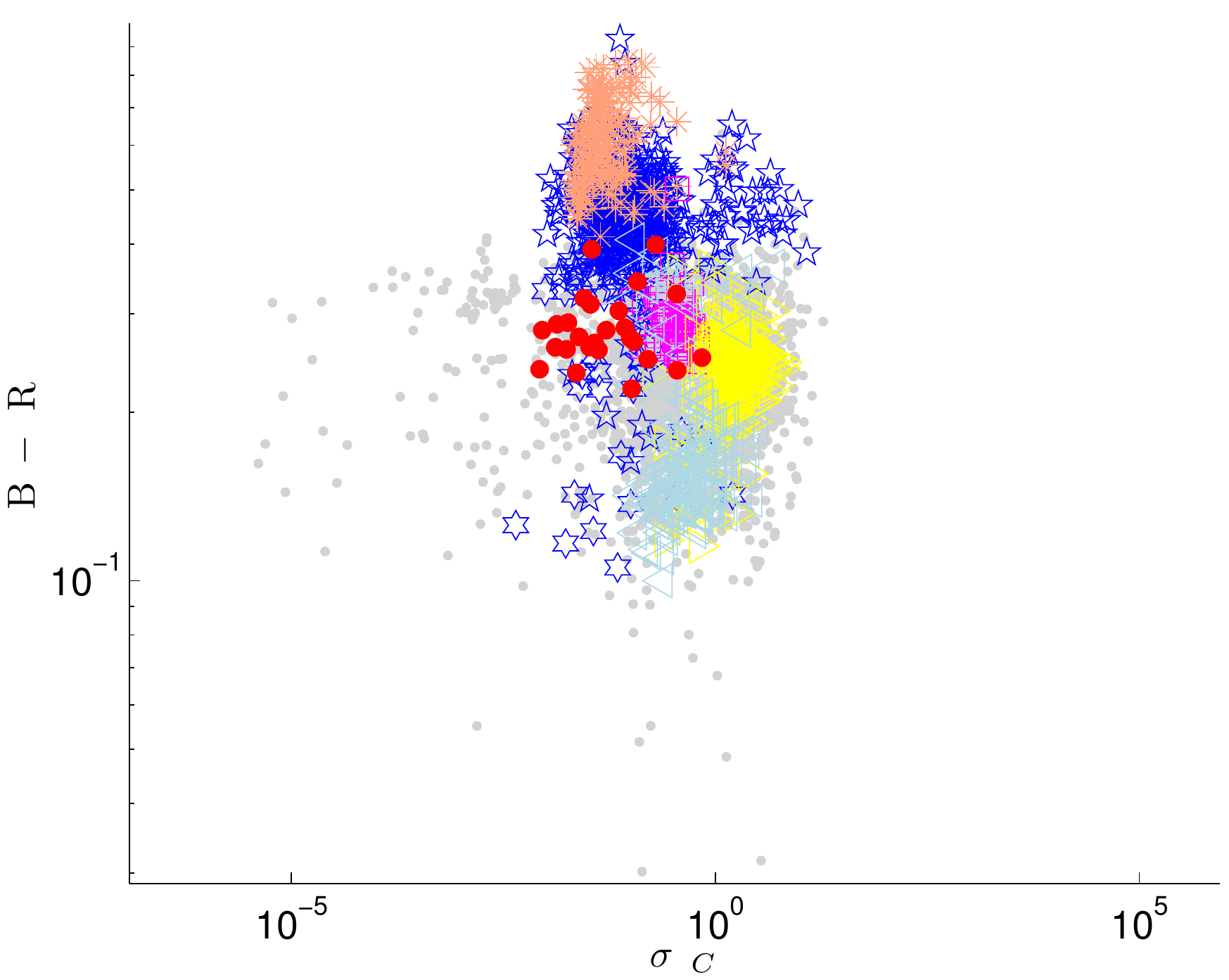}
  \end{minipage}  
  \begin{minipage}[b]{0.48\textwidth}
    \centering
    \includegraphics[width=7cm]{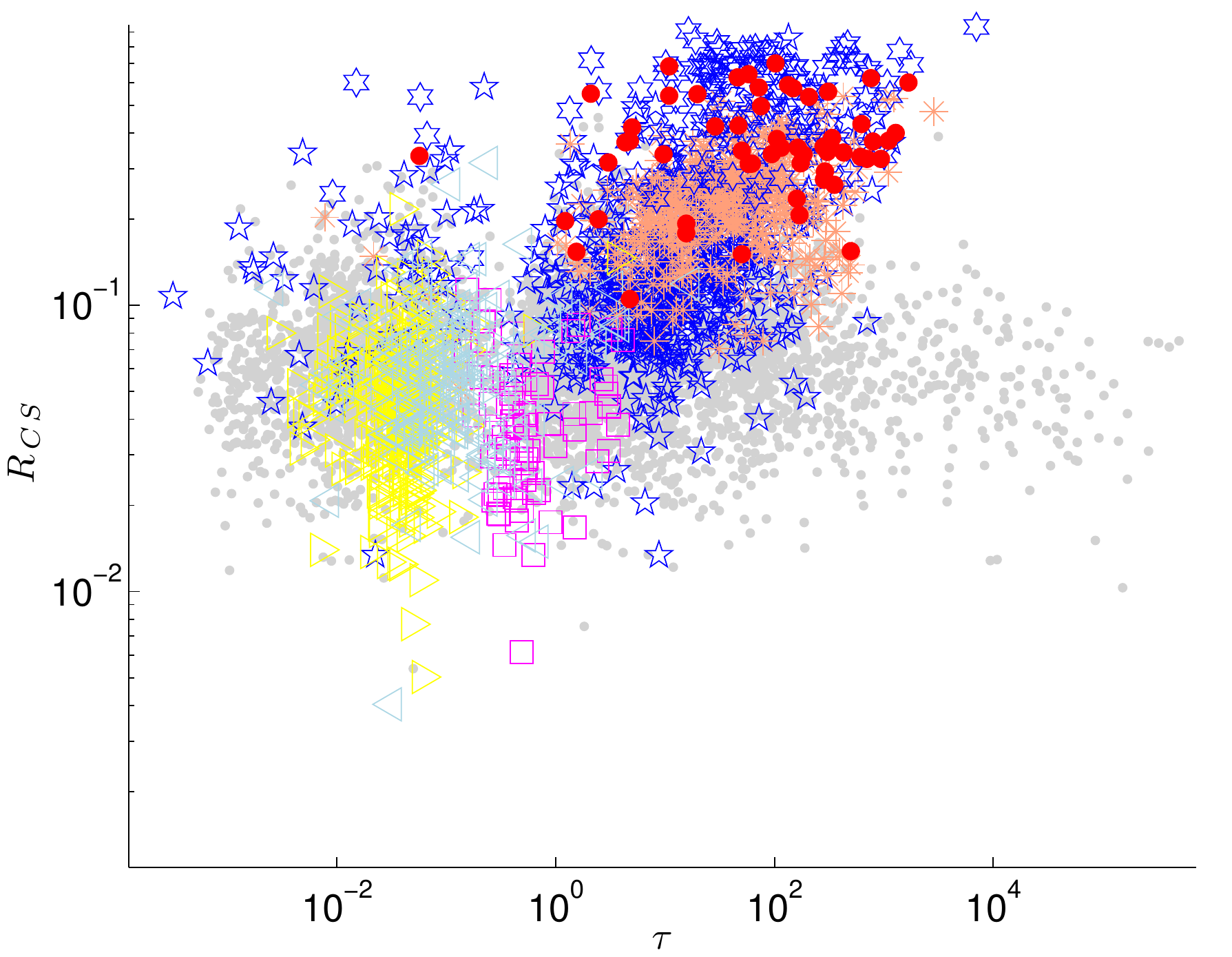}
  \end{minipage}
   \end{center} 
  \caption{Projections on different pairs of features, combining CAR(1) features with time series features for MACHO training data}
  \label{Fig:MACHO_Training_Features_2}  
\end{figure*}

Figures \ref{Fig:MACHO_Training_Features_1} and \ref{Fig:MACHO_Training_Features_2} show the training set projected on a two variables  feature space. We can see that $\sigma_C$ and $\tau$ features show separations between two groups of classes: i) non-variables, Cepheid and Eclipsing Binaries stars and ii) quasars, Microlensings, LPVs and Be stars. Combining $\overline{m}$ and $\tau$ we can see a cluster of quasars, which overlaps with some of the Be stars, non-variables, Microlensing and long period variables, but separates very well quasars from Cepheids, Eclipsing Binaries stars and most of the non-variables. Projecting on  $\sigma_C$ and $\overline{m}$ we can see that quasars separates from LPVs, Cepheids, Eclipsing Binaries, most of Be stars, most of the Microlensings and most of the non-variables. The biggest overlap is with Microlensings. 

By examining these projections we can see that quasars are clustered in high values of $\tau$, with higher values compared to Eclipsing Binaries, Cepheids and RR Lyraes. $\sigma_C$ is very good to separate quasars from non-variables, also from Cepheids, RR Lyraes and Eclipsing Binaries stars. $\sigma_C$ is  not a good feature to separate quasars from Microlensings, Be stars and LPVs, but combining $\sigma_C$ with B $-$ R we get a strong separation between them.

  Table \ref{table:ClassifiersTraining} shows comparative results among different classification models. We included a Support Vector Machine, Random Forest and Radom Forest Boosted with AdaBoost. On each case the classifier is tuned with the optimal set of parameters. 
    
    \begin{table}
    \caption{F-Score for the MACHO training set using 10-fold cross validation for different classification models. Each classifiers is tuned with the optimal set of parameters. We can see that the boosted version of Random Forest with CAR features outperforms other classification models. In all cases
   using CAR features improves the result of the corresponding classifier.}  
   \begin{center}
    \begin{tabular}{c|c|c|c|c|c|}
     \hline 
      SVM     & SVM   &   RF        &    RF  & AB+RF  & AB+RF\\
      No CAR & CAR   & No CAR &  CAR &  No CAR & CAR\\
     \hline 
     0.787 & 0.824 & 0.826 & 0.841 & 0.844 & 0.877\\ 
      \hline 
    \end{tabular}
    \end{center}
     \label{table:ClassifiersTraining}
    \end{table}

  After we select and fit the model to the training set, we run on the whole MACHO data (about 40 million of lightcurves), from where we
  get  2551 quasar candidates. We crossmatch our candidates with the 2566 and 663 strong candidates  in our previous work \citep{Kim2011ApJ} getting 1148 and 494 matches respectively. 


\newcommand{\ancho}{5}

\begin{figure*}
    \includegraphics[width= \ancho cm] {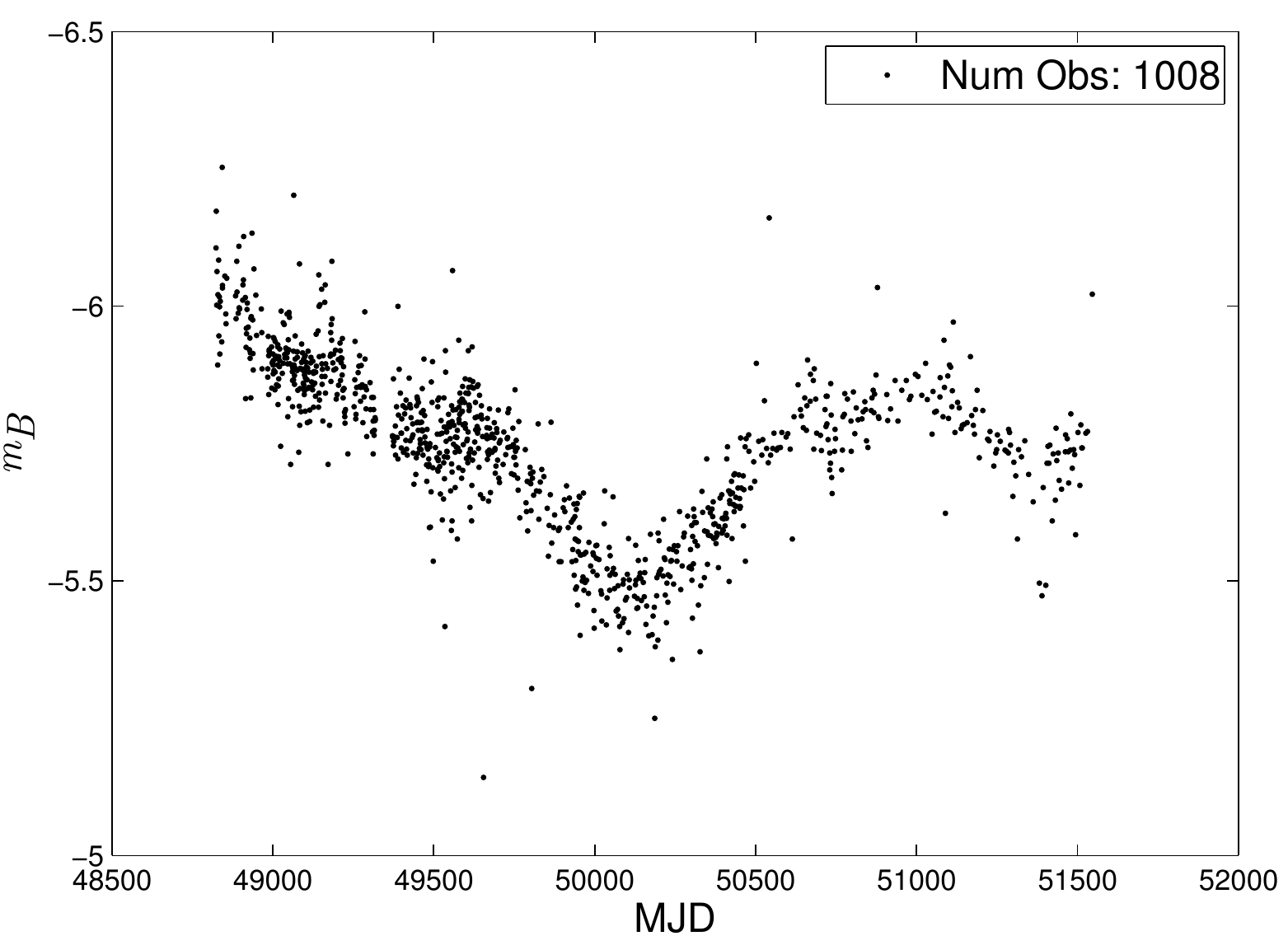}
    \includegraphics[width= \ancho cm] {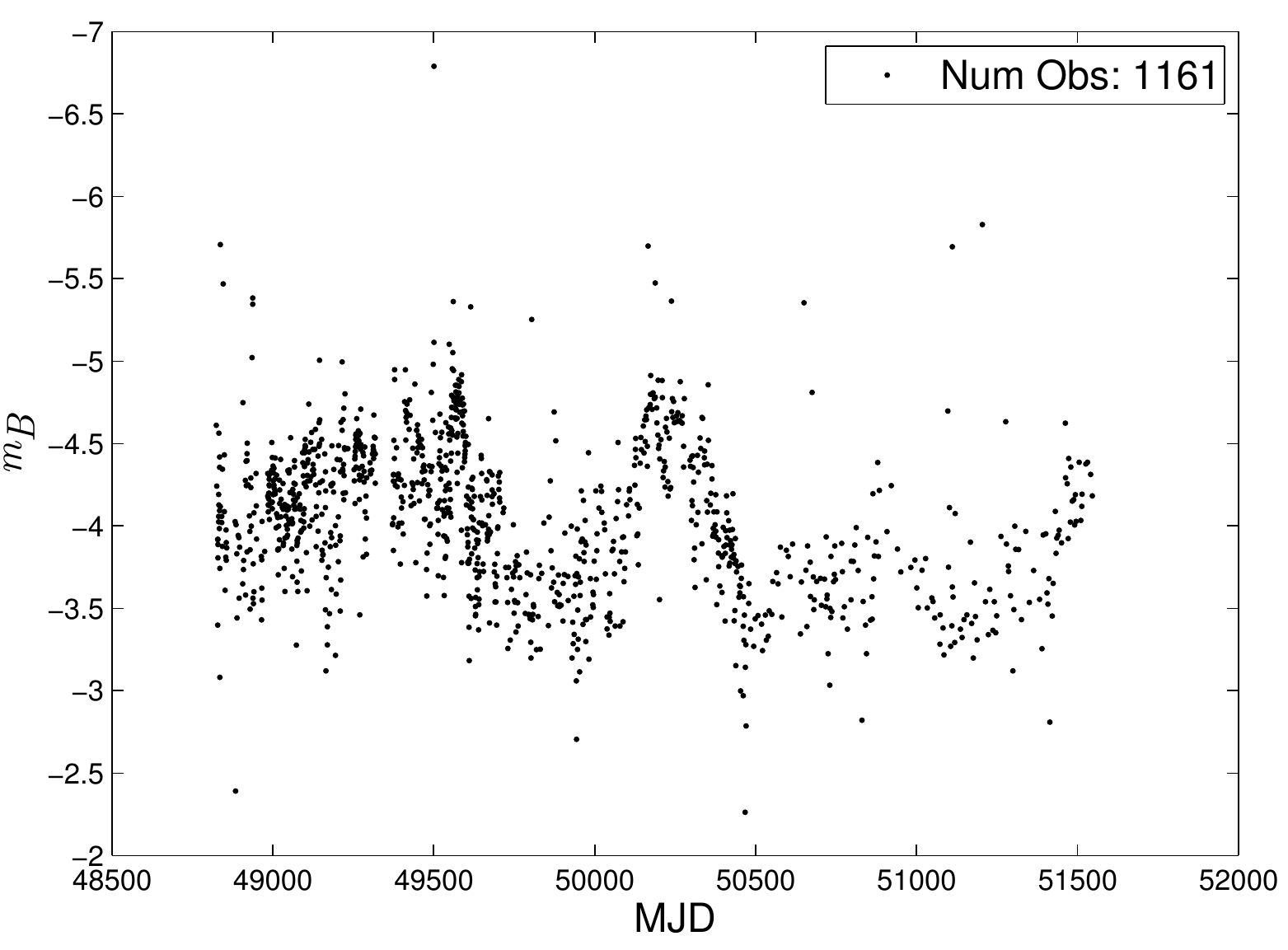}
    \includegraphics[width= \ancho cm] {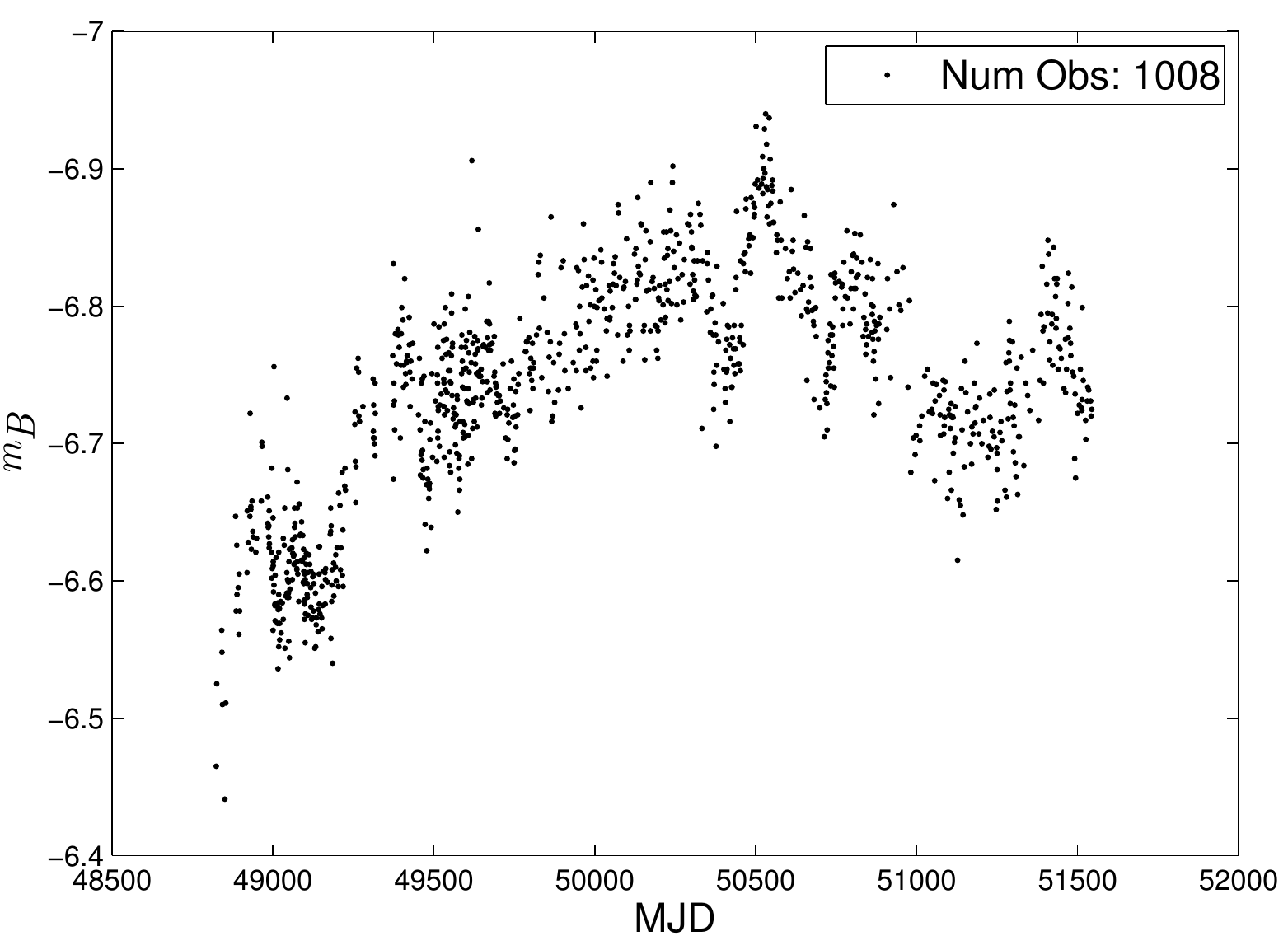}
    \includegraphics[width= \ancho cm] {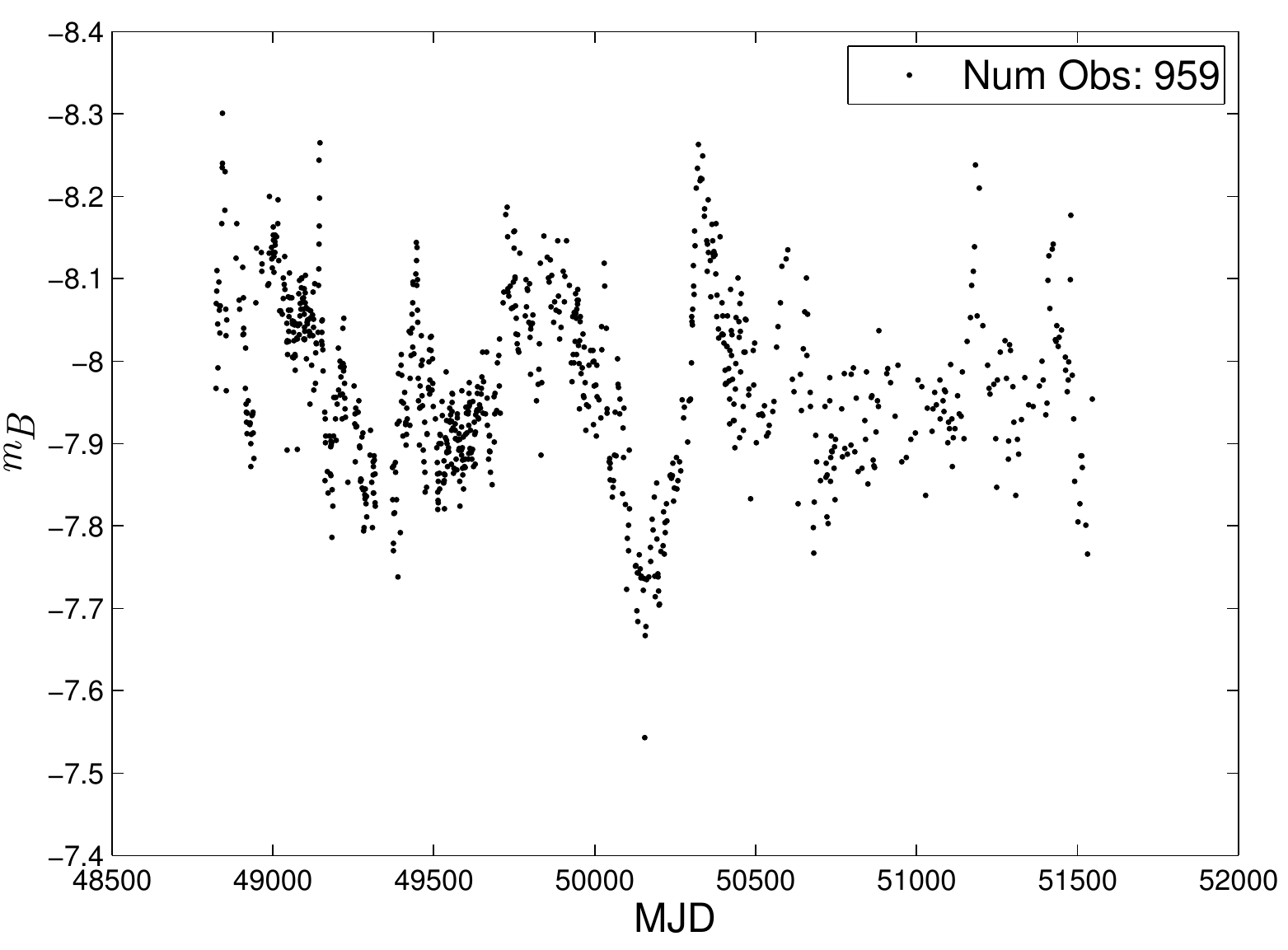}
    \includegraphics[width= \ancho cm] {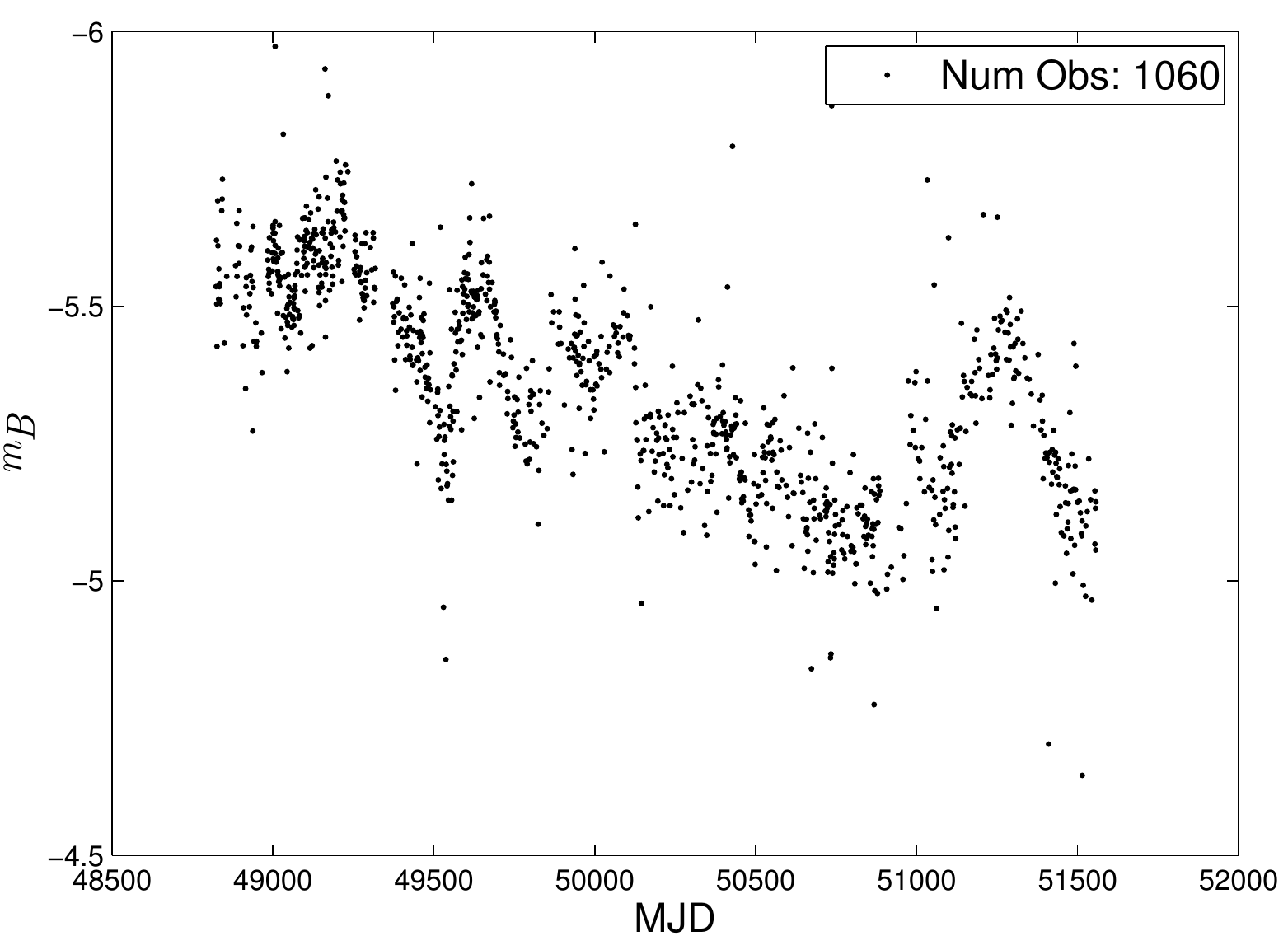}
    \includegraphics[width= \ancho cm] {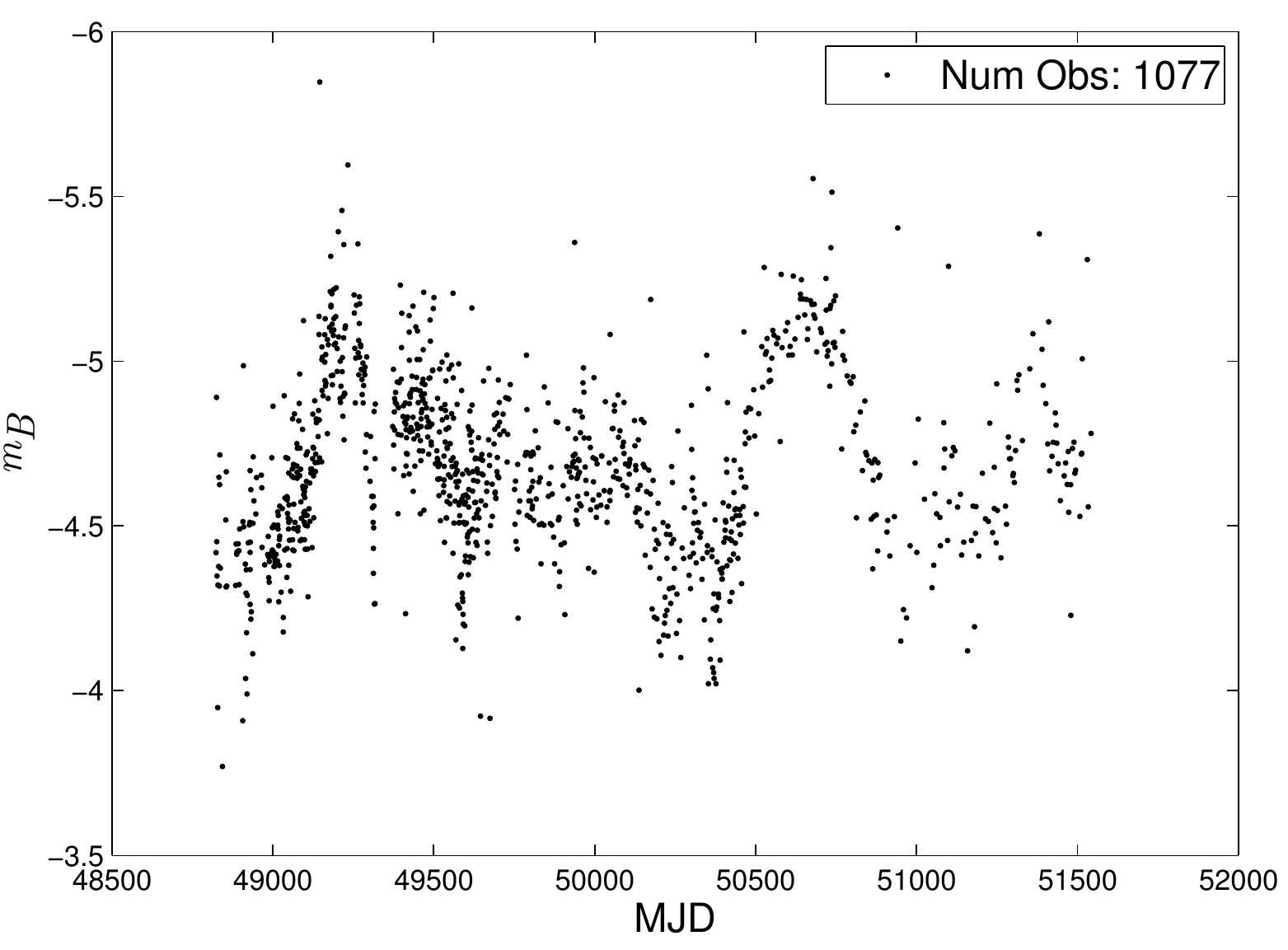}
    \includegraphics[width= \ancho cm] {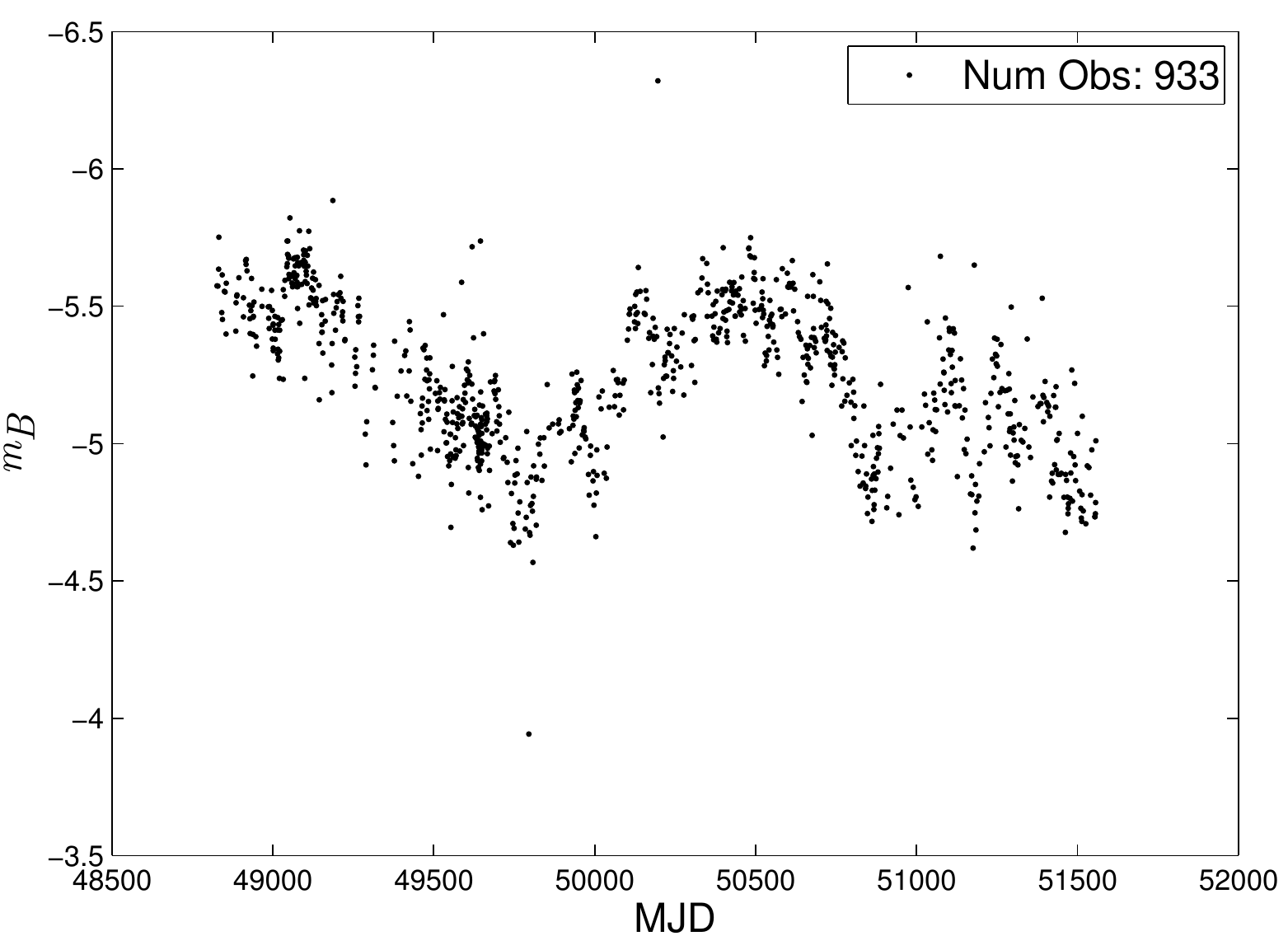}
    \includegraphics[width= \ancho cm] {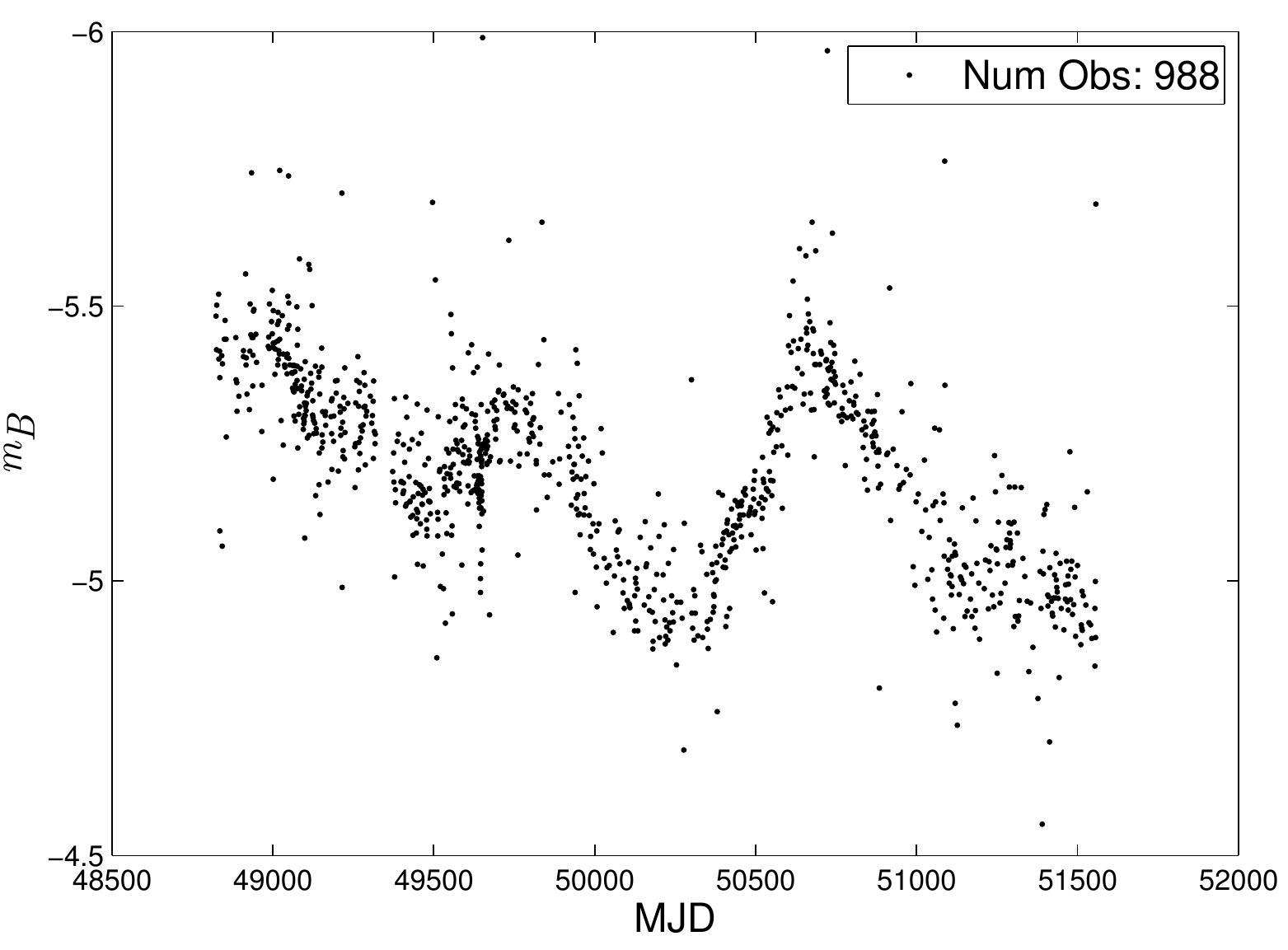}
    \includegraphics[width= \ancho cm] {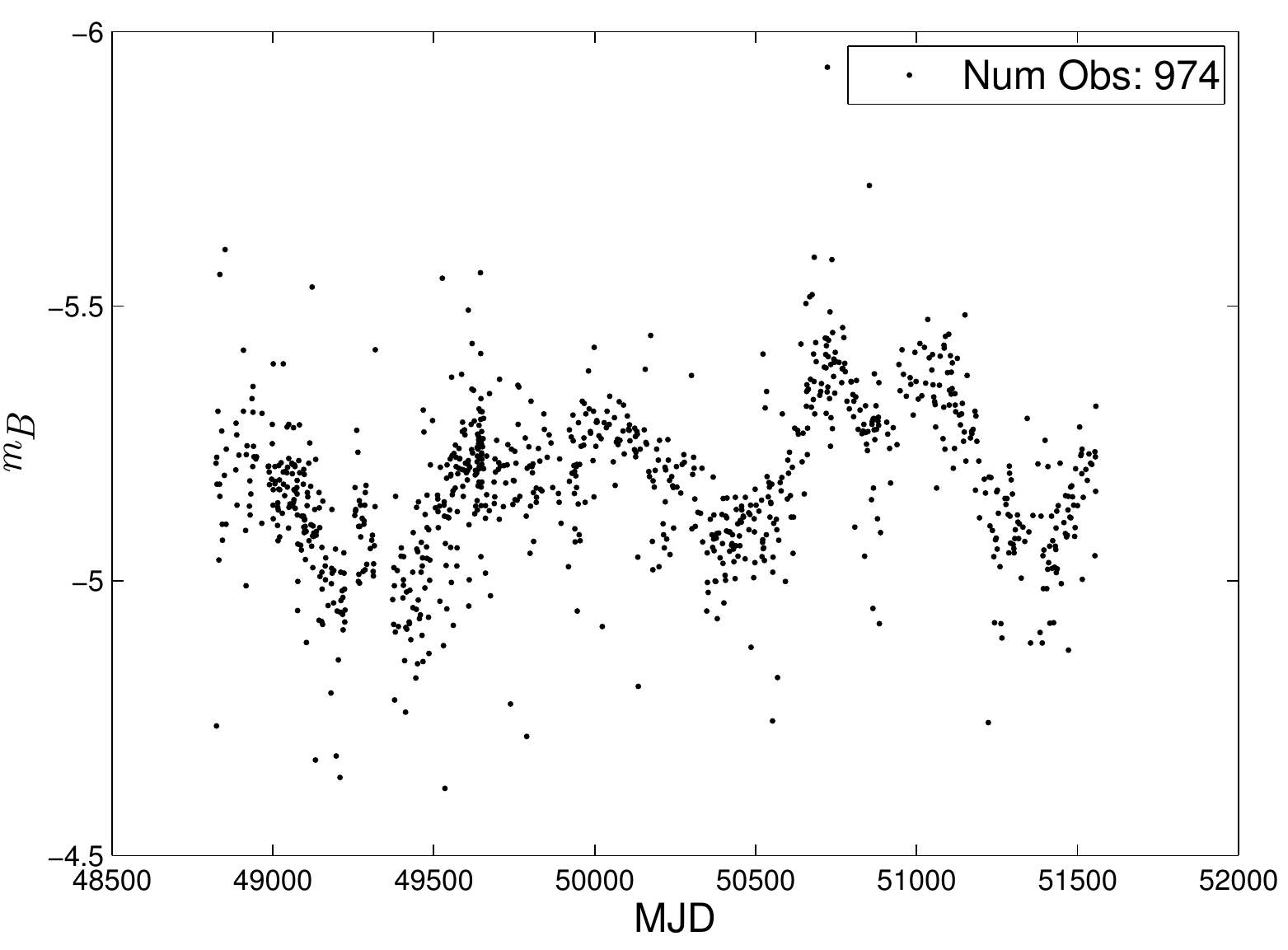}
    \includegraphics[width= \ancho cm] {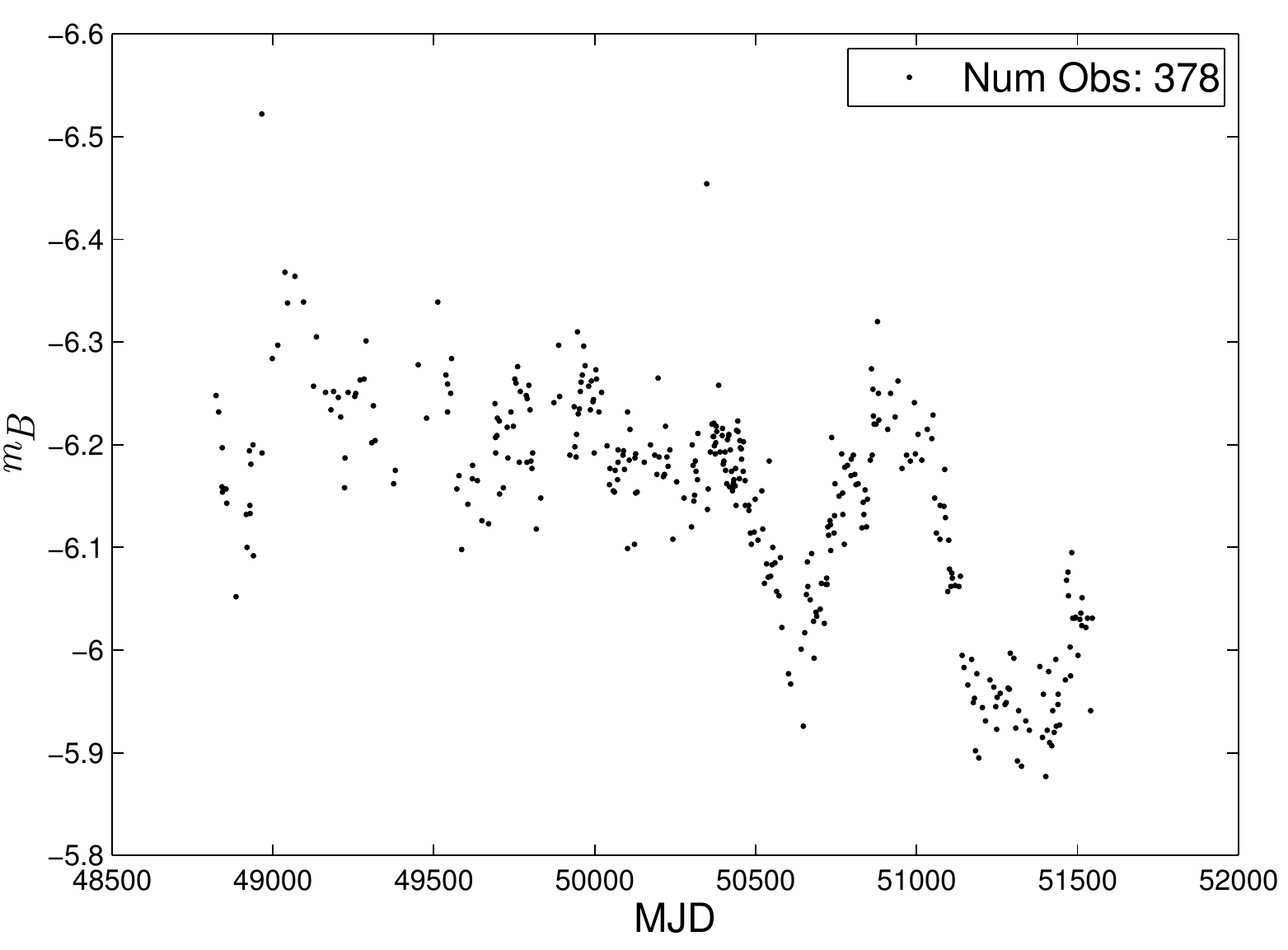}
    \includegraphics[width= \ancho cm] {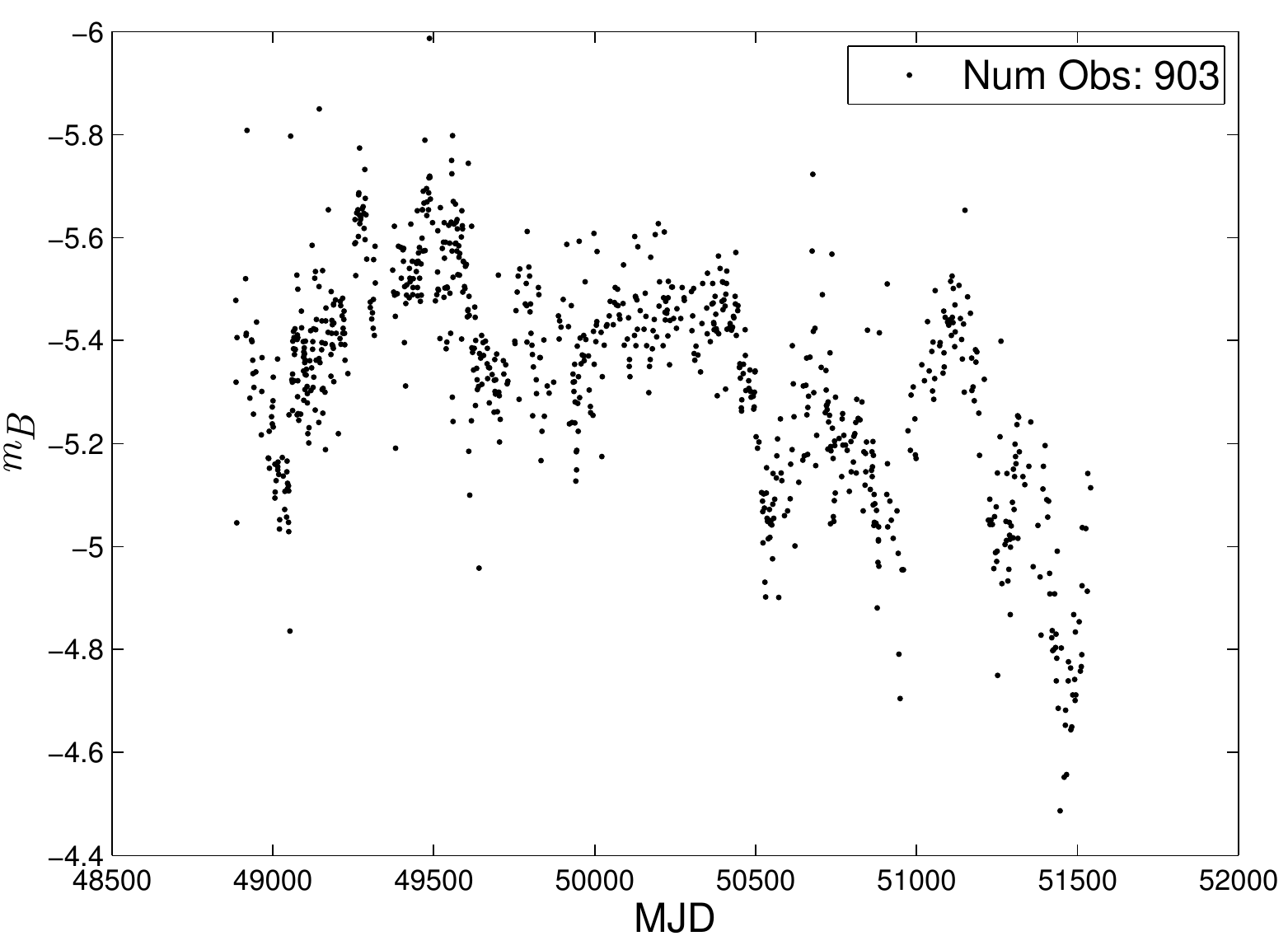}
    \includegraphics[width= \ancho cm] {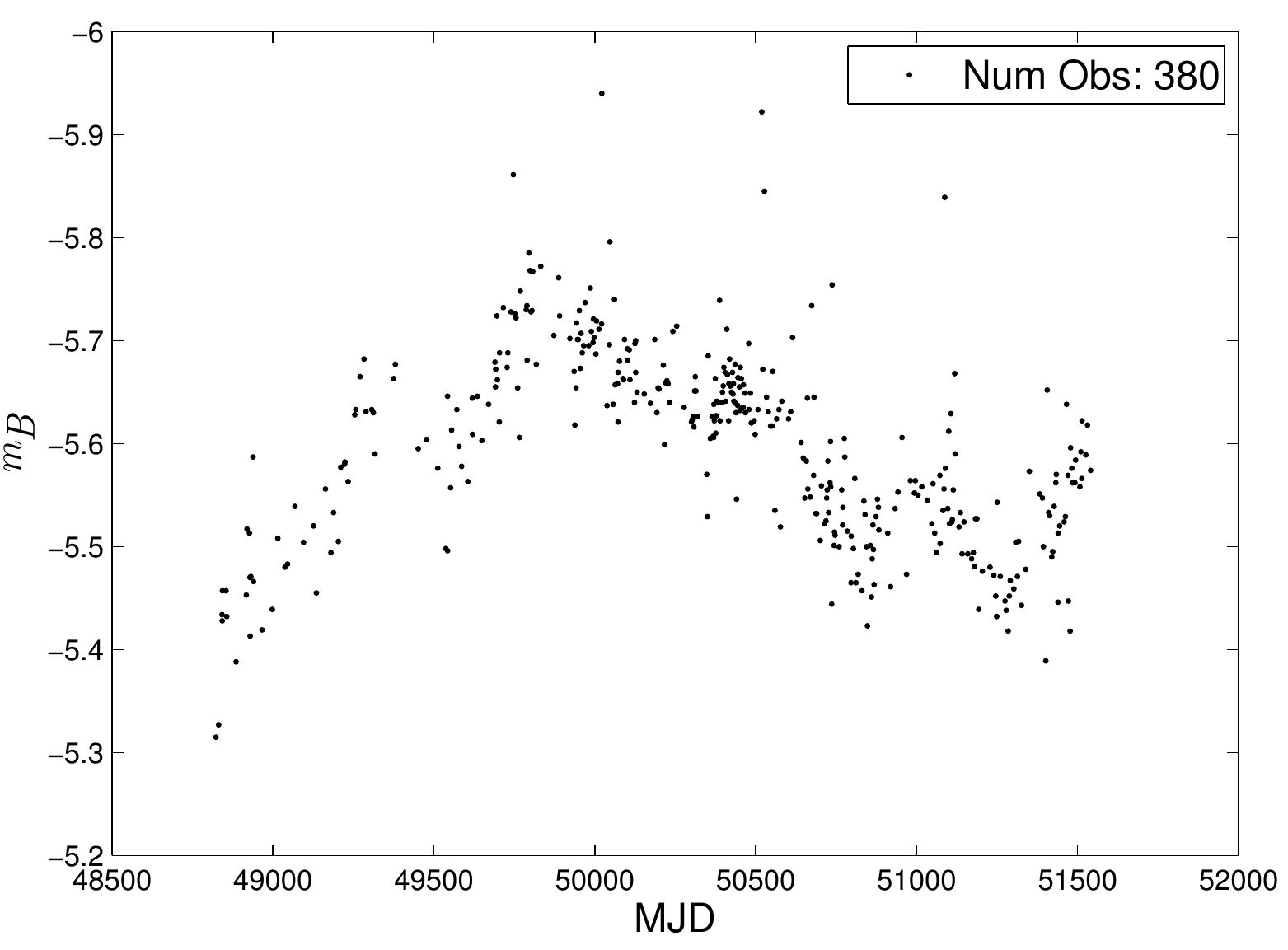}
\caption{Lightcurves of new quasars candidates predicted from MACHO dataset}
\label{Fig:QSO_NEW_Cand_MACHO_pics}
\end{figure*}

 Figure \ref{Fig:QSO_NEW_Cand_MACHO_pics} shows some of the new candidates we find that are not in the previous list for MACHO candidates in \citet{Kim2011ApJ}

\begin{figure*}
    \includegraphics[width= 5 cm] {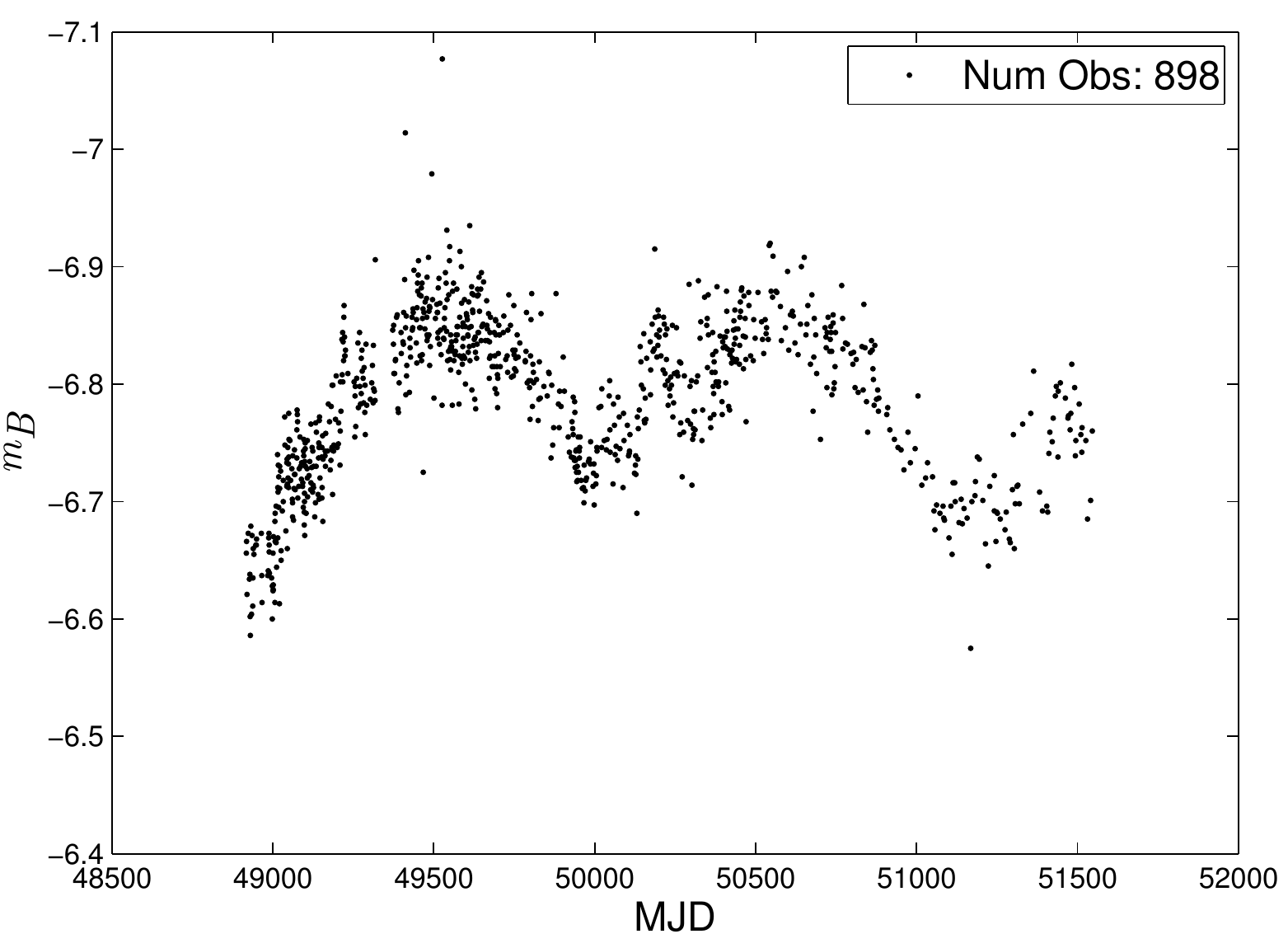}
\caption{Lightcurve of a wrongly predicted quasar in MACHO dataset}
\label{Fig:QSO_wrong_Cand_MACHO}
\end{figure*}

 There are some cases where the model confuses a periodic star with a quasar. Figure \ref{Fig:QSO_wrong_Cand_MACHO} shows one example of this case. 

\begin{figure*}
  \begin{center}
 \fbox{\includegraphics[width=0.8\textwidth]{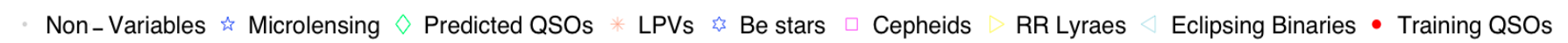}} 
  \begin{minipage}[b]{0.48\textwidth}
    \centering
    \includegraphics[width=7cm]{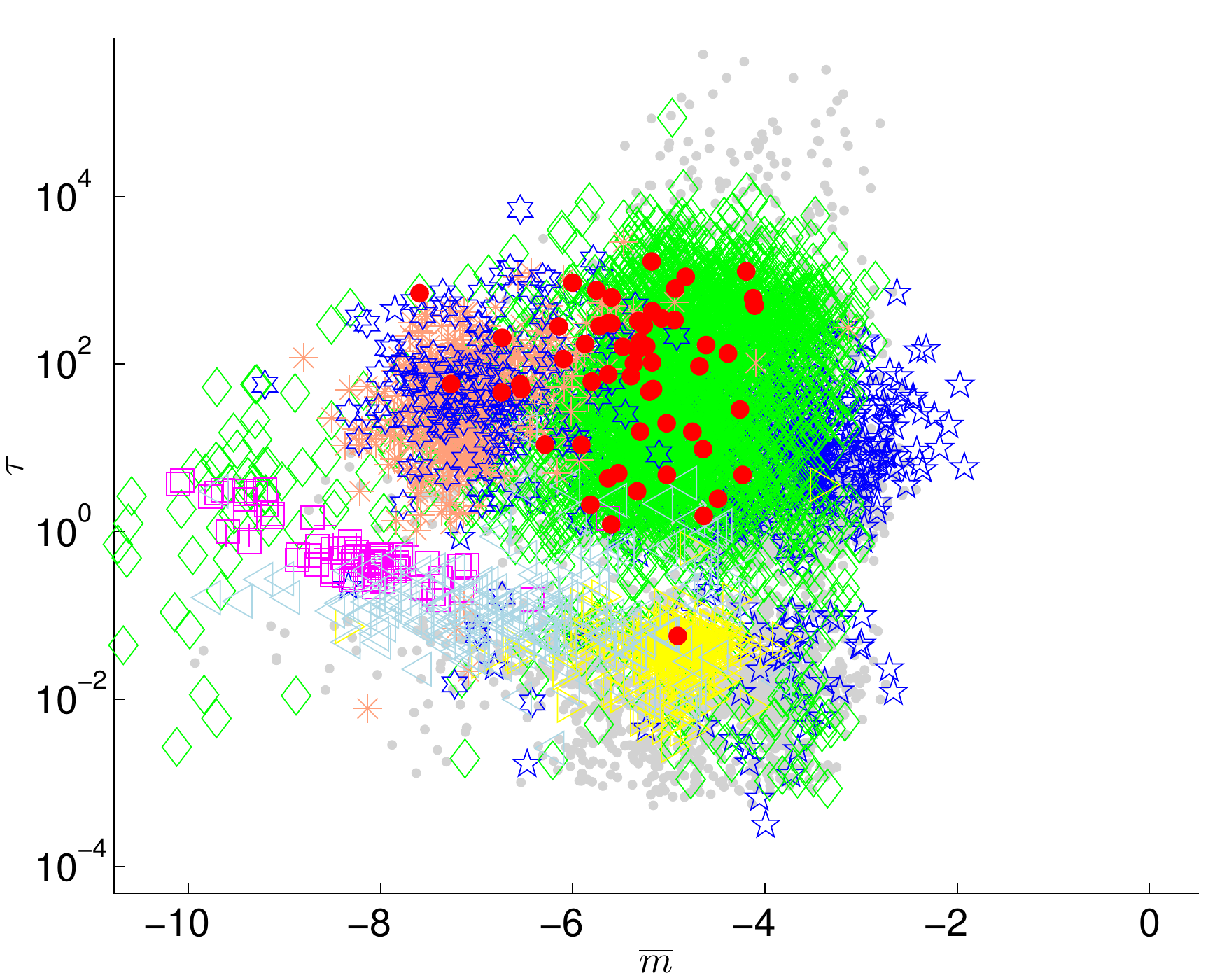}
  \end{minipage}
  \hspace{0.5cm}
  \begin{minipage}[b]{0.48 \textwidth}
    \centering
    \includegraphics[width=7cm]{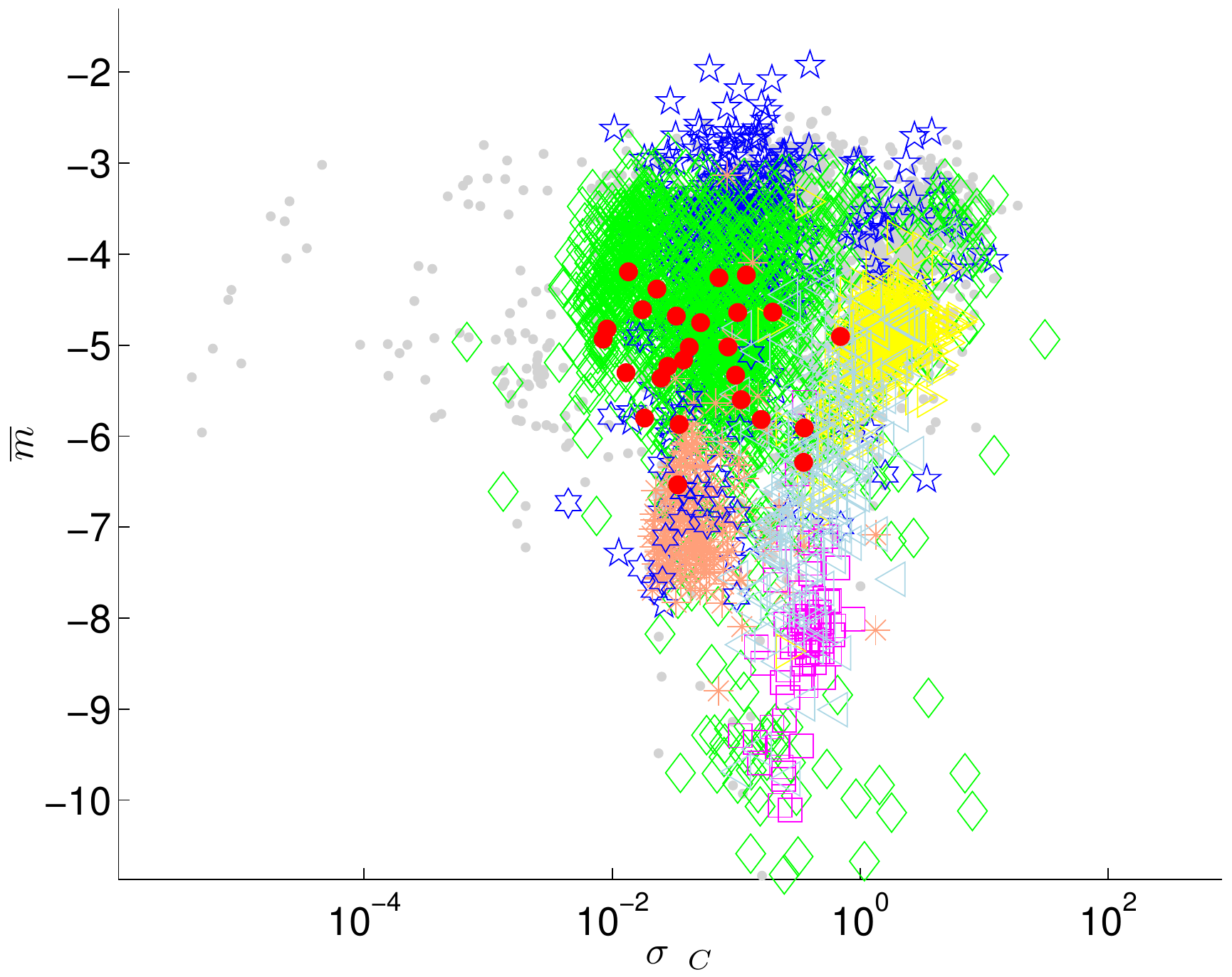}
   \end{minipage}
  \begin{minipage}[b]{0.48\textwidth}
    \centering
    \includegraphics[width=7cm]{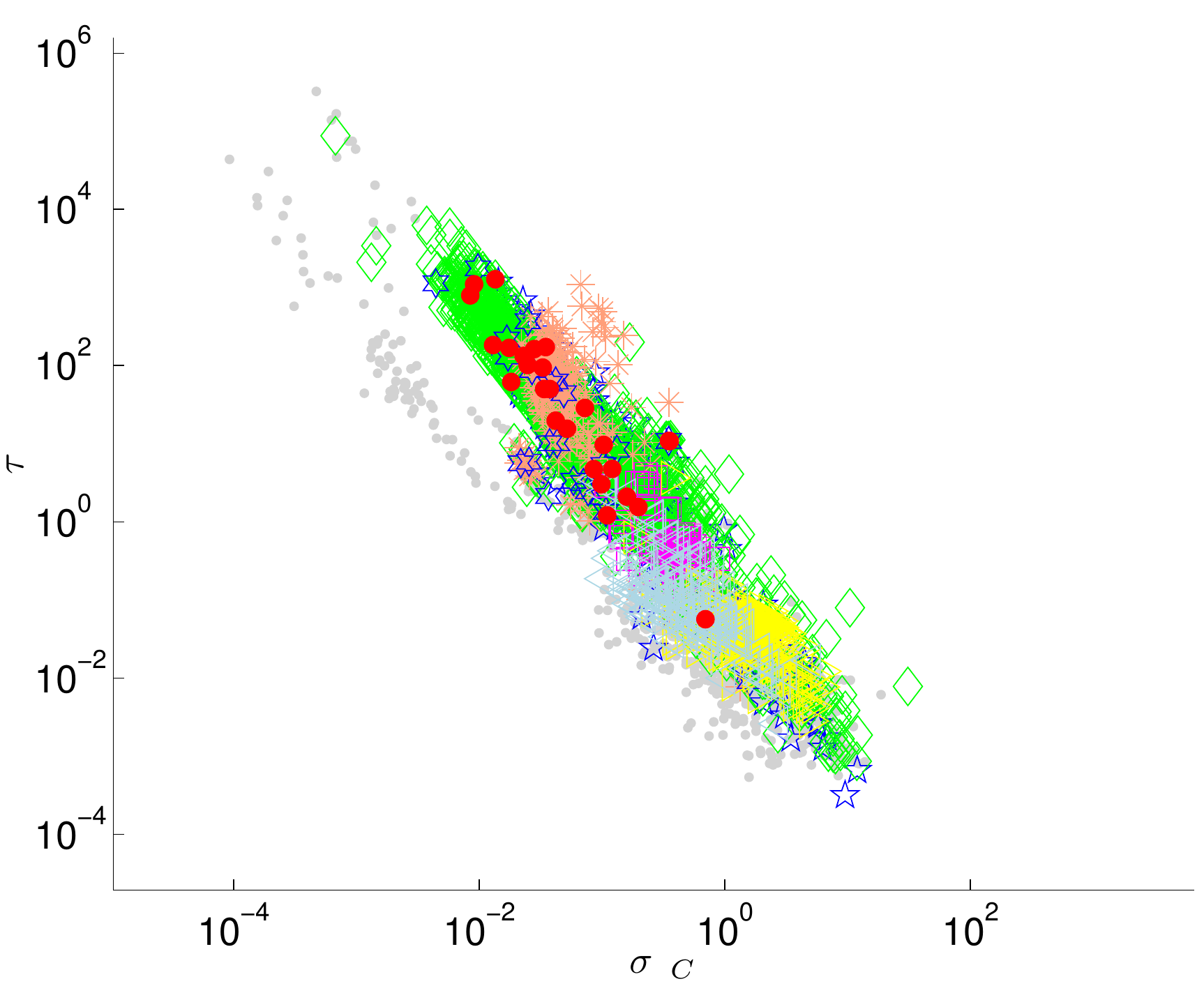}
  \end{minipage}
  \end{center}
   \caption{Predicted quasars and training stars distributions projected on different pairs of CAR(1) features for MACHO data.}
    \label{Fig:MACHO_Pred_Features_1}  
\end{figure*}

\begin{figure*}
  \begin{center}
 \fbox{\includegraphics[width=0.8\textwidth]{./Plots/MACHO/Symbols_MACHO_Pred-eps-converted-to.pdf}} 
  \begin{minipage}[b]{0.48\textwidth}
    \centering
    \includegraphics[width=7cm]{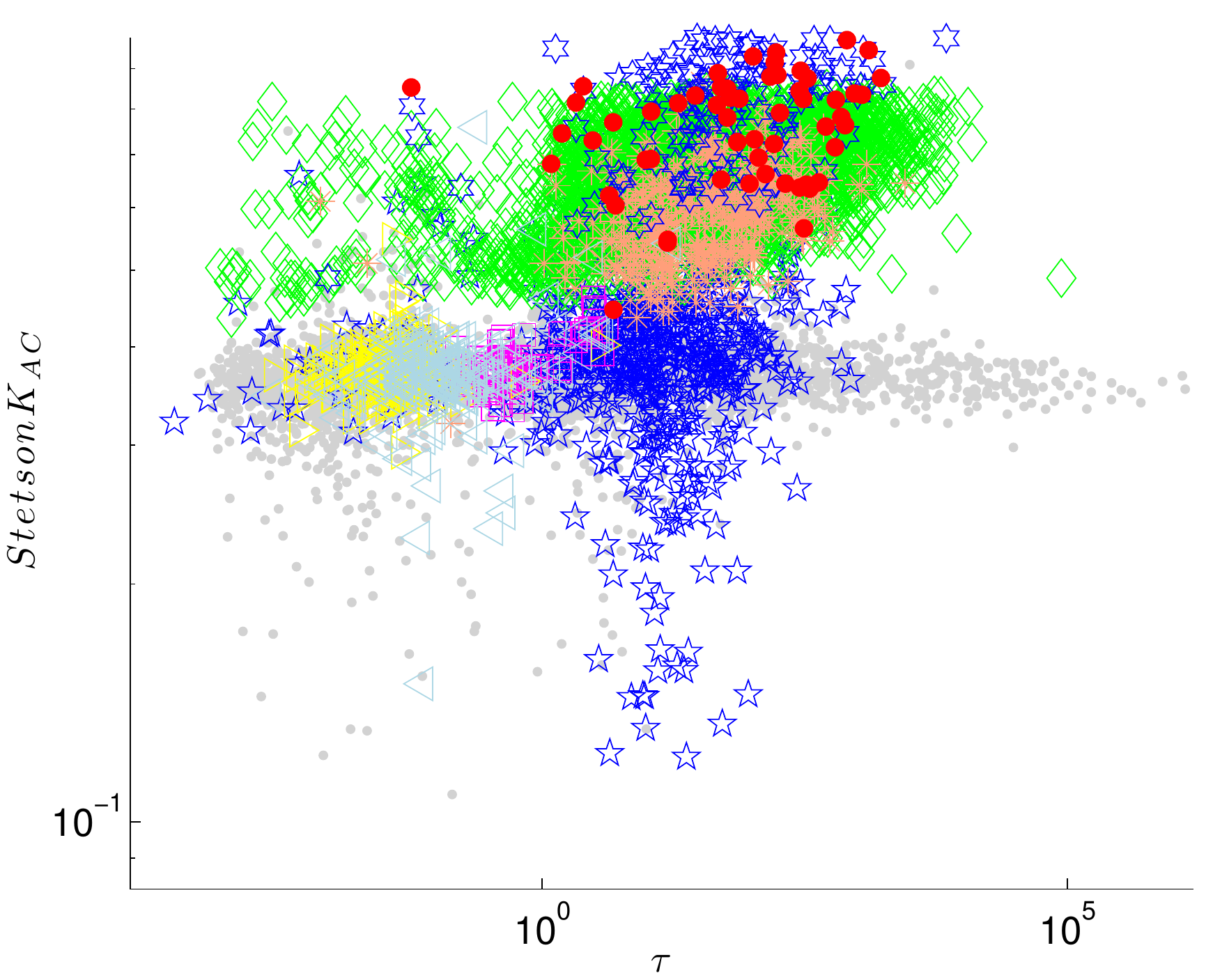}
  \end{minipage}
  \begin{minipage}[b]{0.48\textwidth}
    \centering
    \includegraphics[width=7cm]{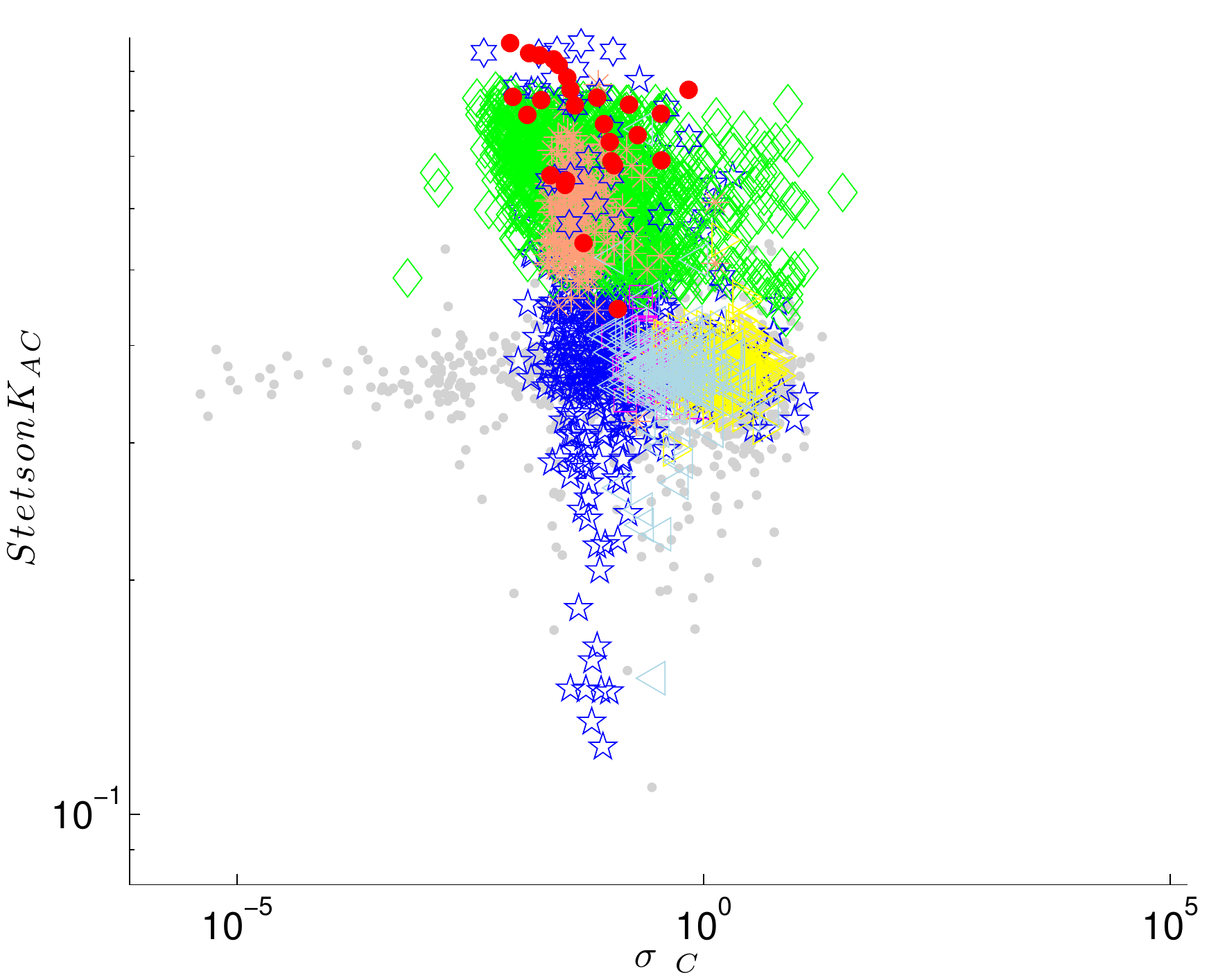}
  \end{minipage}  
  \begin{minipage}[b]{0.48\textwidth}
    \centering
    \includegraphics[width=7cm]{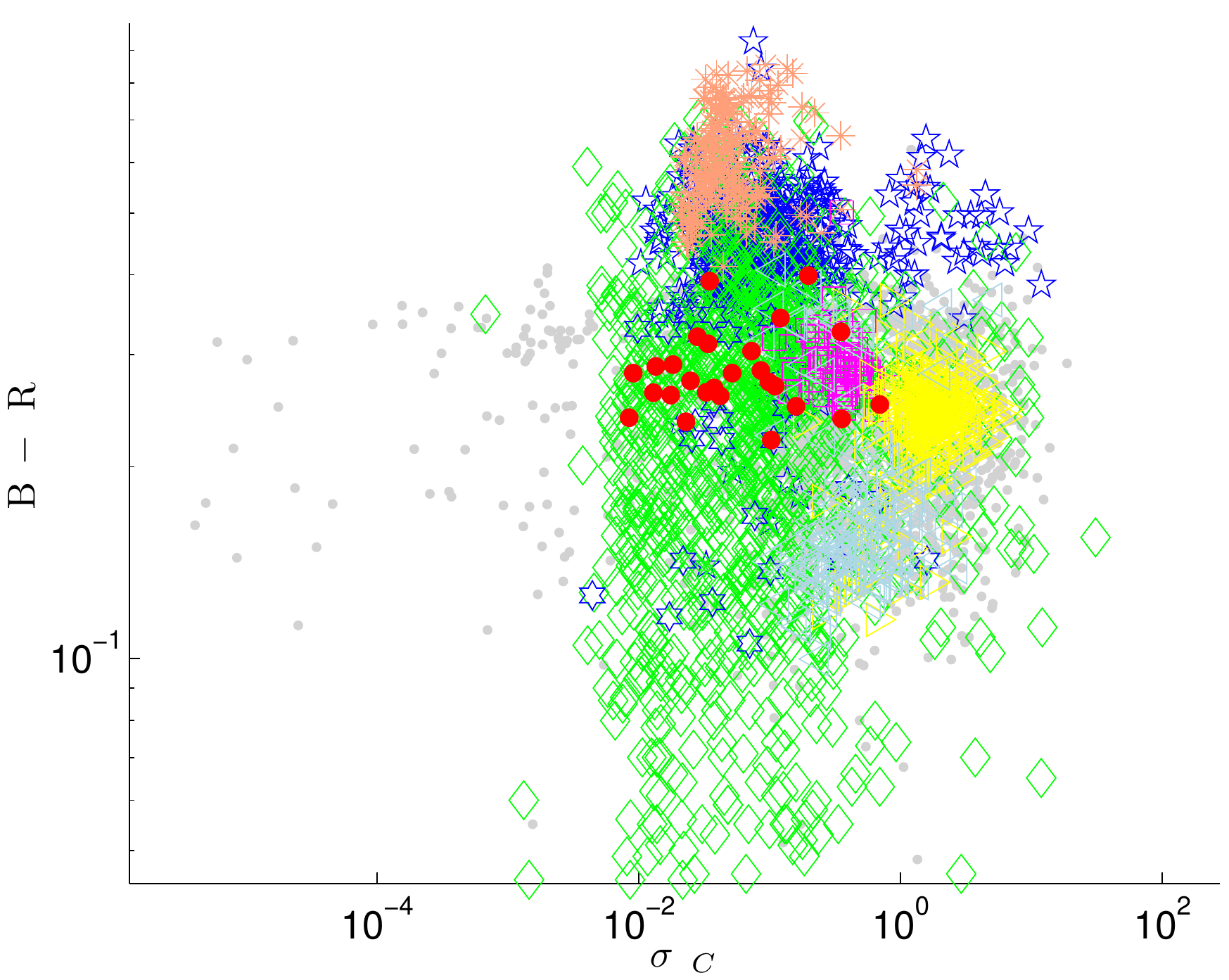}
  \end{minipage}
  \end{center}
   \caption{Predicted quasars and Training stars distributions for $\sigma_C$ and $\tau$ features combined with three time series features for MACHO data.}
    \label{Fig:MACHO_Pred_Features_2}  
\end{figure*}

 To analyze the distribution of predicted quasars in the feature space we show some projections of the training data plus the predicted quasars. Figures \ref{Fig:MACHO_Pred_Features_1} and \ref{Fig:MACHO_Pred_Features_2} show the distribution of predicted quasars , training quasars and all the other classes of stars. As in the EROS-2 case, we can see that in many cases the predicted quasars show similar distributions compared with training quasars. There are some cases where a big portion of the predicted quasars is expanded out of the concentrated cluster of training quasars, for example, combining $\sigma_C$ and B $-$ R 
  
\section{Summary}
\label{sec:summary}

 In this work we present a new list of candidate quasars from MACHO and EROS-2 datasets. This new list is obtained using a new model that uses continuous auto correlation features plus time series features to feed a boosted version of the Random Forest classifier \citep{Breiman:2001}. With this model we obtain a list of \NCanEROS{} candidates for the EROS-2 and \NCanMACHO{} candidates for the MACHO dataset.
From our MACHO candidates we crossmatch them with the old list of candidates from \citet{Kim2011ApJ} and we get \NMatchesMACHOKPDW{} matches. We crossmatch our EROS-2 candidates with the list of \NCanStrong{} MACHO strong candidates in \citet{Kim2011ApJ}. From that list, only \NunExistEROSinMACHODW{} objects exist in the EROS-2 dataset, and we find \NMatchesStrongMACHOEROS{} matches between our EROS-2 candidates and those \NunExistEROSinMACHODW{} objects (see table \ref{table:summary}). We prove that using boosted Random Forest with CAR(1) features we improve the fitting of the model to the training set in both EROS-2 and MACHO datasets. 


We show that quasars are well separated from many other kind of variable stars using CAR(1) features combined with time series features. We also proved that adding CAR(1) features, SVM, Random Forest and Boosted Random Forest improve their training accuracy. There are some challenges to overcome in future work such as the confusion of some periodic stars with quasars. We notice that about 25\% of false positives correspond to periodic stars. We believe that adding a dedicated module to filter periodic stars we can improve the results. 
 
 \begin{table*} 
 \begin{tabular}{|cccc|}
 \hline
     Previous Candidates   &     Previous Strong                            &      New list of MACHO        &   List of EROS-2\\
     MACHO  ($M_1$)        &     Candidates MACHO ($M_2$)     &      Candidates  ($M_3$)   &  Candidates  ($E_1$)\\
\hline     
      \NCanAll{}                    &      \NCanStrong{}                               &       \NCanMACHO{}           &   \NCanEROS{}\\
\hline
\hline 
      Matches between                     &      Matches between                           &   Objects from ($M_2$)                                      &    Matches between\\    
      ($M_3$) and ($M_1$)             &    ($M_3$) and ($M_2$)                      &    Catalogued in  EROS-2                                 &    ($ME$) and ($E_1$)\\     
\hline  
      \NMatchesMACHOKPDW{}    &     \NMatchesMACHOKPDWHC{}      &   \NunExistEROSinMACHODW{}  ($ME$)     &    \NMatchesStrongMACHOEROS{}\\
\hline      
\end{tabular}
\caption{Table summarizing crossmatching results between different lists of quasars candidates}
\label{table:summary}  
 \end{table*} 

\section*{Acknowledgments}

This paper utilizes public domain data obtained by the MACHO Project,
jointly funded by the US Department of Energy through the University
of California, Lawrence Livermore National Laboratory under contract
No. W-7405-Eng-48, by the National Science Foundation through the
Center for Particle Astrophysics of the University of California under
cooperative agreement AST-8809616, and by the Mount Stromlo and Siding
Spring Observatory, part of the Australian National University.
The analysis in this paper has been done using the \href{http://hptc.fas.harvard.edu/}{Odyssey cluster} 
supported by the FAS Research Computing Group at \href{http://harvard.edu/}{Harvard}.
This research has made use of the \href{http://simbad.u-strasbg.fr/simbad/}{SIMBAD} 
database, operated at CDS, Strasbourg, France.\\
We thank everyone from the EROS-2 collaboration for the
access granted to the database. The EROS-2 project was
funded by the CEA and the CNRS through the IN2P3 and INSU institutes.

\bsp
\vspace{0.1cm}

\bibliography{Pichara_QSO_2012_pp,KimQSO2011_A,KimQSO2011_B}

\begin{thebibliography}{51}
\expandafter\ifx\csname natexlab\endcsname\relax\def\natexlab#1{#1}\fi

\bibitem[{{Alcock}(1997{\natexlab{a}})}]{Alcock1997ApJL}
{Alcock}, C., e.~a. 1997{\natexlab{a}}, ApJ, 491, L11+

\bibitem[{{Alcock}(1997{\natexlab{b}})}]{Alcock1997ApJa}
---. 1997{\natexlab{b}}, ApJ, 479, 119

\bibitem[{{Alcock}(2000)}]{Alcock2000ApJ}
---. 2000, ApJ, 542, 281

\bibitem[{{Alcock}(2001)}]{Alcock2001}
---. 2001, Variable Stars in the Large Magellanic Clouds, VizieR Online Data
  Catalog (http://vizier.u-strasbg.fr/viz-bin/VizieR?-source=II/247)

\bibitem[{{Ansari}(1996)}]{1996VA.....40..519A}
{Ansari}, R. 1996, Vistas in Astronomy, 40, 519

\bibitem[{Belcher {et~al.}(1994)Belcher, Hampton, \& Wilson}]{Belcher:1994}
Belcher, J., Hampton, J.~S., \& Wilson, G.~T. 1994, Journal of the Royal
  Statistical Society. Series B (Methodological), 56, 141

\bibitem[{{Bloom} \& {Richards}(2011)}]{Bloom:2011}
{Bloom}, J.~S., \& {Richards}, J.~W. 2011, ArXiv e-prints

\bibitem[{{Bloom} {et~al.}(2011){Bloom}, {Richards}, {Nugent}, {Quimby},
  {Kasliwal}, {Starr}, {Poznanski}, {Ofek}, {Cenko}, {Butler}, {Kulkarni},
  {Gal-Yam}, \& {Law}}]{Bloom2:2011}
{Bloom}, J.~S., {Richards}, J.~W., {Nugent}, P.~E., {Quimby}, R.~M.,
  {Kasliwal}, M.~M., {Starr}, D.~L., {Poznanski}, D., {Ofek}, E.~O., {Cenko},
  S.~B., {Butler}, N.~R., {Kulkarni}, S.~R., {Gal-Yam}, A., \& {Law}, N. 2011,
  ArXiv e-prints

\bibitem[{{Bower} {et~al.}(2006){Bower}, {Benson}, {Malbon}, {Helly}, {Frenk},
  {Baugh}, {Cole}, \& {Lacey}}]{Bower2006MNRAS}
{Bower}, R.~G., {Benson}, A.~J., {Malbon}, R., {Helly}, J.~C., {Frenk}, C.~S.,
  {Baugh}, C.~M., {Cole}, S., \& {Lacey}, C.~G. 2006, MNRAS, 370, 645

\bibitem[{Breiman(1996)}]{Breiman:1996}
Breiman, L. 1996, in Machine Learning, 123--140

\bibitem[{Breiman(2001)}]{Breiman:2001}
Breiman, L. 2001, in Machine Learning, 5--32

\bibitem[{Brockwell \& Davis(2002)}]{Brockwell:2002}
Brockwell, P., \& Davis, R. 2002, Introduction to Time Series and Forecasting
  (Springer New York)

\bibitem[{Carliles {et~al.}(2010)Carliles, Budavri, Heinis, Priebe, \&
  Szalay}]{Carliles:2010}
Carliles, S., Budavri, T., Heinis, S., Priebe, C., \& Szalay, A. 2010, The
  Astrophysical Journal, 712, 511

\bibitem[{Cortes \& Vapnik(1995)}]{Cortes:1995}
Cortes, C., \& Vapnik, V. 1995, Machine Learning, 20, 273

\bibitem[{{Debosscher} {et~al.}(2007){Debosscher}, {Sarro}, {Aerts}, {Cuypers},
  {Vandenbussche}, {Garrido}, \& {Solano}}]{Debosscher:2007}
{Debosscher}, J., {Sarro}, L., {Aerts}, C., {Cuypers}, J., {Vandenbussche}, B.,
  {Garrido}, R., \& {Solano}, E. 2007, Astronomy and Astrophysics, 475, 1159

\bibitem[{{Derue} {et~al.}(2002){Derue}, {Marquette}, {Lupone}, {Afonso},
  {Alard}, {Albert}, {Amadon}, {Andersen}, {Ansari}, {Aubourg}, {Bareyre},
  {Bauer}, {Beaulieu}, {Blanc}, {Bouquet}, {Char}, {Charlot}, {Couchot},
  {Coutures}, {Ferlet}, {Fouqu{\'e}}, {Glicenstein}, {Goldman}, {Gould},
  {Graff}, {Gros}, {Ha{\i}ssinski}, {Hamilton}, {Hardin}, {de Kat}, {Kim},
  {Lasserre}, {Le Guillou}, {Lesquoy}, {Loup}, {Magneville}, {Mansoux},
  {Maurice}, {Milsztajn}, {Moniez}, {Palanque-Delabrouille}, {Perdereau},
  {Pr{\'e}vot}, {Regnault}, {Rich}, {Spiro}, {Vidal-Madjar}, {Vigroux},
  {Zylberajch}, \& {EROS Collaboration}}]{2002A&A...389..149D}
{Derue}, F., {Marquette}, J.-B., {Lupone}, S., {Afonso}, C., {Alard}, C.,
  {Albert}, J.-N., {Amadon}, A., {Andersen}, J., {Ansari}, R., {Aubourg},
  {\'E}., {Bareyre}, P., {Bauer}, F., {Beaulieu}, J.-P., {Blanc}, G.,
  {Bouquet}, A., {Char}, S., {Charlot}, X., {Couchot}, F., {Coutures}, C.,
  {Ferlet}, R., {Fouqu{\'e}}, P., {Glicenstein}, J.-F., {Goldman}, B., {Gould},
  A., {Graff}, D., {Gros}, M., {Ha{\i}ssinski}, J., {Hamilton}, J.-C.,
  {Hardin}, D., {de Kat}, J., {Kim}, A., {Lasserre}, T., {Le Guillou}, L.,
  {Lesquoy}, {\'E}., {Loup}, C., {Magneville}, C., {Mansoux}, B., {Maurice},
  {\'E}., {Milsztajn}, A., {Moniez}, M., {Palanque-Delabrouille}, N.,
  {Perdereau}, O., {Pr{\'e}vot}, L., {Regnault}, N., {Rich}, J., {Spiro}, M.,
  {Vidal-Madjar}, A., {Vigroux}, L., {Zylberajch}, S., \& {EROS Collaboration}.
  2002, Astronomy and Astrophysics, 389, 149

\bibitem[{Dietterich(1995)}]{Dietterich:1995}
Dietterich, T. 1995, ACM Computing Surveys, 27, 326

\bibitem[{Dietterich(2000)}]{Dietterich:2000}
Dietterich, T. 2000, in Proceedings of the First International Workshop on
  Multiple Classifier Systems (Springer Verlag), 1--15

\bibitem[{Duda \& Hart(1973)}]{Duda:Hart:1973}
Duda, R., \& Hart, P. 1973, Pattern Classification and Scene Analysis (John
  Willey \& Sons)

\bibitem[{{Ellaway}(1978)}]{Ellaway1978}
{Ellaway}, P. 1978, Electroencephalography and Clinical Neurophysiology, 45,
  302

\bibitem[{Freund \& Schapire(1997)}]{Freund:1997}
Freund, Y., \& Schapire, R. 1997, Journal of Computer and System Sciences

\bibitem[{{Hamadache}(2004)}]{2004HamadachePhD}
{Hamadache}, C. 2004, PhD thesis, Universit\'e Louis Pasteur - Strasbourg I

\bibitem[{{Heckman} {et~al.}(2004){Heckman}, {Kauffmann}, {Brinchmann},
  {Charlot}, {Tremonti}, \& {White}}]{Heckman2004ApJ}
{Heckman}, T.~M., {Kauffmann}, G., {Brinchmann}, J., {Charlot}, S., {Tremonti},
  C., \& {White}, S.~D.~M. 2004, ApJ, 613, 109

\bibitem[{Jordan(1994)}]{Jordan:1994}
Jordan, M.~I. 1994, Neural Computation, 6, 181

\bibitem[{{Kaiser} {et~al.}(2002){Kaiser}, {Aussel}, {Burke}, {Boesgaard},
  {Chambers}, {Chun}, {Heasley}, {Hodapp}, {Hunt}, {Jedicke}, {Jewitt},
  {Kudritzki}, {Luppino}, {Maberry}, {Magnier}, {Monet}, {Onaka}, {Pickles},
  {Rhoads}, {Simon}, {Szalay}, {Szapudi}, {Tholen}, {Tonry}, {Waterson}, \&
  {Wick}}]{Kaiser2002SPIE}
{Kaiser}, N., {Aussel}, H., {Burke}, B.~E., {Boesgaard}, H., {Chambers}, K.,
  {Chun}, M.~R., {Heasley}, J.~N., {Hodapp}, K.-W., {Hunt}, B., {Jedicke}, R.,
  {Jewitt}, D., {Kudritzki}, R., {Luppino}, G.~A., {Maberry}, M., {Magnier},
  E., {Monet}, D.~G., {Onaka}, P.~M., {Pickles}, A.~J., {Rhoads}, P.~H.~H.,
  {Simon}, T., {Szalay}, A., {Szapudi}, I., {Tholen}, D.~J., {Tonry}, J.~L.,
  {Waterson}, M., \& {Wick}, J. 2002, in Society of Photo-Optical
  Instrumentation Engineers (SPIE) Conference Series, Vol. 4836, Society of
  Photo-Optical Instrumentation Engineers (SPIE) Conference Series, ed.
  {J.~A.~Tyson \& S.~Wolff}, 154--164

\bibitem[{{Keller} {et~al.}(2002){Keller}, {Bessell}, {Cook}, {Geha}, \&
  {Syphers}}]{Keller2002AJ}
{Keller}, S.~C., {Bessell}, M.~S., {Cook}, K.~H., {Geha}, M., \& {Syphers}, D.
  2002, AJ, 124, 2039

\bibitem[{{Keller} {et~al.}(2007){Keller}, {Schmidt}, {Bessell}, {Conroy},
  {Francis}, {Granlund}, {Kowald}, {Oates}, {Martin-Jones}, {Preston},
  {Tisserand}, {Vaccarella}, \& {Waterson}}]{Keller:2007}
{Keller}, S.~C., {Schmidt}, B.~P., {Bessell}, M.~S., {Conroy}, P.~G.,
  {Francis}, P., {Granlund}, A., {Kowald}, E., {Oates}, A.~P., {Martin-Jones},
  T., {Preston}, T., {Tisserand}, P., {Vaccarella}, A., \& {Waterson}, M.~F.
  2007, Publications of the Astronomical Society of Australia, 24

\bibitem[{{Kelly} {et~al.}(2009){Kelly}, {Bechtold}, \&
  {Siemiginowska}}]{Kelly:2009}
{Kelly}, B.~C., {Bechtold}, J., \& {Siemiginowska}, A. 2009, 698, 895

\bibitem[{{Kim} {et~al.}(2011{\natexlab{a}}){Kim}, {Protopapas}, {Byun},
  {Alcock}, {Khardon}, \& {Trichas}}]{Kim:2011}
{Kim}, D.-W., {Protopapas}, P., {Byun}, Y.-I., {Alcock}, C., {Khardon}, R., \&
  {Trichas}, M. 2011{\natexlab{a}}, 735

\bibitem[{{Kim} {et~al.}(2011{\natexlab{b}}){Kim}, {Protopapas}, {Byun},
  {Alcock}, {Khardon}, \& {Trichas}}]{Kim2011ApJ}
---. 2011{\natexlab{b}}, ApJ, 735, 68

\bibitem[{{Kim} {et~al.}(2012){Kim}, {Protopapas}, {Trichas}, {Rowan-Robinson},
  {Khardon}, {Alcock}, \& {Byun}}]{Kim:2012}
{Kim}, D.-W., {Protopapas}, P., {Trichas}, M., {Rowan-Robinson}, M., {Khardon},
  R., {Alcock}, C., \& {Byun}, Y.-I. 2012, 747

\bibitem[{{Lomb}(1976)}]{Lomb1976ApSS}
{Lomb}, N.~R. 1976, Ap\&SS, 39, 447

\bibitem[{Matter(2007)}]{Matter:2007}
Matter, D. 2007, Science, 1

\bibitem[{{Metropolis} {et~al.}(1953){Metropolis}, {Rosenbluth}, {Rosenbluth},
  {Teller}, \& {Teller}}]{Metropolis:1953}
{Metropolis}, N., {Rosenbluth}, A., {Rosenbluth}, M., {Teller}, A., \&
  {Teller}, E. 1953, Journal of Chemical Physics, 21, 1087

\bibitem[{Nelder \& Mead(1965)}]{Nelder:1965}
Nelder, J., \& Mead, R. 1965, Computer Journal, 7, 308

\bibitem[{Plamondon \& Srihari(2000)}]{Plamondon:2000}
Plamondon, R., \& Srihari, S. 2000, Pattern Analysis and Machine Intelligence,
  IEEE Transactions on, 22, 63

\bibitem[{Quinlan(1993)}]{Quinlan:1993}
Quinlan, J. 1993, C4.5: programs for machine learning (Morgan Kaufmann
  Publishers Inc.)

\bibitem[{Rau {et~al.}(2009)Rau, Kulkarni, Law, Bloom, Ciardi, Djorgovski, Fox,
  Gal-Yam, Grillmair, Kasliwal, Nugent, Ofek, Quimby, Reach, Shara, Bildsten,
  Cenko, Drake, Filippenko, Helfand, Helou, Howell, Poznanski, \&
  Sullivan}]{Raw:2009}
Rau, A., Kulkarni, S.~R., Law, N.~M., Bloom, J.~S., Ciardi, D., Djorgovski,
  G.~S., Fox, D.~B., Gal-Yam, A., Grillmair, C.~C., Kasliwal, M.~M., Nugent,
  P.~E., Ofek, E.~O., Quimby, R.~M., Reach, W.~T., Shara, M., Bildsten, L.,
  Cenko, S.~B., Drake, A.~J., Filippenko, A.~V., Helfand, D.~J., Helou, G.,
  Howell, D.~A., Poznanski, D., \& Sullivan, M. 2009, Publications of the
  Astronomical Society of the Pacific, 121, 1334

\bibitem[{{Richards} {et~al.}(2011){Richards}, {Starr}, {Butler}, {Bloom},
  {Brewer}, {Crellin-Quick}, {Higgins}, {Kennedy}, \&
  {Rischard}}]{Richards:2011}
{Richards}, J.~W., {Starr}, D.~L., {Butler}, N.~R., {Bloom}, J.~S., {Brewer},
  J.~M., {Crellin-Quick}, A., {Higgins}, J., {Kennedy}, R., \& {Rischard}, M.
  2011, The Astrophysical Journal, 733

\bibitem[{Rumelhart {et~al.}(1986)Rumelhart, Hinton, \&
  Williams}]{Rumelhart:1986}
Rumelhart, D., Hinton, G., \& Williams, R. 1986, Learning internal
  representations by error propagation (MIT Press), 318--362

\bibitem[{{Scargle}(1982)}]{Scargle1982ApJ}
{Scargle}, J.~D. 1982, ApJ, 263, 835

\bibitem[{{Stetson}(1996)}]{Stetson1996PASP}
{Stetson}, P.~B. 1996, PASP, 108, 851

\bibitem[{{Thomas}(2005)}]{Thomas2005ApJ}
{Thomas}, C.~L., e.~a. 2005, ApJ, 631, 906

\bibitem[{{Tisserand} {et~al.}(2007){Tisserand}, {Le Guillou}, {Afonso},
  {Albert}, {Andersen}, {Ansari}, {Aubourg}, {Bareyre}, {Beaulieu}, {Charlot},
  {Coutures}, {Ferlet}, {Fouqu{\'e}}, {Glicenstein}, {Goldman}, {Gould},
  {Graff}, {Gros}, {Haissinski}, {Hamadache}, {de Kat}, {Lasserre}, {Lesquoy},
  {Loup}, {Magneville}, {Marquette}, {Maurice}, {Maury}, {Milsztajn}, {Moniez},
  {Palanque-Delabrouille}, {Perdereau}, {Rahal}, {Rich}, {Spiro},
  {Vidal-Madjar}, {Vigroux}, {Zylberajch}, \& {EROS-2
  Collaboration}}]{Tisserand:2007}
{Tisserand}, P., {Le Guillou}, L., {Afonso}, C., {Albert}, J.~N., {Andersen},
  J., {Ansari}, R., {Aubourg}, {\'E}., {Bareyre}, P., {Beaulieu}, J.~P.,
  {Charlot}, X., {Coutures}, C., {Ferlet}, R., {Fouqu{\'e}}, P., {Glicenstein},
  J.~F., {Goldman}, B., {Gould}, A., {Graff}, D., {Gros}, M., {Haissinski}, J.,
  {Hamadache}, C., {de Kat}, J., {Lasserre}, T., {Lesquoy}, {\'E}., {Loup}, C.,
  {Magneville}, C., {Marquette}, J.~B., {Maurice}, {\'E}., {Maury}, A.,
  {Milsztajn}, A., {Moniez}, M., {Palanque-Delabrouille}, N., {Perdereau}, O.,
  {Rahal}, Y.~R., {Rich}, J., {Spiro}, M., {Vidal-Madjar}, A., {Vigroux}, L.,
  {Zylberajch}, S., \& {EROS-2 Collaboration}. 2007, Astronomy and
  Astrophysics, 469, 387

\bibitem[{{Trichas} {et~al.}(2009){Trichas}, {Georgakakis}, {Rowan-Robinson},
  {Nandra}, {Clements}, \& {Vaccari}}]{Trichas2009MNRAS}
{Trichas}, M., {Georgakakis}, A., {Rowan-Robinson}, M., {Nandra}, K.,
  {Clements}, D., \& {Vaccari}, M. 2009, MNRAS, 399, 663

\bibitem[{{Trichas} {et~al.}(2010){Trichas}, {Rowan-Robinson}, {Georgakakis},
  {Valtchanov}, {Nandra}, {Farrah}, {Morrison}, {Clements}, \&
  {Waddington}}]{Trichas2010MNRAS}
{Trichas}, M., {Rowan-Robinson}, M., {Georgakakis}, A., {Valtchanov}, I.,
  {Nandra}, K., {Farrah}, D., {Morrison}, G., {Clements}, D., \& {Waddington},
  I. 2010, MNRAS, 405, 2243

\bibitem[{Wachman {et~al.}(2009)Wachman, Khardon, Protopapas, \&
  Alcock}]{Wachman:2009}
Wachman, G., Khardon, R., Protopapas, P., \& Alcock, C. 2009, in Lecture Notes
  in Computer Science, Vol. 5782, Machine Learning and Knowledge Discovery in
  Databases, ed. W.~Buntine, M.~Grobelnik, D.~Mladenic, \& J.~Shawe-Taylor
  (Springer Berlin / Heidelberg), 489--505

\bibitem[{Wang {et~al.}(2010)Wang, Khardon, \& Protopapas}]{Wang:2010}
Wang, Y., Khardon, R., \& Protopapas, P. 2010, in Lecture Notes in Computer
  Science, Vol. 6323, Machine Learning and Knowledge Discovery in Databases
  (Springer Berlin / Heidelberg), 418--434

\bibitem[{{Wood}(2000)}]{Wood2000PASA}
{Wood}, P.~R. 2000, Publications of the Astronomical Society of Australia, 17,
  18

\bibitem[{Xu {et~al.}(1992)Xu, Krzyzak, \& Suen}]{Suen:1992}
Xu, L., Krzyzak, A., \& Suen, C. 1992, Systems, Man and Cybernetics, IEEE
  Transactions, 22, 418

\bibitem[{Zhao {et~al.}(2003)Zhao, Chellappa, Phillips, \&
  Rosenfeld}]{Zhao:2003}
Zhao, W., Chellappa, R., Phillips, P.~J., \& Rosenfeld, A. 2003, ACM Comput.
  Surv., 35, 399

\end{thebibliography}

\label{lastpage}
\end{document}